\documentclass[aps,prl,twocolumn,groupedaddress]{revtex4-1}
\usepackage{amsmath}
\usepackage{color}
\usepackage{commath}
\usepackage{graphicx}
\usepackage{hyperref}
\usepackage[capitalise]{cleveref}
\DeclareMathOperator{\tr}{tr}
\DeclareMathOperator{\Ret}{Re}
\DeclareMathOperator{\Imt}{Im}

\newcommand{\delnospace}[1]{\mathopen{}\del{#1}}
\newcommand{\hE}{\hat{\mathcal{E}}}
\begin{document}
\let\oldaddtocontents\addtocontents
\renewcommand{\addtocontents}[2]{}

\title{Controlled-phase gate for photons based on stationary light}

\author{Ivan Iakoupov$^{1}$}
\author{Johannes Borregaard$^{1,2}$}
\author{Anders S. S{\o}rensen$^{1}$}
\affiliation{$^1$ The Niels Bohr Institute, University of Copenhagen, Blegdamsvej 17, DK-2100 Copenhagen \O, Denmark\\
$^2$ Department of Physics, Harvard University, Cambridge, MA 02138, USA}

\date{\today}

\begin{abstract}
We propose a method to induce strong effective interactions between photons 
mediated by an atomic ensemble. To achieve this, we use the so-called 
stationary light effect to enhance the interaction. 
Regardless of the single-atom coupling to light, the interaction strength 
between the photons 
can be enhanced by increasing the total number of atoms. 
For sufficiently many atoms, the setup can be viable as a controlled-phase 
gate for photons. We derive analytical expressions for the fidelities for two 
modes of gate operation: deterministic and heralded conditioned on the 
presence of two photons at the output.
\end{abstract}


\maketitle

Optical photons are ideal carriers of quantum information over long distances, 
and such quantum communication may enable a wealth of applications 
\cite{kimble_nature08}. Quantum information processing with photonic qubits 
is, however, severely limited by the lack of efficient two-qubit gates. In 
principle, such gates could be realized by strongly coupling photons to a 
single atom~\cite{duan_kimble_2004,chang_naturephys07}. Experiments have 
pushed towards realizing such strong coupling, e.g. in cavity QED
structures~\cite{kim_nphoton13,chen_science13,tiecke_nature14,hacker_nature16,
volz_nphoton2014} and optical
waveguides~\cite{akimov_nature07,claudon_nphoton10,goban_ncomms2014,sollner2015},  
but the realization of two-qubit gates remains challenging. For some 
applications, it is possible to use atomic ensembles where a large number of 
atoms compensates for a weak single-atom coupling 
strength~\cite{hammerer_rmp10}. However, this approach typically does not 
enhance the nonlinear interactions required for quantum gates. Gate operation 
is often pursued by extending the ensemble approach with strong dipole-dipole 
interactions of the atomic Rydberg levels~\cite{saffman_rmp10, roy_rmp17, 
gorshkov_prl11, khazali_pra2015, rydberg-cavity-gate, murray_prx17}. In recent 
years, experiments in e.g. tapered optical
fibers~\cite{le_kien_pra04,vetsch_prl10,goban_prl12,beguin_prl14,gouraud_prl15}
and hollow core photonic-crystal fibers~\cite{bajcsy_prl09,blatt_pra16} have 
realized an intermediate regime where the single-atom coupling to light is 
sizeable, but still not sufficient to realize photonic gates based on single 
atoms. It remains an open question to which degree such moderate couplings 
enable processing of quantum information.

In this letter, we propose a controlled-phase gate that works even if the 
individual atoms are not coupled strongly either to light (e.g. optical 
cavities) or to each other (e.g. Rydberg interactions). We show that by using 
sufficiently many atoms, it is possible to compensate for the limited 
single-atom coupling to light and achieve ideal gate 
operation~\cite{hafezi_pra12a}. The main physical mechanism behind the gate is 
{\it stationary light}~\cite{andre_prl02a,bajcsy_nature03a} where polaritons 
(coupled light-matter excitations) have very low group velocity due to 
counter-propagating classical drives. These polaritions experience reflections 
at the ends of the ensemble. This leads to transmission resonances whenever 
the polaritons form standing waves inside the ensemble~\cite{hafezi_pra12a}, 
akin to an optical cavity. We show that the storage of a single photon 
completely changes the scattering properties of the ensemble because the 
cavity-like structure created by the remaining atoms enhances the interaction 
with the stored excitation. This can be used to mediate a gate between photons 
that can be either deterministic or heralded (successful operation is 
conditioned on subsequent detection of two photons).

\begin{figure}[t]
\begin{center}
\includegraphics{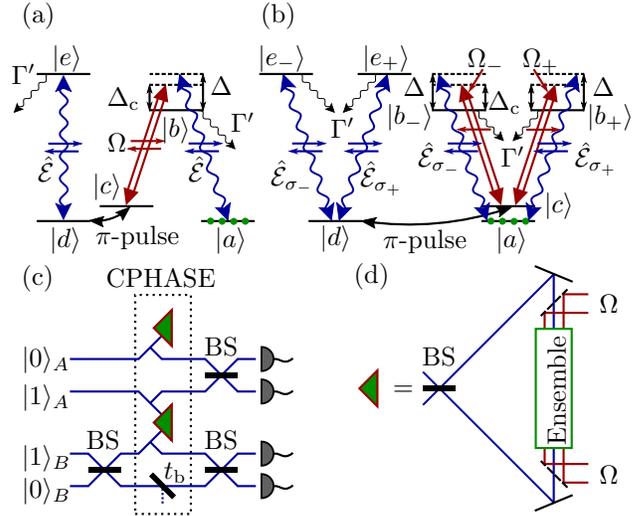}
\end{center}
\caption{(Color online) (a) Level diagram of $\Lambda$\mbox{-}type atoms 
(levels $|a\rangle$, $|b\rangle$, and $|c\rangle$) that can be switched to 
two-level atoms (levels $|d\rangle$ and $|e\rangle$) by the storage of a photon 
followed by a $\pi$\mbox{-}pulse. Green dots indicate the initial state of the 
atoms. (b) Level diagram of dual\mbox{-}V atoms that can be switched to 
V\mbox{-}type atoms. (c) Dual-rail Bell-state measurement setup with the 
controlled-phase gate (CPHASE) as a part of it. An ensemble of 
atoms is placed inside a Sagnac interferometer, shown as a triangle in (c) and 
defined by (d). In the rail corresponding to state $|0\rangle_B$, a beam 
splitter is added with transmission coefficient $t_\text{b}$. All the other 
beam splitters (BS) are~50:50.
\label{fig_setup}}
\end{figure}

\emph{Overview.}
We consider two different level schemes for the atoms in the ensemble: 
$\Lambda$\mbox{-}type and dual\mbox{-}V (Figs. \ref{fig_setup}(a) and 
\ref{fig_setup}(b), respectively). The linear properties of these two schemes 
are described in detail in Ref.~\cite{dispersion-relations}. In the 
$\Lambda$\mbox{-}type scheme, two counter-propagating classical drives have 
the same polarization and frequency. This results in a standing wave of the 
Rabi frequency $\Omega(z)=\Omega_0\cos(k_0 z)$, where $k_0$ is the wave vector 
of the classical drive, assumed to be the same as the wave vector of the probe 
field $\hat{\mathcal{E}}$ (single photon). For the $\Lambda$\mbox{-}type 
scheme, we assume that $N$ atoms are placed at positions $z_j=j\pi/(2k_0)$ 
with $0\leq j\leq N-1$ to achieve the lowest possible group velocity 
(increasing the interatomic distance by integer multiples of $\pi/k_0$ does 
not change the results)~\cite{dispersion-relations}. Low group velocity can also be 
achieved by separating the two counter-propagating classical drives
either in polarization~\cite{zimmer_pra08} or 
frequency~\cite{moiseev_pra2006}. We choose the separation in polarization, i.e. 
the dual\mbox{-}V scheme, but separation in frequency is 
expected to yield similar results~\cite{dispersion-relations}. From a 
practical perspective, the dual\mbox{-}V scheme is desirable since it does not 
require careful placement of the atoms. However, we focus 
on the $\Lambda$\mbox{-}type scheme in the analysis below, since it admits an 
approximate analytical solution. We also perform the numerical analysis for both of the 
schemes and show that the analytical results obtained for the $\Lambda$\mbox{-}type 
scheme provide the correct scaling for the dual\mbox{-}V scheme.

The single-atom coupling to light is characterized by the parameter 
$\Gamma_\text{1D}/\Gamma$ (half of the resonant optical depth per atom), where 
$\Gamma_\text{1D}$ is the decay rate from each of the states $|b\rangle$ and 
$|e\rangle$ (see Fig.~\ref{fig_setup}(a)) into both right-moving and 
left-moving guided modes (assumed to be equal), $\Gamma'$ is the decay rate 
into all the other modes, and $\Gamma=\Gamma_\text{1D}+\Gamma'$ is the total 
decay rate. In the dual-rail 
encoding of photonic qubits shown in Fig.~\ref{fig_setup}(c), two identical 
atomic ensembles are required, where the upper one only functions as a memory. 
Alternatively, the single-rail encoding can also be implemented with one 
atomic ensemble~\cite{rydberg-cavity-gate}, but the dual-rail encoding allows 
heralded operation that has better fidelity. Each ensemble is placed inside 
a Sagnac interferometer (Fig.~\ref{fig_setup}(d)).

The operation of the CP gate is sequential. First, photon~$A$ is stored either 
in the upper ($|0\rangle_A$) or the lower ($|1\rangle_A$) ensemble using 
electromagnetically induced transparency (EIT)~\cite{gorshkov_pra07_2}. Then 
photon~$B$ is scattered from the lower ensemble under stationary light 
conditions ($|1\rangle_B$) or passes through a beam splitter with transmission 
coefficient $t_\text{b}$ ($|0\rangle_B$). The role of this beam splitter will 
be explained below. The Sagnac interferometer can be set up such that most of 
the incident power in each of its two input ports is reflected back through 
the same port, regardless of whether the ensemble is reflective or
transmissive~\cite{bertocchi_jpb2006,bradford_prl2012}. Reflection or 
transmission of the ensemble instead controls the phase of the reflected 
field. The scattering of photon~$B$ can be arranged such that if there is no 
stored photon in the lower ensemble (photon~$A$ is in the state $|0\rangle_A$), 
the atomic ensemble is completely transmissive in the ideal case, and photon~$B$
is reflected from the Sagnac interferometer with no additional phase. If 
there is a stored photon (photon~$A$ is in state $|1\rangle_A$), photon~$B$ is 
reflected from the interferometer with a $\pi$ phase shift. The latter case 
performs the desired controlled-phase gate operation 
$|11\rangle_{AB}\rightarrow-|11\rangle_{AB}$, while the rest of the basis 
states are unchanged. Finally, photon~$A$ is retrieved using EIT.

\emph{Storage and retrieval.}
Before the EIT storage, all atoms are initialized in state $|a\rangle$, and 
after storage, the incident photon is mapped onto an atom being in 
state~$|c\rangle$. To produce an optical non-linearity, we assume that 
state~$|c\rangle$ is subsequently transferred to state~$|d\rangle$ using a
$\pi$\mbox{-}pulse.
Under EIT storage and retrieval, both the incident photon and the classical 
drive are assumed resonant with the respective atomic transitions for simplicity.
The classical drive is incident from one side only. 
Entering the Sagnac interferometer, photon~$A$ 
is split into two parts by the 50:50 beam splitter (see 
Fig.~\ref{fig_setup}(d)). The two parts reach the ensemble from the opposite 
sides with opposite spatial phase factors $e^{ik_0 z}$ and $e^{-ik_0 z}$. 
Inside the ensemble, the two parts will interfere, resulting in a stored spin 
wave with $\cos(k_0 z)$ spatial modulation. Such storage procedure is necessary (for the 
$\Lambda$\mbox{-}type scheme only), since the part of the excitation that is stored on the 
nodes of the standing wave of the classical drive (that is applied during 
scattering of photon~$B$) does not change the scattering properties of the 
ensemble.

\begin{figure}[t]
\begin{center}
\includegraphics{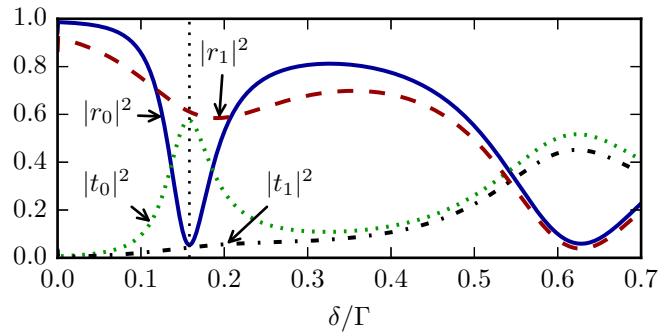}
\end{center}
\caption{(Color online) (a) Reflectances ($|r_0|^2$, $|r_1|^2$) and 
transmittances ($|t_0|^2$, $|t_1|^2$) of an ensemble of $\Lambda$\mbox{-}type 
atoms without ($|r_0|^2$, $|t_0|^2$) and with ($|r_1|^2$, $|t_1|^2$) a stored 
photon for different frequencies (two-photon detunings)~$\delta$. The vertical 
dotted line marks the operation point $\delta_\text{res}$. The parameters are: 
$N=10^4$, $\Gamma_\text{1D}/\Gamma = 0.05$, $\Delta_\text{c}/\Gamma=-10$, and 
$\Omega_0/\Gamma=10$.
}
\label{fig_r_and_t_plot}
\end{figure}

\emph{Reflection and transmission.}
We use the (multi-mode) transfer matrix formalism 
\cite{dispersion-relations,deutsch_pra95a} to model the scattering process. To 
illustrate the scattering behavior, we assume that photon~$A$ was stored in 
the center of the atomic ensemble at an anti-node of the classical drive.  The 
reflectances and transmittances of an ensemble of $\Lambda$\mbox{-}type atoms 
are plotted in Fig.~\ref{fig_r_and_t_plot} as functions of the two-photon detuning 
$\delta=\Delta-\Delta_\text{c}$, where $\Delta$ ($\Delta_\text{c}$) is the 
detuning of the probe field (classical drive). The reflectance $|r_0|^2$ 
($|r_1|^2$) and transmittance $|t_0|^2$ ($|t_1|^2$) are for an ensemble without 
(with) a stored photon. The ensemble is seen to have transmittance resonances 
with a large $|t_0|^2$ and a small $|r_0|^2$. These resonances occur when the standing wave 
condition is fulfilled, i.e. $\sin(qL)=0$, where $q$ is the Bloch vector of 
the stationary light polaritons and $L$ is the length of the 
ensemble~\cite{hafezi_pra12a,dispersion-relations}. When a photon is stored in 
the ensemble, an atom changes from state $|a\rangle$ to $|d\rangle$. In state 
$|d\rangle$, the atom acts as a two-level atom that is resonant with the 
incident photon (see Fig.~\ref{fig_setup}(a)). Since 
the effective interaction is enhanced by the cavity-like behavior of the 
ensemble, this single two-level atom can make the entire ensemble become 
reflective instead of transmissive.

We focus on the behavior at the resonance nearest ${\delta=0}$ (vertical 
dotted line in Fig.~\ref{fig_r_and_t_plot}). In the limit of large atom number 
$N$ and for $|\Delta_\text{c}|\neq 0$, this resonance is at a two-photon 
detuning
${\delta_\text{res}\approx 
-4\pi^2\Delta_\text{c}|\Omega_0|^2/(\Gamma_\text{1D}^2N^2)}$
for which we obtain~\cite{sm}
\begin{align}
\label{r_ensemble_N}
&\begin{aligned}
&r_0\approx\frac{\Gamma_\text{1D}\Gamma'N}{16\Delta_\text{c}^2},
\end{aligned}
\displaybreak[0]\\
\label{t_ensemble_N}
&\begin{aligned}
&t_0\approx 1-r_0,
\end{aligned}
\displaybreak[0]\\
\label{r_imp_ensemble_N}
&\begin{aligned}
&r_1(\tilde{z})\approx
1
-\frac{4\pi^2\Delta_\text{c}^2\Gamma'}{\Gamma_\text{1D}^3 N^2}
-\frac{4i\pi^2\Delta_\text{c}}{\Gamma_\text{1D} N}
\left(\tilde{z}-\frac{1}{2}\right)\\
&-\frac{4\pi^4\Delta_\text{c}^2(2\Gamma_\text{1D}+\Gamma')}{\Gamma_\text{1D}^3 N^2}
\left(\tilde{z}-\frac{1}{2}\right)^2,
\end{aligned}
\displaybreak[0]\\
\label{t_imp_ensemble_N}
&\begin{aligned}
&t_1(\tilde{z})\approx 
\frac{4\pi^2\Delta_\text{c}^2\Gamma'}{\Gamma_\text{1D}^3 N^2}
+\frac{8\pi^4\Delta_\text{c}^2\Gamma'}{\Gamma_\text{1D}^3 N^3}
\left(\tilde{z}-\frac{1}{2}\right)\\
&+\frac{4\pi^4\Delta_\text{c}^2\Gamma'}{\Gamma_\text{1D}^3 N^2}
\left(\tilde{z}-\frac{1}{2}\right)^2.
\end{aligned}
\end{align}
Here, $t_1$ and $r_1$ were obtained by solving the discrete problem where a 
photon is stored in a single discrete atom, and then taking the continuum limit 
such that the index of the atom is replaced by its position inside the 
ensemble $\tilde{z}=z/L$. By aligning the interferometer, the reflection 
coefficients of the combined interferometer-ensemble system are given by 
$R_0=-(r_0-t_0)$ and $R_1(\tilde{z})=-(r_1(\tilde{z})-t_1(\tilde{z}))$~\cite{sm}. 
If we take $\tilde{z}=1/2$ and a detuning 
$|\Delta_\text{c}|\sim\Gamma_\text{1D}N^{3/4}$, we have 
$r_0, t_1 \sim \Gamma'/(\Gamma_\text{1D} \sqrt{N})$, $r_1\approx 1-t_1$, and 
$t_0 \approx 1-r_0$. Hence, 
even for small $\Gamma_\text{1D}/\Gamma'$, we can achieve an ideal 
CP gate ($R_0=1$, $R_1=-1$) with sufficiently many atoms.

\emph{Fidelity.} To quantify the errors of the gate, we calculate the 
Choi-Jamiolkowski (CJ) 
fidelity~\cite{rydberg-cavity-gate,real-and-ideal-quantum-processes}.  The EIT 
storage is described using the storage $K_\text{s}$ and retrieval $K_\text{r}$ 
kernels derived in Ref.~\cite{gorshkov_pra07_2} (suitably modified to take 
into account storage from both directions~\cite{sm}). When photon~$A$ is 
stored and retrieved without scattering of photon~$B$, the output wave 
function of photon~$A$ is
$\phi_{A,\text{out},0}(t)=\iint 
K_\text{r}(\tilde{z},t)K_\text{s}(\tilde{z},t')\phi_{A,\text{in}}(t')
\dif t'\dif \tilde{z}$, where $\phi_{A,\text{in}}$ 
is the input wave function. The efficiency of the storage 
and retrieval is $\eta_\text{EIT}=\int |\phi_{A,\text{out},0}(t)|^2\dif t$. 
If photon~$B$ was reflected from the interferometer while photon~$A$ was 
stored in the ensemble (state $|11\rangle_{AB}$), the output wave 
function is instead
$\phi_{A,\text{out},1}(t)=\iint 
K_\text{r}(\tilde{z},t)R_{1}(\tilde{z})
K_\text{s}(\tilde{z},t')\phi_{A,\text{in}}(t')
\dif t'\dif \tilde{z}$.
Neglecting bandwidth effects of photon~$B$ to find the upper limit set by the 
atomic ensemble, we obtain the CJ 
fidelity~\cite{real-and-ideal-quantum-processes,
iakoupov_phd_thesis,fidelity}
\begin{gather}
\label{F_CJ_general}
F_\text{CJ}=\frac{\eta_\text{EIT}}{16}\envert{2t_\text{b}+R_0-R_{1,1}}^2,
\end{gather}
where
$R_{1,1}=(1/\eta_\text{EIT})
\int \phi_{A,\text{out},0}^*(t)\phi_{A,\text{out},1}(t)\dif t$.
If the gate is conditioned on the presence of two photons after the gate 
operation~\cite{rydberg-cavity-gate,duan_pra05,borregaard_prl15}, we find 
the success probability
\begin{gather}
\label{P_suc_general}
P_\text{suc}=\frac{\eta_\text{EIT}}{4}\del{2|t_\text{b}|^2+|R_0|^2
+R_{1,2}},
\end{gather}
with
$R_{1,2}=(1/\eta_\text{EIT})\int |\phi_{A,\text{out},1}(t)|^2\dif t$.
The conditional CJ fidelity is $F_\text{CJ,cond}=F_\text{CJ}/P_\text{suc}$.  

\begin{figure}[t]
\begin{center}
\includegraphics{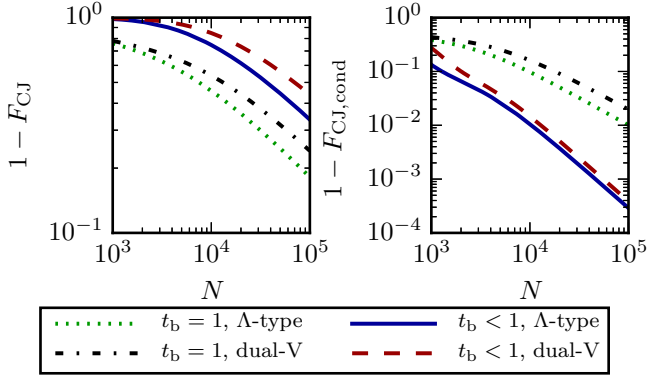}
\end{center}
\caption{(Color online) 
Numerically calculated $F_\text{CJ}$ and $F_\text{CJ,cond}$. For the 
$t_\text{b}<1$ curves, $t_\text{b}$ is
optimized numerically such that $F_\text{CJ,cond}$ is maximal. The dual\mbox{-}V scheme uses 
regular interatomic distance ${d=0.266\pi/k_0}$. 
The common parameters are ${\Gamma_\text{1D}/\Gamma = 0.05}$, and 
${\Omega_0/\Gamma=1}$. Under EIT (storage and retrieval), $\Omega(z)=\Omega_0$. 
Under stationary light (scattering), $\Omega(z)=\Omega_0\cos(k_0 z)$ and 
$\Omega_\pm(z)=\Omega_0e^{\pm ik_0 z}$ for $\Lambda$\mbox{-}type and 
dual\mbox{-}V respectively.}
\label{fig_cphase_fidelity_NAtoms}
\end{figure}

To optimize the performance of the gate, we set $t_\text{b}=1$ and optimize 
$\Delta_\text{c}$ and the width of the stored spin wave 
$\tilde{\sigma}=\sigma/L$ such that $F_\text{CJ}$ is maximal. In 
Fig.~\ref{fig_cphase_fidelity_NAtoms} we plot the numerically calculated 
$F_\text{CJ}\approx P_\text{suc}$ and $F_\text{CJ,cond}$, where photon~$A$ was 
chosen to have a Gaussian
temporal profile, and photon~$B$ is centered on $\delta=\delta_\text{res}$ and 
assumed to be narrow in frequency compared to the resonance width. As seen in 
the figure, both $F_\text{CJ}$ and $F_\text{CJ,cond}$ approach their ideal 
value of unity for large $N$, but $F_\text{CJ,cond}$ approaches it much 
faster.

For large $N$, we can find analytical expressions for the $\Lambda$\mbox{-}type 
curves in Fig.~\ref{fig_cphase_fidelity_NAtoms} if we neglect distortions of 
photon~$A$ under storage and retrieval, but still account for the errors due 
to the spatial extent of the stored excitation. The stored spin wave is
approximately Gaussian of the form
$S(\tilde{z})=(2\pi\tilde{\sigma}^2)^{-1/4}
\exp\delnospace{-(\tilde{z}-1/2)^2/(4\tilde{\sigma}^2)}$.
Consequently,
$\eta_\text{EIT}\approx
1-\Gamma'/(2N\Gamma_\text{1D}\tilde{\sigma}^2)$~\cite{chang_njp11a,sm},
$R_{1,1}\approx \int R_{1,\text{s}}(\tilde{z})|S(\tilde{z})|^2\dif
\tilde{z}$, and
$R_{1,2}\approx\int |R_{1,\text{s}}(\tilde{z})|^2|S(\tilde{z})|^2\dif
\tilde{z}$. Here, $R_{1,\text{s}}(\tilde{z})=(R_1(\tilde{z})+R_1(1-\tilde{z}))/2$ 
is the symmetrized version of $R_{1}$ that accounts for storage and scattering 
from both sides of the ensemble.

For fixed $\Gamma_\text{1D}$ and large $N$, after choosing 
$\tilde{\sigma}^2=1/(\pi^{3/2}N^{1/4})
\sqrt{\Gamma'/(\Gamma_\text{1D}+\Gamma')}$,
$\Delta_\text{c}^2=(\Gamma_\text{1D}^2N^{3/2})/(8\pi)$,
and
${t_\text{b}=1}$, 
$F_\text{CJ}$ is maximal, and
\begin{align}
\label{F_CJ_1}
&F_{\text{CJ},t_\text{b}=1}
\approx P_{\text{suc},t_\text{b}=1}
\approx 1-\frac{\pi\Gamma'}{\Gamma_\text{1D}\sqrt{N}},\\
\label{F_CJ_cond_1}
&F_{\text{CJ,cond},t_\text{b}=1}
\approx 1-\frac{\pi ^2 \Gamma'^2}{4 \Gamma_\text{1D}^2 N}.
\end{align}
These expressions confirm that the gate fidelity improves with $N$ and that 
the conditional fidelity has better scaling.

With $t_\text{b}=1$, the losses are different for the different computational 
basis states. By setting $t_\text{b}\approx R_0$ we approximately equalize the losses, 
leading to a substantial improvement of $F_\text{CJ,cond}$ at the cost of 
increasing $1-P_\text{suc}$ by a constant factor. Whether this is a desirable trade off, 
depends on the particular application. Taking the same values of 
$\Delta_\text{c}$~and~$\tilde{\sigma}$,
\begin{align}
\label{F_CJ_2}
&F_{\text{CJ},t_\text{b}=R_0}
\approx P_{\text{suc},t_\text{b}=R_0}
\approx 1-\frac{2\pi\Gamma'}{\Gamma_\text{1D}\sqrt{N}},\\
\label{F_CJ_cond_2}
&F_{\text{CJ,cond},t_\text{b}=R_0}
\approx 1-\frac{11\pi^3\left(\Gamma_\text{1D}+\Gamma'\right)\Gamma'}
{16 \Gamma_\text{1D}^2 N^{3/2}}.
\end{align}
Here, $1-F_\text{CJ,cond}$ is limited by the non-zero $\tilde{\sigma}$.

Numerical simulations suggest that the fidelities are independent of $\Omega_0$ 
over a wide range of values. E.g. for $\Gamma_\text{1D}/\Gamma=0.05$ and 
$N=10^4$, $|\Omega_0|$ up to $30\Gamma$ with a negligible change in the 
optimal $\Delta_\text{c}$, and to at least $100\Gamma$ with some increase in 
the optimal $\Delta_\text{c}$~\cite{sm}.

\emph{Dual-V scheme.} First, some technical differences from the $\Lambda$\mbox{-}type 
scheme. The decay rate $\Gamma_\text{1D}$ is from each of the states 
$|b_\pm\rangle$ and $|e_\pm\rangle$  (see Fig.~\ref{fig_setup}(b)). When 
switched to state $|d\rangle$, the atom becomes a resonant V\mbox{-}type atom. For 
storage and retrieval, the $\Lambda$\mbox{-}type and dual\mbox{-}V schemes behave the same, 
since only one classical drive is incident. For the numerical calculation of 
the fidelities for the dual\mbox{-}V scheme in 
Fig.~\ref{fig_cphase_fidelity_NAtoms}, the distance $d$ between the atoms was 
set to be incommensurate with the wavelength of the classical drive, 
$d=0.266\pi/k_0$. The results are, however, almost independent of $d$, and the 
gate can function even with completely random placement of the 
atoms~\cite{sm}. The dual\mbox{-}V scheme is seen to have the same scaling as 
the $\Lambda$\mbox{-}type scheme.

\emph{Gate time.}
The total gate time is split between EIT storage and retrieval, two $\pi$\mbox{-}pulses, 
and scattering. The EIT time $t_\text{EIT}$ is equal to the time to pass the 
ensemble, i.e. $t_\text{EIT}\sim L/v_\text{g}$, where 
$v_\text{g}=(2L|\Omega_0|^2)/(N\Gamma_\text{1D})$ is the EIT group velocity~\cite{chang_njp11a}.
The $\pi$-pulse time $t_\pi$ is set by the splitting between states $|a\rangle$ 
and $|d\rangle$. In the supplemental material~\cite{sm}, we discuss a specific 
implementation in $^{87}$Rb where that splitting 
is proportional to 
$\Delta_\text{c}$, resulting in $t_\pi\gtrsim 1/|\Delta_\text{c}|$.

To discuss scattering time, we need to model a non-zero bandwidth of 
photon~$B$. The reflection coefficient $R_0$ (at 
$\delta=\delta_\text{res}$) in Eq.~\eqref{F_CJ_general} should be replaced by
$\int R_0(\delta)|\phi_B(\delta)|^2\dif \delta$, where $\phi_B$ is the 
frequency distribution of photon~$B$. Since $r_1$ and $t_1$ vary much slower 
than $r_0$ and $t_0$ around $\delta=\delta_\text{res}$ (see 
Fig.~\ref{fig_r_and_t_plot}), we ignore a similar modification to $R_{1,1}$.
By expanding, we get 
${r_0(\delta)\approx r_0(\delta_\text{res})}+(2/w^2)(\delta-\delta_\text{res})^2$ 
with the resonance width
$w=(32\sqrt{2}\pi^2\Delta_\text{c}^2|\Omega_0|^2)/(\Gamma_\text{1D}^3N^3)$.
Defining  
$\sigma_B^2=\int (\delta-\delta_\text{res})^2 |\phi_B(\delta)|^2\dif \delta$ 
(spectral width of photon~$B$) and using the optimal 
$\Delta_\text{c}^2=(\Gamma_\text{1D}^2N^{3/2})/(8\pi)$, 
this gives a modification of the fidelity
$F_{\text{CJ},t_\text{b}=1,\sigma_B}
\approx F_{\text{CJ},t_\text{b}=1}
-(\Gamma_\text{1D}^2N^3\sigma_B^2)/(16|\Omega_0|^4\pi^2)$.
Requiring the error from non-zero $\sigma_B$ to be the same as the error in 
Eq.~\eqref{F_CJ_1}, we find that the scattering time is 
$1/\sigma_B=(\Gamma_\text{1D}^{3/2}N^{7/4})/(4\pi^{3/2}\sqrt{\Gamma'}|\Omega_0|^2)$.

For $\Gamma_\text{1D}/\Gamma=0.05$, $N=10^4$, $|\Omega_0|/\Gamma=10$, we have 
$t_\text{EIT}\sim 5/\Gamma$, $t_\pi\gtrsim 1/|\Delta_\text{c}|=0.1/\Gamma$, and 
$1/\sigma_B\sim 52/\Gamma$. Hence, the scattering time is dominant in 
the total gate time (${\sim1.6}$~$\mu$s for $^{87}$Rb~\cite{steck_rb87}). This is short 
compared to the coherence time expected for cooled and trapped 
atoms (e.g. few hundreds of microseconds in Ref.~\cite{beguin_arxiv2017}).

\emph{Other imperfections.} Classical drives may couple states~$|a\rangle$~and~$|d\rangle$ 
off-resonantly to the excited states. The coupling of the
former results in four-wave mixing noise, but this can be suppressed by a careful 
choice of the energy levels~\cite{walther_ijqi07,sm}. The 
coupling of the latter introduces loss of the stored photon 
with the effective rate 
$\Gamma_\text{eff}\sim \Gamma'|\Omega_0|^2/\Delta_\text{hfs}^2$~\cite{sm,reiter_pra12}, 
where $\Delta_\text{hfs}$ is the 
hyperfine splitting of the ground states ($|a\rangle$ and $|c\rangle$). 
Hence, the total reduction in success probability during scattering is
$\Gamma_\text{eff}/\sigma_B\sim(\Gamma_\text{1D}^{3/2}\sqrt{\Gamma'}N^{7/4})/(4\pi^{3/2}\Delta_\text{hfs}^2)$.
E.g. in $^{87}$Rb, $\Delta_\text{hfs}/\Gamma\sim 10^3$~\cite{steck_rb87}, and
this error is negligble compared to other losses for 
$\Gamma_\text{1D}/\Gamma=0.05$ and $N=10^4$ but becomes significant for $N\sim 10^5$.

If the path lengths of the Sagnac interferometer are not completely stabilized, there is 
an additional error ${\sim(k_0 l)^2}$, where $l$ is the deviation of the 
propagation length from the beam splitter to either end of the ensembles due 
to misalignment~\cite{sm}. Finally, the heralded gate is rather insensitive to 
imperfections in the $\pi$\mbox{-}pulses. The conditional fidelity will only be affected 
by the part of the excitation that still remains in states $|c\rangle$ after 
both $\pi$\mbox{-}pulses and is subsequently read out with a wrong phase. 
This error thus only enters to a higher order and can be eliminated completely by doing EIT retrieval before the 
second $\pi$\mbox{-}pulse~\cite{sm}.

\emph{Conclusion.} We have shown, how stationary light can be used to create a
CP gate between photons. Most importantly, the gate uses a large number 
of atoms~$N$ to compensate for a limited single-atom coupling to light. 
In particular, the gate can have a rapid convergence as $N^{-3/2}$ towards unit fidelity if it is operated in a heralded fashion. 
The gate is ideally suited for the setups currently 
under development
\cite{vetsch_prl10,goban_prl12,beguin_prl14,gouraud_prl15,bajcsy_prl09,
blatt_pra16}, 
where there is a moderate coupling efficiency to light 
$\Gamma_\text{1D}/\Gamma\sim 10^{-3}-10^{-1}$ and total number of atoms $N\sim 10^3-10^5$. In the supplemental material~\cite{sm}, we describe how the gate can be directly employed to 
improve the communication rate of quantum repeaters based on atomic ensembles. 
In general, the gate may serve as a tool for photonics based quantum information 
processing.

\begin{acknowledgments}
The research leading to these results was funded by the European Union Seventh 
Framework Programme through SIQS (Grant No. 600645) and ERC Grant QIOS (Grant 
No. 306576). J.B. acknowledges funding from the Carlsberg foundation.

\emph{Note added.} Recently, we became aware of a related study~\cite{lahad_prl17}.
\end{acknowledgments}

\onecolumngrid

\renewcommand{\theequation}{S\arabic{equation}}
\renewcommand{\thefigure}{S\arabic{figure}}
\renewcommand{\bibnumfmt}[1]{[S#1]}
\renewcommand{\citenumfont}[1]{S#1}
\setcounter{secnumdepth}{3}
\setcounter{section}{0}
\makeatletter
\def\p@subsection{}
\def\p@subsubsection{}
\makeatother
\renewcommand\thesection{\arabic{section}}
\renewcommand\thesubsection{\thesection.\arabic{subsection}}
\renewcommand\thesubsubsection{\thesubsection.\arabic{subsubsection}}

\newpage
\begin{center}
\textbf{\large Supplemental Material for\\
``Controlled-phase gate for photons based on stationary light''}
\end{center}
\tableofcontents
\let\addtocontents\oldaddtocontents
\section{Application to quantum repeaters}
\label{sec_quantum_repeaters}
\begin{figure}[hbt]
\begin{center}
\includegraphics{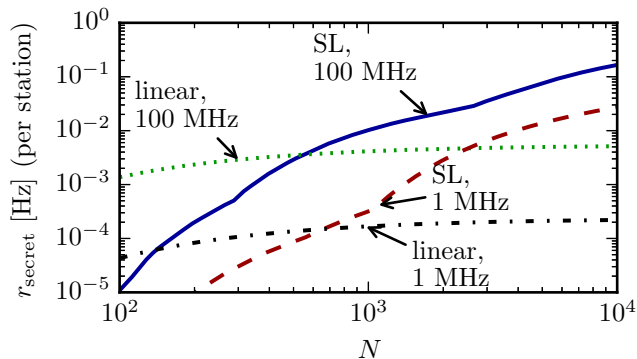}
\end{center}
\caption{
Secret key rate $r_\text{secret}$ per repeater station as a function
of the number of atoms $N$ with fixed $\Gamma_\text{1D}/\Gamma=0.5$ for dual-V 
atoms with regular interatomic distance ${d=0.266\pi/k_0}$. The communication distance
is 1000 km. We compare the protocol of Ref.~\cite{sangouard_pra08_sm} (``linear'') with
a modified protocol where the entanglement swapping (and also initial 
entanglement generation if it improves $r_\text{secret}$) is performed with 
the proposed stationary light CP gate (``SL''). We consider two different source repetition rates: 100~MHz and 1~MHz. We assume an 
attenuation length of 22 km in the fibers and an optical signal speed of 
$2 \times 10^5$ km/s. The ensemble storage and retrieval efficiency 
increases with $N$ and is set to the same value in the original 
protocol as for the modified one. 
The photodetector efficiency is assumed to be 90\%. The stationary light gate 
is assumed to have a constant gate time of 10 $\mu$s. The steps in the curves 
occur when the fidelity of 
the CP gate allows additional swap levels.
}
\label{fig_secret_key_rate}
\end{figure}

As a direct application of the proposed CP gate, we consider quantum key 
distribution using quantum repeaters based on atomic 
ensembles~\cite{dlcz_nature2001_sm,sangouard_rmp11_sm}. We modify one of the fastest 
known repeater protocols for atomic ensembles~\cite{sangouard_pra08_sm} by 
implementing the proposed CP gate instead of linear optics for entanglement 
swapping using the setup in Fig.~\ref{fig_setup}(c) of the main 
article. The secret key rate per repeater station is calculated as described 
in Ref.~\cite{borregaard_pra2015_sm} and compared to the results of the original 
protocol (see Fig.~\ref{fig_secret_key_rate}). The secret key rate depends 
strongly on the repetition rate of the probabilistic single photon sources used at the lowest 
level of the considered repeater protocol. In the figure, we make the comparison for similar 
repetition rates as assumed in Ref.~\cite{sangouard_pra08_sm}: an optimistic fast 
rate of 100 MHz and a more realistic of 1 MHz. For completeness, we assume a 
conservative gate time of 10~$\mu$s. The gate time is, however, negligible 
compared to the time of the single photon generation and the signaling time 
between stations.

The analysis is similar to the analysis done for the CP gate in 
Ref.~\cite{rydberg-cavity-gate_sm} with the difference that we also consider the 
possibility of generating the initial entanglement using the CP gate. 
Generation of initial entanglement is only better than linear optics for very 
high conditional fidelities of the controlled-phase gate (for 
$\Gamma_\text{1D}/\Gamma=0.5$ and $N\gtrsim 10^3$ in 
Fig.~\ref{fig_secret_key_rate}). For low conditional fidelities of the gate, the 
generation of initial entanglement based on the linear optics as described in 
Ref.~\cite{sangouard_pra08_sm} is better since it essentially has 
perfect conditional fidelity. For a fair comparison, we consider 
equal storage and retrieval efficiencies for both protocols. As seen in 
Fig.~\ref{fig_secret_key_rate} for $\Gamma_\text{1D}/\Gamma=0.5$, the proposed 
gate allows improving the rate of quantum repeaters if $N\gtrsim 1000$ for 
both the considered source repetition rates.

\section{Implementation in $^{87}$Rb atoms}
\label{sec_rb_implementation}

\begin{figure}[t]
\begin{center}
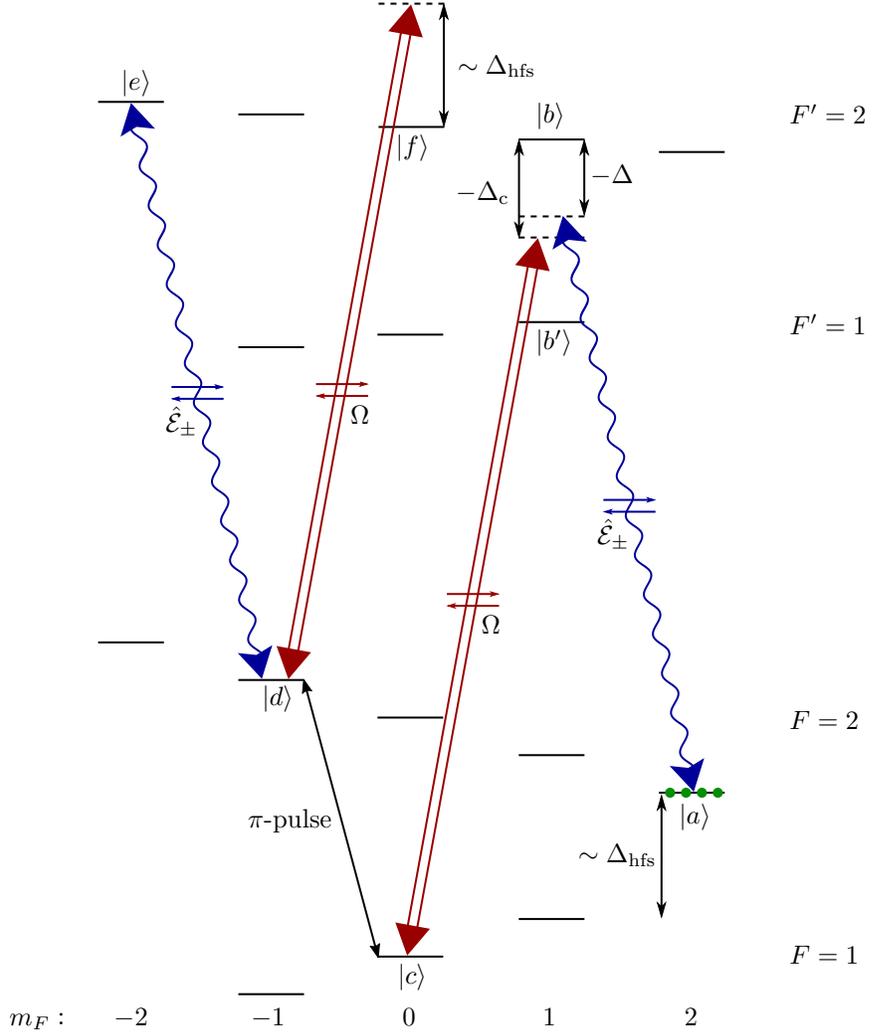
\end{center}
\caption{Implementation of the $\Lambda$-type scheme on the D$_1$ line of 
$^{87}$Rb~\cite{steck_rb87_sm}. The manifolds $F=1$ and $F=2$ belong to the 
ground state $5^2\text{S}_{1/2}$, and the manifolds $F'=1$ and $F'=2$ belong to the 
excited state $5^2\text{P}_{1/2}$. The $m_F$-levels in 
each manifold are split by the applied magnetic field. The detunings $\Delta$ 
and $\Delta_{\rm c}$ are shown with a minus sign to make their definition 
consistent with Fig.~\ref{fig_setup}(a) of 
the main article and illustrate the case when $\Delta,\Delta_{\rm c}<0$.  
The energy splittings are not drawn to scale. During scattering, we assume 
$|\Delta_\text{hfs}|\gg|\Delta|,|\Delta_{\rm c}|$.
During storage and retrieval, we assume $\Delta=\Delta_{\rm c}=0$.
Atoms are assumed to be 
pumped into the $F=2$, $m_F=2$ level ($|a\rangle$) before the beginning of the protocol (shown 
by the green circles on this level). During the EIT storage, the incident photon is 
stored in the $F=1$, $m_F=0$ level ($|c\rangle$). After a microwave $\pi$-pulse, 
this excitation is transferred to the $F=2$, $m_F=-1$ level ($|d\rangle$). Together 
with the $F'=2$, $m_{F'}=-2$ level ($|e\rangle$), this constitutes a two-level 
atom that provides the optical non-linearity. The classical drive that 
creates stationary light will also couple the $F=2$, $m_F=-1$ level ($|d\rangle$) 
to the $F'=2$, $m_{F'}=0$ ($|f\rangle$) level with a detuning approximately 
equal to $\Delta_\text{hfs}$.} 
\label{fig_Rb_implementation}
\end{figure}

In Fig.~\ref{fig_Rb_implementation}, we show a possible implementation of the 
$\Lambda$-type scheme (Fig.~\ref{fig_setup}(a) of the main article) 
on the D$_1$ line of $^{87}$Rb~\cite{steck_rb87_sm}. The $m_F$-levels in 
each manifold are split by the applied magnetic field, so that we can select 3
ground states ($|a\rangle$, $|c\rangle$, and $|d\rangle$) with different 
energies. The frequency separation between states $|a\rangle$ and $|c\rangle$ is given 
by the hyperfine splitting $\Delta_\text{hfs}$. The frequency separations of the 
other states follow from the requirements
\begin{align}
\label{condition_ab_off_resonant}
&\Delta=\omega_0-\omega_{ab},\\
\label{condition_de_resonant}
&0=\omega_0-\omega_{de},
\end{align}
where $\omega_0$ is the carrier frequency of 
the quantum field, and $\omega_{ab}$ and $\omega_{de}$ are the atomic 
transition frequencies. The equations above express the facts that the
transition $|a\rangle\leftrightarrow|b\rangle$ has to be off-resonant with 
detuning $\Delta$, and the transition $|d\rangle\leftrightarrow|e\rangle$ has 
to be on resonance.

Below, we relate $\Delta$ to the required magnetic field strength and 
determine the splitting between the adjacent $m_F$\mbox{-}levels in the $F=2$ 
manifold that is important for determining the required time for the $\pi$-pulse 
between the levels $|c\rangle$ and $|d\rangle$. The starting point is the 
expression for the shift in frequency $\Delta_{|F,m_F\rangle}$ of the state 
$|F,m_F\rangle$ due to a weak magnetic field $B_z$~\cite{steck_rb87_sm}
\begin{gather}\label{shift_due_to_magnetic_field}
\Delta_{|F,m_F\rangle}=\frac{\mu_B}{\hbar} g_F m_F B_z,
\end{gather}
where $\mu_\text{B}/\hbar=2\pi\cdot 1.4\text{ MHz/G}$ is the Bohr magneton in units of the 
Planck constant, and $g_F$ is the hyperfine Land{\'e} $g$-factor. We follow 
the convention of using $F$ and $m_F$ for the ground state $5^2\text{S}_{1/2}$ 
and $F'$ and $m_{F'}$ for the excited state $5^2\text{P}_{1/2}$. The atomic 
transition frequencies for the transitions $|a\rangle\leftrightarrow|b\rangle$ 
and $|d\rangle\leftrightarrow|e\rangle$ can be written
\begin{align}
\label{omega_ab_in_terms_of_omega_ab_B_z_zero}
&\omega_{ab}=\omega_{ab,B_z=0}
+\Delta_{|F'=2,m_{F'}=1\rangle}
-\Delta_{|F=2,m_{F}=2\rangle},\\
\label{omega_de_in_terms_of_omega_de_B_z_zero}
&\omega_{de}=\omega_{de,B_z=0}
+\Delta_{|F'=2,m_{F'}=-2\rangle}
-\Delta_{|F=2,m_{F}=-1\rangle},
\end{align}
in terms of the transition frequencies $\omega_{ab,B_z=0}$ and 
$\omega_{de,B_z=0}$ for $B_z=0$. Note further that 
$\omega_{ab,B_z=0}=\omega_{de,B_z=0}$, so that we can define 
$\Delta_{a}=\omega_0-\omega_{ab,B_z=0}=\omega_0-\omega_{de,B_z=0}$. Using this 
definition together with 
\cref{shift_due_to_magnetic_field,omega_ab_in_terms_of_omega_ab_B_z_zero,omega_de_in_terms_of_omega_de_B_z_zero} 
allows to write 
Eqs.~\eqref{condition_ab_off_resonant}~and~\eqref{condition_de_resonant}
\begin{align}
\label{condition_ab_off_resonant_2}
&\Delta
=\Delta_{a}
-\Delta_{|F'=2,m_{F'}=1\rangle}
+\Delta_{|F=2,m_{F}=2\rangle}
=\Delta_{a}
-\frac{\mu_B}{\hbar} g_{F'=2} B_z
+2\frac{\mu_B}{\hbar} g_{F=2} B_z,\\
\label{condition_de_resonant_2}
&0
=\Delta_{a}
-\Delta_{|F'=2,m_{F'}=-2\rangle}
+\Delta_{|F=2,m_{F}=-1\rangle}
=\Delta_{a}
+2\frac{\mu_B}{\hbar} g_{F'=2} B_z
-\frac{\mu_B}{\hbar} g_{F=2} B_z.
\end{align}
Solving Eqs.~\eqref{condition_ab_off_resonant_2}~and~\eqref{condition_de_resonant_2} 
with $g_{F'=2}=1/6$ and 
$g_{F=2}=1/2$~\cite{steck_rb87_sm}
gives
\begin{align}
&\Delta_\text{a}
=\frac{1}{6}\frac{\mu_B B_z}{\hbar},\\
&\Delta
=\frac{\mu_B B_z}{\hbar}.
\end{align}

To determine the time for the $\pi$-pulse transferring state 
$|c\rangle$ to state $|d\rangle$ (see more discussion in 
Sec.~\ref{sec_pi_pulse}), we need to find the splitting of state $|d\rangle$ ($F=2$, $m_F=-1$) from the adjacent
state with $F=2$, $m_F=0$. This splitting is
\begin{gather}
\Delta_{|F=2,m_{F}=-1\rangle}
-\Delta_{|F=2,m_{F}=0\rangle}
=-\frac{1}{2}\frac{\mu_B B_z}{\hbar}
=-\frac{1}{2}\Delta.
\end{gather}
To provide a numerical example for the required magnetic field, we can 
assume $\Gamma_\text{1D}/\Gamma=0.05$ and $N=10^4$. This gives an optimal detuning
$\Delta\approx\Delta_\text{c}\approx-\sqrt{(\Gamma_\text{1D}^2N^{3/2})/(8\pi)}\approx-10\Gamma$.
Since $\Gamma=2\pi\cdot 5.75\text{ MHz}$ for the D$_1$ line~\cite{steck_rb87_sm}, 
the required magnetic field is $B_z = \hbar \Delta/\mu_B\approx -41\text{ G}$.

Due to the use of the extreme $m_F$-level for state $|a\rangle$ (see 
Fig.~\ref{fig_Rb_implementation}), it is not coupled to any excited states in 
the D$_1$ line by the classical drive, suppressing four-wave mixing noise~\cite{walther_ijqi07_sm}. There 
is, however, a possible drawback of this level structure~\cite{vurgaftman_pra13_sm}: 
reduction of the effective coupling 
strength of the $|a\rangle\leftrightarrow|c\rangle$ transition due to the fact 
that state $|a\rangle$ is also coupled to the excited state $F'=2$, $m_{F'}=1$ 
($|b'\rangle$). This reduction happens due to the 
destructive interference of the 
paths 
$|a\rangle\leftrightarrow|b\rangle\leftrightarrow|c\rangle$ and
$|a\rangle\leftrightarrow|b'\rangle\leftrightarrow|c\rangle$ 
if the detuning of the probe field from transitions
$|a\rangle\leftrightarrow|b\rangle$ and $|a\rangle\leftrightarrow|b'\rangle$ has the same sign. 
Hence, one has to make sure that the fields are either resonant 
with one of the transitions (the case during storage and retrieval) or that the detunings have different 
signs (as shown in Fig.~\ref{fig_Rb_implementation}). The latter can be 
satisfied for scattering, if $|\Delta|$ is smaller than the splitting of 
$|b\rangle$ and $|b'\rangle$. Since the latter splitting is $\sim 100\Gamma$~\cite{steck_rb87_sm}, 
the condition is easily satisfiable for e.g. $\Gamma_\text{1D}/\Gamma=0.05$ 
and $N=10^4$, where $\Delta\approx -10\Gamma$ (for optimal fidelity).
Even though the above approach suppresses the noise during EIT storage and 
retrieval, the presence of the classical drive can still induce noise during 
scattering due to the off-resonant coupling of state $|d\rangle$ to state $|f\rangle$ (see 
Fig.~\ref{fig_Rb_implementation}). See 
Sec.~\ref{sec_dephasing_decay_of_stored_photon} below for a discussion of this 
coupling.

Finally, we note that by setting the two counter-propagating classical drives 
at different frequencies such that they have detunings $\Delta_{\text{c}+}$ 
and $\Delta_{\text{c}-}$ (we are only interested in the off-resonant case, 
i.e. $\Delta_{\text{c}+},\Delta_{\text{c}-}\neq 0$), this setup 
becomes an implementation of a the dual-color 
scheme~\cite{moiseev_pra2006_sm}. This scheme is equivalent to the dual-V~\cite{zimmer_pra08_sm} scheme 
discussed in the main article, if~\cite{dispersion-relations_sm}
\begin{gather}\label{dual_color_dual_V_equivalence_condition_1}
|\Delta_{\text{c}+}-\Delta_{\text{c}-}|< \frac{|\Delta_{\text{c}+}+\Delta_{\text{c}-}|}{2}
\end{gather}
(i.e. splitting between the two frequencies is smaller than their average magnitude), 
and 
\begin{gather}\label{dual_color_dual_V_equivalence_condition_2}
|\Delta_{\text{c}+}-\Delta_{\text{c}-}|\gtrsim \frac{|\Omega_0|^2}{|\Delta_{\text{c}+}+\Delta_{\text{c}-}|}
\end{gather}
(i.e. splitting between the two frequencies is bigger than the ac Stark 
shift induced by the classical drives). These two 
conditions set additional restrictions on the Rabi frequency and, in 
particular, may require the gate time (dominated by the scattering time 
$(\Gamma_\text{1D}^{3/2}N^{7/4})/(4\pi^{3/2}\sqrt{\Gamma'}|\Omega_0|^2)$ 
derived in Sec.~\ref{sec_analytical_fidelity} below) to 
be longer for the dual-color scheme than for the $\Lambda$-type scheme. E.g. for $\Gamma_\text{1D}/\Gamma=0.05$ 
and $N=10^4$, $|\Delta_{\text{c}+}+\Delta_{\text{c}-}|/2\approx 10\Gamma$ 
needs to be chosen for optimal fidelity. From 
Eqs.~\eqref{dual_color_dual_V_equivalence_condition_1}~and~\eqref{dual_color_dual_V_equivalence_condition_2}, 
we thus have $|\Omega_0|\lesssim 10\Gamma$. A full investigation of this is beyond 
the scope of the present analysis.

\section{Fidelity}
\label{sec_fidelity_general}

Here, we state the expressions required for evaluating the Choi-Jamiolkowski 
fidelity and success probability accounting for non-zero bandwidth of the 
scattered photon $B$. We will also discuss, how the unconditional and 
conditional fidelities are related to figures of merit for quantum repeaters. 
A more detailed discussion can be found in 
Refs.~\cite{iakoupov_phd_thesis_sm,fidelity_sm}.

The ideal evolution of the 
controlled-phase gate is defined by its action on the computational basis states
\begin{align}
&|00\rangle_\text{in} \rightarrow |00\rangle_\text{out},\\
&|01\rangle_\text{in} \rightarrow |01\rangle_\text{out},\\
&|10\rangle_\text{in} \rightarrow |10\rangle_\text{out},\\
&|11\rangle_\text{in} \rightarrow -|11\rangle_\text{out}.
\end{align}
To calculate the Choi-Jamiolkowski fidelity, we have to define the input basis 
states ($|jj'\rangle_\text{in}$ for $j,j'\in\cbr{0,1}$) and their ideal 
evolution into the output states ($|jj'\rangle_\text{out}$). Since photon~$A$ 
is stored and retrieved, and photon~$B$ is scattered, it is most natural to 
describe the former as a temporal wave packet and the latter through a 
frequency distribution over $\delta_B$, where $\delta_B=\Delta-\Delta_\text{c}$ 
(see Figs. \ref{fig_setup}(a) and \ref{fig_setup}(b) of the 
main article) is the two-photon detuning. We define the ideal evolution such 
that the output frequency distribution of photon~$B$ is equal to the input 
frequency distribution, and the output wave packet of photon~$A$ is set to be 
the one obtained after storage and retrieval in the absence of scattering. 
These choices of the ideal output wave functions $\phi_{A,\text{out},0}$ and 
$\phi_B$ are natural, but there may exist more optimal choices that give 
better fidelities. In terms of the storage and retrieval kernels, 
$K_{\text{s},j}(t_A)$ and $K_{\text{r},j}(t_A)$, respectively (discussed in 
Sec.~\ref{eit_storage_retrieval} below), the output wave packet of photon~$A$ 
is
\begin{gather}\label{phi_out_0_definition}
\phi_{A,\text{out},0}(t_A)=\sum_j\int K_{\text{r},j}(t_A)K_{\text{s},j}(t_A')
\phi_{A,\text{in}}(t_A')\dif t_A',
\end{gather}
where $\phi_{A,\text{in}}$ is the input wave packet. We define 
$\phi_{A,\text{out},0}$ to be unnormalized, and the absolute square of its norm 
$\eta_\text{EIT}=\int |\phi_{A,\text{out},0}(t_A)|^2\dif t_A$ is the EIT 
storage and retrieval efficiency, which is, in general, less than unity 
due to imperfections. When defining the output basis states 
($|jj'\rangle_\text{out}$), we ensure that they are normalized by dividing them 
by $\sqrt{\eta_\text{EIT}}$.
In the numerical calculations, we use the discrete
definition~\eqref{phi_out_0_definition}
of $\phi_{A,\text{out},0}$ (and a similar one for $\phi_{A,\text{out},1}$ below) instead of the continuum version that was stated 
in the main article for simplicity. 

We also need to find the output wave packet for photon~$A$ for the case 
when photon~$B$ was scattered from the atomic ensemble while photon~$A$ was 
stored in it. It is
\begin{gather}\label{phi_out_1_definition}
\phi_{A,\text{out},1}(t_A,\delta_B)
=\sum_j\int K_{\text{r},j}(t_A)R_{1,j}(\delta_B)K_{\text{s},j}(t_A')
\phi_{A,\text{in}}(t_A')\dif t_A'.
\end{gather}
Note that, compared to $\phi_{A,\text{out},0}$, the reflection 
coefficient $R_{1,j}(\delta_B)$ appears in the definition. This reflection 
coefficient depends on the position where photon~$A$ was stored, hence it 
cannot be taken outside of the summation over $j$. Perfect controlled-phase gate 
operation (neglecting the bandwidth of photon~$B$) is achieved when 
$\phi_{A,\text{out},1}(t_A,\delta_\text{res})=-\phi_{A,\text{out},0}(t_A)$ for 
some $\delta_\text{res}$.

The Choi-Jamiolkowski fidelity is computed by assuming a particular entangled 
state as the input and applying both the ideal and real evolution to 
it~\cite{iakoupov_phd_thesis_sm,fidelity_sm,real-and-ideal-quantum-processes_sm}. The 
fidelity between the output states is then the Choi-Jamiolkowski fidelity. It 
is given by
\begin{gather}\label{CJ_fidelity_computational_basis_final}
F_\text{CJ}=
\frac{\eta_\text{EIT}}{16} \envert{
2t_\text{b}
+\int R_0(\delta_B)|\phi_B(\delta_B)|^2\dif \delta_B
-\frac{1}{\eta_\text{EIT}}
\iint \phi_{A,\text{out},0}^*(t_A)\phi_{A,\text{out},1}(t_A,\delta_B)
|\phi_B(\delta_B)|^2\dif t_A\dif \delta_B
}^2.
\end{gather}
If we neglect the bandwidth of photon~$B$ by setting $|\phi_B(\delta_B)|^2$ to 
be the Dirac delta function, $|\phi_B(\delta_B)|^2=\delta(\delta_B-\delta_\text{res})$, 
in the above expression,
we obtain Eq.~\eqref{F_CJ_general} of the main article.

The success probability for having two photons at the output is
\begin{gather}
\begin{aligned}\label{P_suc_computational_basis_final}
P_\text{suc}
=\frac{\eta_\text{EIT}}{4}\del{
2|t_\text{b}|^2+
\int |R_0(\delta_B)|^2|\phi_B(\delta_B)|^2\dif \delta_B
+\frac{1}{\eta_\text{EIT}}
\iint |\phi_{A,\text{out},1}(t_A,\delta_B)|^2
|\phi_B(\delta_B)|^2\dif t_A\dif \delta_B
}.
\end{aligned}
\end{gather}
Setting $|\phi_B(\delta_B)|^2=\delta(\delta_B-\delta_\text{res})$, this reduces to 
Eq.~\eqref{P_suc_general} of the main article. The conditional 
Choi-Jamiolkowski fidelity is $F_\text{CJ,cond}=F_\text{CJ}/P_\text{suc}$.

\begin{figure}[hbt]
\begin{center}
\includegraphics{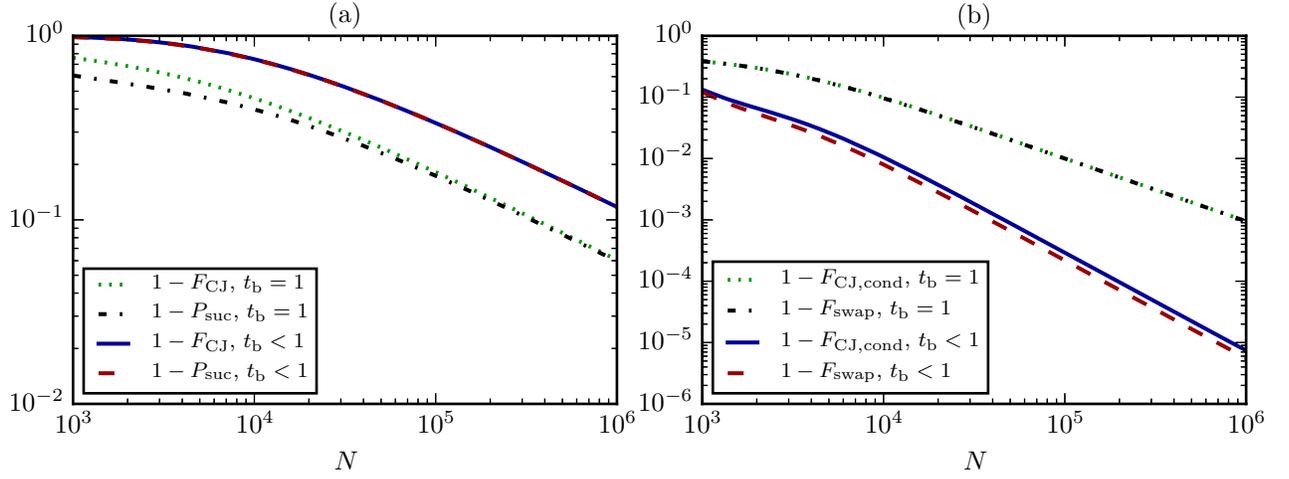}
\end{center}
\caption{(a) Comparison of the Choi-Jamiolkowski fidelity 
$F_\text{CJ}$ for deterministic operation of the controlled-phase gate 
with the success probability $P_\text{suc}$ for heralded operation of 
the gate. (b) Comparison of the conditional Choi-Jamiolkowski fidelity 
$F_\text{CJ,cond}$ with the entanglement swap fidelity $F_\text{swap}$ for 
heralded operation of the gate. For both 
(a) and (b), $\Lambda$-type scheme is used, but the comparison for the dual\mbox{-}V 
scheme is similar. Dotted green and dash-dotted black curves are calculated with 
$t_\text{b}=1$, and the solid blue and dashed red curves are calculated with 
$t_\text{b}$ chosen such that the entanglement swap fidelity $F_\text{swap}$ 
is maximal. All quantities are plotted as functions of the number of atoms 
$N$ with fixed $\Gamma_\text{1D}/\Gamma=0.05$ and $\Omega_0/\Gamma=1$. Under EIT (storage and retrieval), $\Omega(z)=\Omega_0$. 
Under stationary light (scattering), $\Omega(z)=\Omega_0\cos(k_0 z)$.
For storage and retrieval, we use the discretized continuum storage and 
retrieval kernels discussed 
in Sec.~\ref{continuum_eit_storage_retrieval}.
}
\label{fig_cphase_fidelity_F_swap}
\end{figure}

The abstract Choi-Jamiolkowski fidelity considered above can be related to 
more concrete figures of merit, such as entanglement swap fidelity and success 
probability in the setting of quantum repeaters. In 
Refs.~\cite{iakoupov_phd_thesis_sm,fidelity_sm} it is shown that the success 
probability of the entanglement swap operation (see 
Fig.~\ref{fig_setup}(c) of the main article and Sec.~\ref{sec_quantum_repeaters} above) is 
exactly the same as the success probability $P_\text{suc}$ defined by Eq.~\eqref{P_suc_computational_basis_final}. Furthermore, since $F_\text{CJ}$ measures the probability of 
the photons to be in the right modes with the right phase, whereas 
$P_\text{suc}$ measures whether the photons are coming out, it holds that 
$F_\text{CJ}\leq P_\text{suc}$. The expressions for $F_\text{CJ}$ stated in 
the main article thus give a lower bound for $P_\text{suc}$. For our 
particular implementation of the controlled-phase gate, most of the error in 
$F_\text{CJ}$ is caused by photon loss, and hence we have 
$F_\text{CJ}\approx P_\text{suc}$. At the same time, $F_\text{CJ,cond}$ is a 
lower bound and an approximation for the entanglement swap 
fidelity $F_\text{swap}$~\cite{iakoupov_phd_thesis_sm,fidelity_sm}. In 
Fig.~\ref{fig_cphase_fidelity_F_swap}, we illustrate the approximate 
equalities of $F_\text{CJ}$ with $P_\text{suc}$ and $F_\text{CJ,cond}$ with 
$F_\text{swap}$.

\section{Scattering coefficients for the ensemble}
\label{sec_scattering_coefficients}
\subsection{Transfer matrix formalism}
To find the scattering coefficients for an ensemble of $\Lambda$\mbox{-}type atoms, the single-mode transfer matrix 
formalism~\cite{deutsch_pra95a_sm} is sufficient. For the dual\mbox{-}V scheme, 
two different polarization modes of the electric field are coupled to the 
atoms, and hence a generalization to a two-mode transfer matrix formalism is 
required. The general multi-mode transfer matrix formalism is developed in 
Ref.~\cite{dispersion-relations_sm}. Here, we briefly summarize it before applying it 
to calculate the scattering coefficients of atomic ensembles.

In the multi-mode transfer matrix formalism, electric fields are described by vectors of 
$2n_\text{m}$ elements, where $n_\text{m}$ is the number of the different 
modes of the electric field. The fields propagating to the right and the 
fields propagating to the left are treated as being distinct, hence there are 
$n_\text{m}$ elements for each propagation direction. We can write the electric field 
vectors
\begin{gather}\label{multimode_E_field_vector}
\mathbf{E}(z)=\begin{pmatrix}
\mathbf{E}_+(z)\\
\mathbf{E}_-(z)
\end{pmatrix},
\end{gather}
where $\mathbf{E}_+$ is the part of the electric field that propagates to 
the right (in the positive direction), $\mathbf{E}_-$ is the part of the electric field that propagates to 
the left (in the negative direction).

Atoms and free propagation between atoms are described by $2n_\text{m}\times 2n_\text{m}$ matrices that relate the vectors of electric field at one 
position to vectors of electric field at a different position. In general, the transfer 
matrix for the whole ensemble is obtained by multiplying the 
transfer matrices for the atoms and free propagation. We use this approach for 
the dual-V atoms, for which we assume a placement of the 
atoms that is incommensurate with the wavelength of the classical drives (or even completely 
random). For the $\Lambda$-type scheme, the atoms are regularly 
placed with distance $d=\pi/(2k_0)$, as shown in
Fig.~\ref{fig_Lambda_type_kd_pi_half_setup}. Hence, the ensemble consists of 
repeated unit cells, and one can instead exponentiate 
the transfer matrix for a single unit cell to find the transfer matrix for the 
whole ensemble. For $\Lambda$\mbox{-}type atoms 
described by $2\times 2$ transfer matrices, closed-form expression can be 
obtained for the transfer matrix for the whole ensemble given in terms of the 
transfer matrix for the unit cell~\cite{dispersion-relations_sm}. If the transfer 
matrix for the unit cell is
\begin{gather}
T_\text{cell}=
\begin{pmatrix}
T_{11} & T_{12}\\
T_{21} & T_{22}
\end{pmatrix}
\end{gather}
then the transfer matrix for an ensemble of $n$ such unit cells is
\begin{gather}\label{matrix_T_pow_n}
T_\text{cell}^n=e^{in\theta A}=(\cos n\theta)I+i(\sin n\theta)A
\end{gather}
where $I$ is the identity matrix,
\begin{gather}
A=\frac{1}{\sin\theta}
\begin{pmatrix}
i(T_{22}-T_{11})/2 & -iT_{12}\\
-iT_{21} & -i(T_{22}-T_{11})/2
\end{pmatrix},
\end{gather}
and $\theta$ fulfills $\cos(\theta)=\tr(T_\text{cell})/2=(T_{11}+T_{22})/2$. Physically, $\theta$ is equal to the Bloch 
vector $q$ multiplied by the length occupied by the unit cell. In the case of 
the $\Lambda$\mbox{-}type scheme with 
inter-atomic spacing $d=\pi/(2k_0)$ below, the unit cell has length $2d$ and 
hence we have $\theta=2qd$.

Once the transfer matrix for the ensemble $T_\text{e}$ has been obtained, one needs to 
extract the scattering coefficients from it. The transfer matrix for the 
ensemble $T_\text{e}$ fulfills the relation
\begin{gather}
\label{transfer_matrix_ensemble_E_relation}
\begin{pmatrix}
\mathbf{E}_+(L^+)\\
\mathbf{E}_-(L^+)
\end{pmatrix}
=
\begin{pmatrix}
T_{\text{e},{11}} & T_{\text{e},{12}}\\
T_{\text{e},{21}} & T_{\text{e},{22}}
\end{pmatrix}
\begin{pmatrix}
\mathbf{E}_+(0^-)\\
\mathbf{E}_-(0^-)
\end{pmatrix}.
\end{gather}
For the dual\mbox{-}V atoms, the parts $\mathbf{E}_\pm$ have two elements, 
corresponding to the two polarizations ($\sigma_+$ and $\sigma_-$). To 
calculate the reflection coefficient of the ensemble, when the field is 
incident from the left, we assume input fields
\begin{gather}\label{input_E_for_T}
\mathbf{E}_+(0^-)=
\begin{pmatrix}
1\\
0
\end{pmatrix},\quad
\mathbf{E}_-(L^+)=
\begin{pmatrix}
0\\
0
\end{pmatrix}
\end{gather}
(i.e. only a right-moving $\sigma_+$ field incident from the left). Then from 
Eq.~\eqref{transfer_matrix_ensemble_E_relation} we see that
\begin{align}
&\mathbf{E}_-(0^-)=-T_{\text{e},{22}}^{-1}T_{\text{e},{12}}\mathbf{E}_+(0^-),\\
&\mathbf{E}_+(L^+)
=(T_{\text{e},{11}}
-T_{\text{e},{12}}T_{\text{e},{22}}^{-1}T_{\text{e},{21}})\mathbf{E}_+(0^-).
\end{align}
The first ($\sigma_+$) element of 
$\mathbf{E}_+(L^+)$ is the transmission coefficient $t_\text{e}$, and the second 
($\sigma_-$) element of $\mathbf{E}_-(0^-)$ is the reflection coefficient 
$r_\text{e}$. In the single-mode case, the reflection and transmission 
coefficients simplify to $r_\text{e}=-T_{\text{e},21}/T_{\text{e},22}$ and 
$t_\text{e}=1/T_{\text{e},{22}}$, respectively. Note that this result gives a 
full description of the scattering problem including possible phase shifts 
induced by the scattering~\cite{peters_ol10_sm}.

The only thing left to describe in the general case, is how the matrix 
$T_\text{cell}$ (and $T_\text{e}$ if it is not simply equal to 
$T_\text{cell}^n$) is calculated. It is a product of the matrices $T_\text{a}(\beta)$ describing scattering of the electric fields by the atoms and the matrices $T_\text{f}(k_0 d)$ 
describing the free propagation of the electric fields. They are given by
\begin{gather}\label{multi_mode_transfer_matrices_a_f}
T_\text{a}(\beta)=
\begin{pmatrix}
I-\beta & -\beta\\
\beta & I + \beta
\end{pmatrix},\quad
T_\text{f}(k_0 d)=
\begin{pmatrix}
e^{ik_0 d}I & 0\\
0 & e^{-ik_0 d}I
\end{pmatrix},
\end{gather}
where $I$ means the $n_\text{m}\times n_\text{m}$ identity matrix (a scalar 
equal to unity in the single-mode case), $k_0$ is the wave vector, and $d$ is the propagation distance.
 Each matrix 
$T_\text{a}(\beta)$ is given in terms a $n_\text{m}\times n_\text{m}$ matrix 
$\beta$ that will be defined for the specific cases of $\Lambda$\mbox{-}type 
and dual\mbox{-}V atoms considered below.

\subsection{Reflection and transmission for the $\Lambda$\mbox{-}type scheme}
\subsubsection{Without a stored photon}
\begin{figure}[t]
\begin{center}
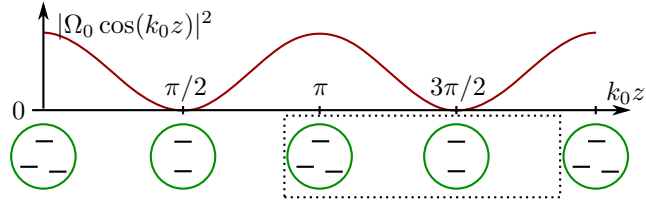
\end{center}
\caption{Standing wave of the Rabi frequency of classical drive for the 
$\Lambda$-type scheme.
The circles with 2 or 3 energy levels below the 
plot represent how the atoms will effectively behave in the different 
positions, i.e. either as two-level atoms on the nodes or as $\Lambda$\mbox{-}type 
atoms on the anti-nodes. The unit cell is shown by the dotted rectangle.}
\label{fig_Lambda_type_kd_pi_half_setup}
\end{figure}

Here, we derive the scattering coefficients $r_0$ and $t_0$ given by Eqs. \eqref{r_ensemble_N} and 
\eqref{t_ensemble_N} of the main article. For the $\Lambda$-type atoms, 
we use single-mode transfer matrices, and hence the parameter $\beta$ in 
Eq.~\eqref{multi_mode_transfer_matrices_a_f} is a scalar. We consider a 
unit cell that consists of two atoms and two lengths of free 
propagation (see 
Fig.~\ref{fig_Lambda_type_kd_pi_half_setup}). One of the atoms is placed on the anti-node of the 
standing wave of the classical drive, and the other is placed on the node. The 
scattering from the former is described by the 
parameter~\cite{dispersion-relations_sm}
\begin{gather}\label{beta3}
\beta_3
=\frac{\Gamma_\text{1D}\delta}
{(\Gamma'-2i\Delta)\delta+2i|\Omega_0|^2},
\end{gather}
and the scattering for the latter (an effective two-level atom) is described 
by the parameter
\begin{gather}\label{beta2}
\beta_2=\frac{\Gamma_\text{1D}}{\Gamma'-2i\Delta}.
\end{gather}
The transfer matrix for the unit cell is
\begin{gather}\label{matrix_T_lambda_over_4}
T_\text{cell}=T_\text{f}(\pi/2) T_\text{a}(\beta_2) T_\text{f}(\pi/2) T_\text{a}(\beta_3).
\end{gather}
Carrying out the above matrix multiplications results in
\begin{gather}
T_\text{cell}=\begin{pmatrix}
-(1-\beta_2)(1-\beta_3)-\beta_2\beta_3
& \beta_3(1-\beta_2)-\beta_2(1+\beta_3)\\
\beta_2(1-\beta_3)-\beta_3(1+\beta_2)
& -\beta_2\beta_3-(1+\beta_2)(1+\beta_3)
\end{pmatrix}.
\end{gather}
From Eq.~\eqref{matrix_T_pow_n}, we then have
\begin{gather}
\label{matrix_T_pow_n_Lambda}
T_\text{e}=T_\text{cell}^n=\cos(n\theta)
\begin{pmatrix}
1 & 0\\
0 & 1
\end{pmatrix}
+\frac{\sin(n\theta)}{\sin(\theta)}
\begin{pmatrix}
\beta_2+\beta_3 & -\beta_2+\beta_3-2\beta_2\beta_3\\
\beta_2-\beta_3-2\beta_2\beta_3 & -\beta_2-\beta_3
\end{pmatrix}
\end{gather}
with $\theta$ given by $\cos(\theta)=\tr(T_\text{cell})/2=-1-2\beta_2\beta_3$. 
From this matrix, we obtain the reflection and transmission coefficients
\begin{align}\label{r_T_n}
&r_0=-\frac{T_{\text{e},{21}}}{T_{\text{e},{22}}}
=\frac{-\beta_2+\beta_3+2\beta_2\beta_3}
{\frac{\cos(n\theta)}{\sin(n\theta)}\sin(\theta)-(\beta_2+\beta_3)},\\
\label{t_T_n}
&t_0=\frac{1}{T_{\text{e},{22}}}
=\frac{1}{\cos(n\theta)-\frac{\sin(n\theta)}{\sin(\theta)}(\beta_2+\beta_3)}.
\end{align}

The minima of $r_0$ and maxima of $t_0$ (see 
Fig.~\ref{fig_r_and_t_plot} of the main article) occur when 
$\sin(n\theta)$ in Eq.~\eqref{matrix_T_pow_n_Lambda} is approximately equal to zero. 
However, exact equality is never satisfied, since $\theta$ is complex (a 
consequence of $\Gamma'>0$). In the regime where losses are 
small ($\Imt[\tr(T_\text{cell})]\ll 1$), the approximate resonance condition is 
\begin{gather}\label{approximate_resonance_condition}
\sin\delnospace{n\arccos(\Ret[\tr(T_\text{cell})]/2)}=0.
\end{gather}

In the coefficients \eqref{r_T_n} and \eqref{t_T_n} we can approximate
\begin{gather}\label{theta_arccos_expansion}
\theta=\arccos(\tr(T_\text{cell})/2)\approx\arccos(\Ret[\tr(T_\text{cell})]/2)
-\frac{i\Imt[\tr(T_\text{cell})]/2}{\sqrt{(1-\Ret[\tr(T_\text{cell})]/2)(1+\Ret[\tr(T_\text{cell})]/2)}}.
\end{gather}
From Eq.~\eqref{approximate_resonance_condition} we have
\begin{gather}
n\arccos(\Ret[\tr(T_\text{cell})]/2)=\pi k
\end{gather}
for some integer $k$. Since the are interested in the first reflection minimum 
closest to $\delta=0$, we choose $k=n-1$. For large $n$, we have
\begin{gather}\label{approximate_resonance_condition2}
\Ret[\tr(T_\text{cell})]/2=\cos(\pi(n-1)/n)\approx -1+\pi^2/(2n^2).
\end{gather}
Hence, Eq.~\eqref{theta_arccos_expansion} can be approximated
\begin{gather}
\theta\approx\pi(n-1)/n
-in\Imt[\tr(T_\text{cell})]/(2\pi),
\end{gather}
and we also obtain the approximate expressions:
\begin{align}
&\sin(n\theta)\approx (-1)^{n-1} \left(-in^2\Imt[\tr(T_\text{cell})]/(2\pi)\right),\\
&\cos(n\theta)\approx(-1)^{n-1},\\
&\sin(\theta)\approx \pi/n.
\end{align}

With these approximations and using the fact that 
$\Imt[\tr(T_\text{cell})]=-4\Imt[\beta_2\beta_3]$, Eqs. \eqref{r_T_n} and 
\eqref{t_T_n} become
\begin{align}\label{r_T_n_approx1}
&r_0\approx\frac{-\beta_2+\beta_3+2\beta_2\beta_3}
{-\frac{i\pi^2}{2n^3\Imt[\beta_2\beta_3]}-\beta_2-\beta_3},\\
\label{t_T_n_approx1}
&t_0\approx
\frac{(-1)^{n-1}}{1-\frac{2in^3\Imt[\beta_2\beta_3]}{i\pi^2}(\beta_2+\beta_3)}.
\end{align}
To determine the dominant terms in Eqs. 
\eqref{r_T_n_approx1} and \eqref{t_T_n_approx1}, we write the 
approximate expressions for Eqs. \eqref{beta3} and \eqref{beta2} in the limit 
where $\delta$ is small, and $|\Delta_\text{c}|$ is large 
($\Delta=\Delta_\text{c}+\delta$). We thereby get
\begin{align}\label{beta3_approx1}
&\beta_3\approx 
-i\frac{\Gamma_\text{1D}\delta}{2|\Omega_0|^2},\\
\label{beta2_approx1}
&\beta_2\approx
i\frac{\Gamma_\text{1D}}{2\Delta}
+\frac{\Gamma_\text{1D}\Gamma'}{4\Delta^2}.
\end{align}
The second term on the right hand side of Eq.~\eqref{beta2_approx1} is 
included, since it is the lowest order term in $\Ret[\beta_2]$. The latter is 
used for finding an approximation for 
$\Imt[\beta_2\beta_3]\approx \Ret[\beta_2]\Imt[\beta_3]$. An expression for 
the detuning $\delta$ can be found using 
Eq.~\eqref{approximate_resonance_condition2}. Expanding its left hand side to 
second order in $\delta$ around $0$, results in the quadratic equation 
that determines the two-photon detuning $\delta=\delta_\text{res}$ of the 
transmission resonance nearest $\delta=0$,
\begin{gather}
\frac{\Gamma_\text{1D}^2}{2\Delta_\text{c}|\Omega_0|^2}\delta_\text{res}
-\frac{\Gamma_\text{1D}^2(|\Omega_0|^2-\Delta_\text{c}^2)}
{2\Delta_\text{c}^2|\Omega_0|^4}\delta_\text{res}^2
+\frac{\pi^2}{2n^2}=0.
\end{gather}
We choose the solution of this equation where $\delta_\text{res}$ and 
$\Delta_\text{c}$ have opposite signs (we assume $\delta_\text{res}>0$ and 
$\Delta_\text{c}<0$, but the opposite case should also work), i.e.
\begin{gather}
\frac{\delta_\text{res}}{|\Omega_0|^2}
=\frac{\Delta_\text{c}\left(-\Gamma_\text{1D}
+\sqrt{\Gamma_\text{1D}-4(\Delta_\text{c}^2-|\Omega_0|^2)\pi^2/n^2}\right)}
{2\Gamma_\text{1D}(\Delta_\text{c}^2-|\Omega_0|^2)}
\end{gather}
In the limit of large $n$, we find
\begin{gather}\label{Delta1_resonance_solution}
\frac{\delta_\text{res}}{|\Omega_0|^2}
\approx -\frac{\pi^2\Delta_\text{c}}{\Gamma_\text{1D}^2n^2}
-\frac{\pi^4\Delta_\text{c}^3}{\Gamma_\text{1D}^4n^4}
+\frac{\pi^4\Delta_\text{c}|\Omega_0|^2}{\Gamma_\text{1D}^4n^4}.
\end{gather}
The first term on the right hand side of Eq.~\eqref{Delta1_resonance_solution} 
could also be derived using the quadratic approximation of the dispersion relation, i.e.
$\delta\approx(1/2m)(qd)^2=(1/2m)(\theta d/2)^2$ with the effective mass 
${m=-\Gamma_\text{1D}^2/
(8(\Delta_\text{c}+i\Gamma'/2)|\Omega_0|^2)}$~\cite{dispersion-relations_sm},
but the other two terms result from higher 
order corrections to this approximation. The resonance detuning 
$\delta_\text{res}\approx 
-4\pi^2\Delta_\text{c}|\Omega_0|^2/(\Gamma_\text{1D}^2N^2)$ 
that is quoted in the main article results from only keeping the first term 
on the right hand side of Eq.~\eqref{Delta1_resonance_solution} and using $N=2n$.

When we calculate the fidelity 
$F_\text{CJ}$ in Sec.~\ref{sec_analytical_fidelity} below, we find that it is 
maximal for a detuning
\begin{gather}\label{optimal_Deltac_proportional}
|\Delta_\text{c}|\propto \Gamma_\text{1D}n^{3/4}.
\end{gather}
If we insert this expression into \eqref{Delta1_resonance_solution}, we find 
that the first term on the right hand side is proportional to $n^{-5/4}$, and 
the second one is proportional to $n^{-7/4}$. Hence, the second one is smaller 
for large $n$ and can be neglected. We keep the third term, since it depends 
on $\Omega_0$ and will be important when accounting for the non-zero bandwidth 
of the scattered photon.

Using only the first term in Eq.~\eqref{Delta1_resonance_solution} and 
inserting (i.e. setting $\Delta=\Delta_\text{c}+\delta_\text{res}$) it into 
the first term of Eq.~\eqref{beta2_approx1}, we find
\begin{gather}\label{beta2_approx2}
\beta_2\approx \Imt[\beta_2]\approx i\frac{\Gamma_\text{1D}}
{2\Delta_\text{c}\left(1-(\pi^2|\Omega_0|^2)/(\Gamma_\text{1D}^2n^2)\right)}
\approx i\frac{\Gamma_\text{1D}}{2\Delta_\text{c}}
+i\frac{\pi^2|\Omega_0|^2}{2\Delta_\text{c}\Gamma_\text{1D}n^2}.
\end{gather}
Inserting into the second term of Eq.~\eqref{beta2_approx1} gives
\begin{gather}\label{Re_beta2_approx2}
\Ret[\beta_2]\approx \frac{\Gamma_\text{1D}\Gamma'}
{4\Delta_\text{c}^2
\left(1-(\pi^2|\Omega_0|^2)/(\Gamma_\text{1D}^2n^2)\right)^2}
\approx \frac{\Gamma_\text{1D}\Gamma'}{4\Delta_\text{c}^2}
+\frac{\pi^2\Gamma'|\Omega_0|^2}{2\Delta_\text{c}^2\Gamma_\text{1D}n^2}.
\end{gather}
Using the first and the third terms of 
Eq.~\eqref{Delta1_resonance_solution} and inserting them into 
Eq.~\eqref{beta3_approx1}, we find
\begin{gather}\label{beta3_approx2}
\beta_3\approx \Imt[\beta_3]\approx -i\frac{\Gamma_\text{1D}}{2}
\left(-\frac{\pi^2\Delta_\text{c}}{\Gamma_\text{1D}^2n^2}
+\frac{\pi^4\Delta_\text{c}|\Omega_0|^2}{\Gamma_\text{1D}^4n^4}\right)
\approx i\frac{\pi^2\Delta_\text{c}}{2\Gamma_\text{1D} n^2} 
-i\frac{\pi^4\Delta_\text{c}|\Omega_0|^2}{2\Gamma_\text{1D}^3n^4}.
\end{gather}
Combining Eqs. \eqref{Re_beta2_approx2} and \eqref{beta3_approx2} and 
neglecting a term of order $n^{-6}$ results in
\begin{gather}
\Imt[\beta_2\beta_3]\approx\Ret[\beta_2]\Imt[\beta_3]
\approx \frac{\pi^2\Gamma'}{8\Delta_\text{c}n^2}
+\frac{\pi^4\Gamma'|\Omega_0|^2}{8\Delta_\text{c}\Gamma_\text{1D}^2n^4}.
\end{gather}
Invoking Eq.~\eqref{optimal_Deltac_proportional} again and neglecting the $\Omega_0$ 
dependent terms for a moment, we see that $\beta_2\propto n^{-3/4}$, 
$\beta_3\propto n^{-5/4}$ and $n^3\Imt[\beta_2\beta_3]\propto n^{1/4}$. Hence, 
$\beta_3 \ll \beta_2$; 
$\beta_2,\beta_3 \ll (n^3\Imt[\beta_2\beta_3])^{-1}$, and we can approximate 
Eqs. \eqref{r_T_n_approx1} and \eqref{t_T_n_approx1} (now including the $\Omega_0$ 
dependent terms) by

\begin{align}
\label{r_0_sm}
&r_0\approx -i\frac{2n^3}{\pi^2}\beta_2\Imt[\beta_2\beta_3]
\approx \frac{\Gamma_\text{1D}\Gamma'n}{8\Delta_\text{c}^2}
+\frac{\pi^2\Gamma'|\Omega_0|^2}{4\Delta_\text{c}^2\Gamma_\text{1D}n},\\
\label{t_0_sm}
&t_0\approx(-1)^{n-1}\del{1+i\frac{2n^3}{\pi^2}\beta_2\Imt[\beta_2\beta_3]}
\approx (-1)^{n-1}\del{1-\frac{\Gamma_\text{1D}\Gamma'n}{8\Delta_\text{c}^2}
-\frac{\pi^2\Gamma'|\Omega_0|^2}{4\Delta_\text{c}^2\Gamma_\text{1D}n}}.
\end{align}
Since the number of atoms is $N=2n$, the above expressions correspond to Eqs. 
\eqref{r_ensemble_N} and \eqref{t_ensemble_N} of the main 
article, except for the removal of the overall phase factor $(-1)^{n-1}$ for 
the transmission coefficient $t_0$ (discussed in 
Sec.~\ref{Sec_sagnac_interferometer_and_phases} below) and ignoring the shown 
$\Omega_0$ dependent terms (they get canceled in the fidelity calculations 
in Sec.~\ref{sec_analytical_fidelity} below). In 
Fig.~\ref{fig_r_and_t_plot_Omega}(a) we 
plot $|t_0|^2$ as a function of $\Omega_0$ and show that the approximate analytical 
expression in Eq.~\eqref{t_0_sm} matches the full expression in 
Eq.~\eqref{t_T_n} evaluated at the resonance frequency $\delta_\text{res}$ 
(found numerically).

\begin{figure}[bt]
\begin{center}
\includegraphics{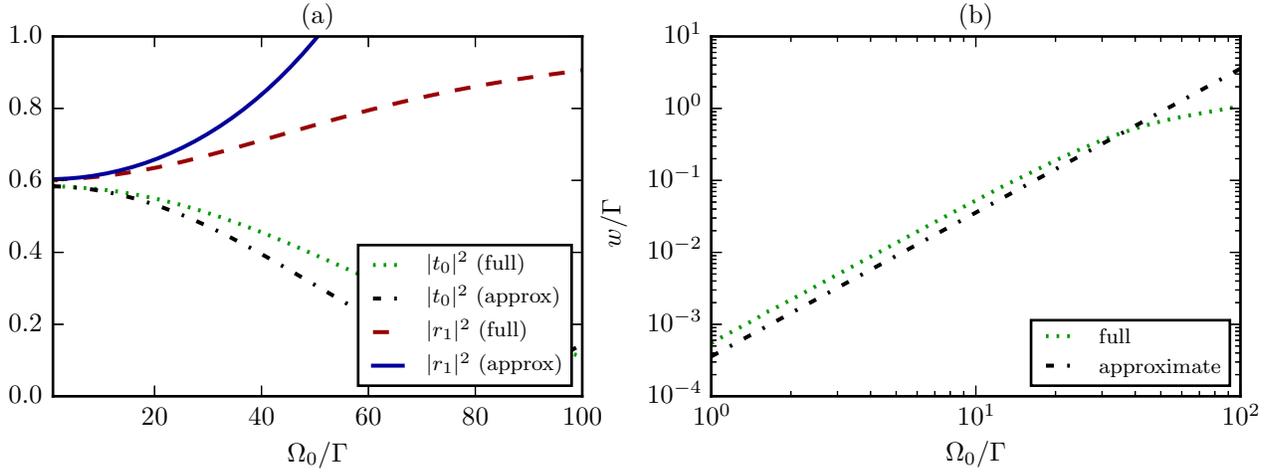}
\end{center}
\caption{(a) Reflectance with a stored photon $|r_1|^2$ and 
transmittance without a stored photon $|t_0|^2$ of an ensemble of $\Lambda$\mbox{-}type atoms plotted as 
functions of the 
Rabi frequency of the classical drive $\Omega_0$ and evaluated at the 
transmission resonance closest to $\delta=0$ (see 
Fig.~\ref{fig_r_and_t_plot} of the main article), i.e. at 
$\delta=\delta_\text{res}$. Both 
$t_0$ and $r_1$ are calculated either numerically by directly using 
Eqs.~\eqref{t_T_n}~and~\eqref{r_T_e}, respectively (``full''), or
from the approximate expressions in Eqs.~\eqref{t_0_sm}~and~\eqref{r_1_sm_Omega},
respectively (``approx''). For the approximate expressions, the curves are 
shifted such that the values for $\Omega_0=0$ are the same for both the full 
and approximate curves. This is done to show the dependency on $\Omega_0$ 
more clearly. For the unshifted curves, there is 
still visible difference even for $\Omega_0=0$. This difference disappears as 
$N$ increases (and $\Delta_\text{c}$ is changed accordingly). (b) The width of the transmission resonance 
calculated either from the full expression as 
$w=\eval[2]{\text{Re}\sbr{\sqrt{4/\partial_\delta^2 t_0(\delta)}}}_{\delta=\delta_\text{res}}$
 or using the approximate 
expression \eqref{delta_res_width} 
(dash-dotted black).
The parameters for both (a) and (b) are: $N=10^4$, 
$\Gamma_\text{1D}/\Gamma=0.05$, and $\Delta_\text{c}/\Gamma=-10$. (The same 
as in Fig.~\ref{fig_r_and_t_plot} of the main article except for 
$\Omega_0$ that is varied here.)}
\label{fig_r_and_t_plot_Omega}
\end{figure}

To account for the non-zero bandwidth of the scattered photon, we also need 
the width of the resonance. After expanding the reflection coefficient $r_0$ 
around $\delta_\text{res}$, we get
\begin{gather}\label{r_0_expansion_around_delta_res}
r_0(\delta)\approx(2/w^2)(\delta-\delta_\text{res})^2+r_0(\delta_\text{res}),
\end{gather}
where $r_0(\delta_\text{res})$ is given by Eq.~\eqref{r_0_sm} and
\begin{gather}\label{delta_res_width}
w=\frac{32\sqrt{2}\pi^2\Delta_\text{c}^2|\Omega_0|^2}
{\Gamma_\text{1D}^3N^3},
\end{gather}
is the width. Since $t_0\approx 1-r_0$, the width of the 
transmission resonance is $w$. In Fig.~\ref{fig_r_and_t_plot_Omega}(b), we 
compare Eq.~\eqref{delta_res_width} with the numerically computed width.

\subsubsection{With a stored photon}
\label{Lambda_r_t_with_stored_photon}
Here, we derive the scattering coefficients $r_1$ and $t_1$ given by Eqs. 
\eqref{r_imp_ensemble_N} and \eqref{t_imp_ensemble_N} of the 
main article. The starting point is the assumption that the photon has been 
stored in a single atom that is placed at the anti-node of the standing wave 
of the classical drive (storing in an atom that is on the node will have a 
negligible change in the scattering properties, unless 
$\Gamma_\text{1D}/\Gamma$ is close to unity). The storage of a photon in the 
atom transfers it from state $|a\rangle$ to state $|d\rangle$, such that it 
behaves like a resonant two-level atom (given by the 
$|d\rangle\leftrightarrow|e\rangle$ transition) that is described by the 
transfer matrix $T_\text{a}(\beta_{2,de})$ with
\begin{gather}\label{beta2_de}
\beta_{2,de}=\frac{\Gamma_\text{1D}}{\Gamma'}.
\end{gather}
Hence, the transfer matrix for the unit cell containing the stored photon is
\begin{gather}\label{matrix_T_lambda_over_4_photon}
T_\text{cell,ph}
=T_\text{f}(\pi/2)
T_\text{a}(\beta_2)
T_\text{f}(\pi/2)
T_\text{a}(\beta_{2,de}),
\end{gather}
instead of Eq.~\eqref{matrix_T_lambda_over_4}. If we assume that the photon is 
stored in the unit cell with index $n_\text{ph}$, the transfer matrix for the 
whole ensemble $T_{\text{e},n_\text{ph}}$ is given by
\begin{gather}\label{ensemble_transfer_matrix_with_impurity}
T_{\text{e},n_\text{ph}}=T_\text{cell}^{n-n_\text{ph}}T_\text{cell,ph}
T_\text{cell}^{n_\text{ph}-1}.
\end{gather} 
From 
Eq.~\eqref{ensemble_transfer_matrix_with_impurity} we can find the 
scattering coefficients as 
\begin{align}\label{r_T_e}
&r_{1,n_\text{ph}}
=-\frac{T_{\text{e},n_\text{ph},{21}}}{T_{\text{e},n_\text{ph},{22}}},\\
\label{t_T_e}
&t_{1,n_\text{ph}}
=\frac{1}{T_{\text{e},n_\text{ph},{22}}},
\end{align}
where $T_{\text{e},n_\text{ph},{jj'}}$ are the elements of the matrix 
$T_{\text{e},n_\text{ph}}$. 

For the numerical calculations, 
Eqs.~\eqref{r_T_e}~and~\eqref{t_T_e} are used directly.
For the analytical calculations, we can find approximate expressions for the 
scattering coefficients, but the procedure is rather involved. We shall 
therefore restrict ourselves to a brief discussion of the main steps. We do 
several simplifications on the (very complicated) expressions resulting from 
Eqs. \eqref{r_T_e} and \eqref{t_T_e}. We use the fact that 
${\beta_2\beta_3=(-1-\cos(\theta))/2}$ and the approximate expression 
$\theta\approx\pi(n-1)/n$. Also, while expanding the numerator and denominator 
around large $n$, we use that $\beta_2\propto n^{-3/4}$ (a consequence of
Eqs.~\eqref{beta2}~and~\eqref{optimal_Deltac_proportional}) to determine which 
terms can be neglected. Then we replace the index of the unit cell with the 
stored photon $n_\text{ph}$ by $n\tilde{z}$, where $\tilde{z}=z/L$ is the 
rescaled position coordinate. After further approximating $1/n\approx 0$ and 
$n\pm 1\approx n$, we get
\begin{align}
&r_1(\tilde{z})=-\frac{\beta_{2,de} (\pi \cos (\pi \tilde{z})-2 \beta_2 n \sin (\pi \tilde{z}))^2}{\sin ^2(\pi \tilde{z}) \left(\pi ^2-4 \beta_2^2 \beta_{2,de} n^2\right)-2 \pi \beta_2^2 \beta_{2,de} n \sin (2 \pi \tilde{z})+\pi ^2 (\beta_{2,de}+1) \cos ^2(\pi \tilde{z})},\\
&t_1(\tilde{z})
=\frac{(-1)^{n-1}\pi ^2 }{\sin ^2(\pi \tilde{z}) \left(\pi ^2-4 \beta_2^2 \beta_{2,de} n^2\right)-2 \pi \beta_2^2 \beta_{2,de} n \sin (2 \pi \tilde{z})+\pi ^2 (\beta_{2,de}+1) \cos ^2(\pi \tilde{z})}.
\end{align}

Next, we insert the expressions for $\beta_2$ and $\beta_{2,de}$ with the 
approximation $\Delta\approx\Delta_\text{c}$, expand around $\tilde{z}=1/2$, 
and use $|\Delta_\text{c}|\propto \Gamma_\text{1D}n^{3/4}$ to identify which 
terms are dominant for large $n$. This results in
\begin{align}
\label{r_1_sm_z}
&r_1(\tilde{z})\approx
1
-\frac{\pi^2\Delta_\text{c}^2\Gamma'}{\Gamma_\text{1D}^3 n^2}
-\frac{2i\pi^2\Delta_\text{c}}{\Gamma_\text{1D} n}
\left(\tilde{z}-\frac{1}{2}\right)
-\frac{\pi^4\Delta_\text{c}^2(2\Gamma_\text{1D}+\Gamma')}{\Gamma_\text{1D}^3 n^2}
\left(\tilde{z}-\frac{1}{2}\right)^2,\\
\label{t_1_sm_z}
&t_1(\tilde{z})\approx (-1)^{n-1}\del{
\frac{\pi^2\Delta_\text{c}^2\Gamma'}{\Gamma_\text{1D}^3 n^2}
+\frac{\pi^4\Delta_\text{c}^2\Gamma'}{\Gamma_\text{1D}^3 n^3}
\left(\tilde{z}-\frac{1}{2}\right)
+\frac{\pi^4\Delta_\text{c}^2\Gamma'}{\Gamma_\text{1D}^3 n^2}
\left(\tilde{z}-\frac{1}{2}\right)^2
}.
\end{align}

We see that the resulting expressions do not depend on $\Omega_0$. This is a 
consequence of approximating $\Delta\approx\Delta_\text{c}$. If we use 
$\Delta=\Delta_\text{c}+\delta$ together with 
Eq.~\eqref{Delta1_resonance_solution}, we find corrections from the dependence 
on $\Omega_0$. We only need the first term in 
Eq.~\eqref{Delta1_resonance_solution} to find the lowest order correction due 
to $\Omega_0$. At $\tilde{z}=1/2$, and expanding around large $n$, we have
\begin{align}
\label{r_1_sm_Omega}
&r_1
\approx 1
-\frac{\pi^2\Delta_\text{c}^2\Gamma'}{\Gamma_\text{1D}^3 n^2}
+\frac{2\pi^4 \Delta_\text{c}^2 \Gamma' |\Omega_0|^2}{\Gamma_\text{1D}^5 n^4},\\
\label{t_1_sm_Omega}
&t_1\approx
(-1)^{n-1}\del{\frac{\Delta_\text{c}^2\pi^2\Gamma'}{\Gamma_\text{1D}^3n^2}
-\frac{2\pi^4 \Delta_\text{c}^2 \Gamma' |\Omega_0|^2}{\Gamma_\text{1D}^5 n^4}}.
\end{align}
In Fig.~\ref{fig_r_and_t_plot_Omega}(a), we plot $|r_1|^2$ as a 
function of $\Omega_0$ and show that the analytical expression in 
Eq.~\eqref{r_1_sm_Omega} matches the full expression in 
Eq.~\eqref{r_T_e} evaluated at the resonance frequency. In Eq.~\eqref{r_imp_ensemble_N} of the main article, we 
include all error terms of Eq.~\eqref{r_1_sm_z} but ignore $\Omega_0$ 
dependent error term from Eq.~\eqref{r_1_sm_Omega} (since it gets canceled in the fidelity calculations 
in Sec.~\ref{sec_analytical_fidelity} below). Likewise, in 
Eq.~\eqref{t_imp_ensemble_N} of the main article, we include all error 
terms of Eq.~\eqref{t_1_sm_z} but ignore the $\Omega_0$ dependent error term 
from Eq.~\eqref{t_1_sm_Omega}.

\subsection{Reflection and transmission for the dual\mbox{-}V scheme}

For the dual\mbox{-}V scheme, we need two-mode ($4\times 4$) transfer matrices 
to describe the $\sigma_+$ and $\sigma_-$ polarized modes. The transfer 
matrices for the atoms in Eq.~\eqref{multi_mode_transfer_matrices_a_f} have 
the form $T_a(\beta_j)$ ($j$ is the index of the atom)
where~\cite{dispersion-relations_sm} 
\begin{gather}
\beta_j=-(I+S_{j,r})^{-1}S_{j,r},
\end{gather}
and
\begin{gather}\label{S_jr_def}
S_{j,r}=
\begin{pmatrix}
r_{j,++} & r_{j,-+}\\
r_{j,+-} & r_{j,--}
\end{pmatrix}.
\end{gather}
In the expressions above, we have used the same convention as in 
Eqs.~\eqref{input_E_for_T}, i.e. that the first element of $E_\pm$ is the 
$\sigma_+$ component and the second element is the $\sigma_-$ component. 
Hence, in Eq.~\eqref{S_jr_def}, $r_{j,+-}$ for example means the reflection 
coefficient of the process where an incident $\sigma_+$ field is reflected 
also into an outgoing $\sigma_-$ field. Similarly for the other reflection 
coefficients.

If the atom is in state $|a\rangle$ (without a stored photon), the 
elements of $S_{j,r}$ are given by~\cite{dispersion-relations_sm}
\begin{align}
&r_{j,--}=r_{j,++}
=-\frac{i(\Gamma_\text{1D}/2)\del{\Delta_\Gamma\delta
-|\Omega_0|^2}}
{\Delta_\Gamma^2\delta
-2\Delta_\Gamma|\Omega_0|^2},\displaybreak[0]\\
&r_{j,-+}
=-\frac{i(\Gamma_\text{1D}/2)|\Omega_0|^2}
{\Delta_\Gamma^2\delta
-2\Delta_\Gamma|\Omega_0|^2}e^{2ik_0 z_j},\\
&r_{j,+-}
=-\frac{i(\Gamma_\text{1D}/2)|\Omega_0|^2}
{\Delta_\Gamma^2\delta
-2\Delta_\Gamma|\Omega_0|^2}e^{-2ik_0 z_j},
\end{align}
where $\Delta_\Gamma=\Delta+i(\Gamma/2)$.
If the atom is in state $|d\rangle$ (with a stored photon), it acts as a 
resonant V-type atom, and hence the elements of $S_{j,r}$ are given by
\begin{align}
&r_{j,--}=r_{j,++}=-\frac{\Gamma_\text{1D}}{\Gamma},\\
&r_{j,-+}=r_{j,+-}=0.
\end{align}
We only calculate the reflection and transmission coefficients of ensembles of dual-V 
atoms numerically.

\section{Sagnac interferometer and adjustment of the phases}
\label{Sec_sagnac_interferometer_and_phases}
Here, we calculate the result of scattering from the Sagnac interferometer 
shown in Fig.~\ref{fig_setup}(d) of the main article. The sequential 
picture of the scattering is that the incident field on one of the ports is 
split by the 50:50 beam splitter, gets scattered by the ensemble, and then the 
transmitted and reflected parts will again interfere on the same beam 
splitter. Thus the matrix that relates the outputs to the inputs can be written
\begin{gather}\label{M_Sagnac_decomposition}
M_\text{Sagnac}=HM_\text{f}SM_\text{f}H,
\end{gather}
where the matrix $H$ describes the beam splitter, the matrix $M_\text{f}$ 
describes the free propagation, and the matrix $S$ describes scattering from 
the ensemble.

We choose the phases of the beam splitter such that it performs 
the Hadamard operation on the field, i.e.
\begin{gather}
H=\frac{1}{\sqrt{2}}
\begin{pmatrix}
1 & 1\\
1 & -1
\end{pmatrix}.
\end{gather}
The free propagation matrix is
\begin{gather}
M_\text{f}=\begin{pmatrix}
e^{ik_0 l_1} & 0\\
0 & e^{-ik_0 l_2}
\end{pmatrix},
\end{gather}
where $l_1$ and $l_2$ are lengths of propagation from the beam splitter until 
either end of the ensemble.

The ensemble can, in general, have different transmission and reflection 
coefficients depending on, whether the field is incident from the left or 
right. Therefore, we write
\begin{gather}
S=\begin{pmatrix}
r_+ & t_+\\
t_- & r_-
\end{pmatrix},
\end{gather}
where $r_+$ and $t_+$ are respectively the reflection and transmission 
coefficients when the field is incident from the left (propagating in the 
positive direction), and $r_-$ and $t_-$ are respectively the reflection and 
transmission coefficients when the field is incident from the right 
(propagating in the negative direction).

Multiplying the matrices, we get
\begin{gather}\label{M_Sagnac}
M_\text{Sagnac}=\frac{1}{2}
\begin{pmatrix}
r_{+}e^{2ik_0 l_1}+(t_{+}+t_{-})e^{ik_0 (l_1+l_2)}+r_{-}e^{2ik_0 l_2}
&r_{+}e^{2ik_0 l_1}-(t_{+}-t_{-})e^{ik_0 (l_1+l_2)}-r_{-}e^{2ik_0 l_2}\\
r_{+}e^{2ik_0 l_1}+(t_{+}-t_{-})e^{ik_0 (l_1+l_2)}-r_{-}e^{2ik_0 l_2}
&r_{+}e^{2ik_0 l_1}-(t_{+}+t_{-})e^{ik_0 (l_1+l_2)}+r_{-}e^{2ik_0 l_2}
\end{pmatrix}.
\end{gather}
For an empty non-rotating Sagnac interferometer where $r_+=r_-=0$ and 
$t_+=t_-=1$, we recover the well known result, that light always leaves the port 
in which it is incident~\cite{bertocchi_jpb2006_sm,bradford_prl2012_sm}. Once we 
have put a scatterer inside the interferometer, this result is still true if 
$r_+=r_-$, $t_+=t_-$, and $l_1=l_2$.
Because the equalities $r_+=r_-$, $t_+=t_-$, and $l_1=l_2$ need not be true, the off-diagonal entries of 
$M_\text{Sagnac}$ are, in general, non-zero and describe the leakage of the 
incident power to the other port of the Sagnac interferometer. However, due to 
the sequential operation of the gate, this leakage does not introduce any 
logic errors: since scattering of photon~$B$ happens while photon~$A$ is 
stored inside the ensemble, leakage of photon~$B$ into the rail that encodes 
state $|1\rangle_A$ (see Fig.~\ref{fig_setup}(c) of the main article) 
is separated in time from the subsequent retrieval of photon~$A$ and hence can 
be either absorbed or rerouted along a different path.

Even though the off-diagonal entries of $M_\text{Sagnac}$ are non-zero in 
general, they are strongly suppressed in the ideal limit, since e.g. both $r_+$ 
and $r_-$ approach the same value (either $1$ or $0$ depending on whether a 
photon was stored in the ensemble or not). As a concrete example, for the 
$\Lambda$\mbox{-}type scheme, we can use Eq.~\eqref{matrix_T_pow_n_Lambda} and 
find
\begin{align}
\label{r_0_plus}
&r_{0+}=-\frac{T_{\text{e},{21}}}{T_{\text{e},{22}}}
=\frac{-\beta_2+\beta_3+2\beta_2\beta_3}
{\frac{\cos(n\theta)}{\sin(n\theta)}\sin(\theta)-(\beta_2+\beta_3)},\\
\label{r_0_minus}
&r_{0-}=\frac{T_{\text{e},{12}}}{T_{\text{e},{22}}}
=\frac{-\beta_2+\beta_3-2\beta_2\beta_3}
{\frac{\cos(n\theta)}{\sin(n\theta)}\sin(\theta)-(\beta_2+\beta_3)},
\end{align}
where $r_{0+}$ is the same as $r_0$ in Eq.~\eqref{r_T_n}, and $r_{0-}$ is the 
reflection coefficient where the field is incident from the right instead 
of the left. The difference between 
Eqs.~\eqref{r_0_plus}~and~\eqref{r_0_minus} is only in the sign of the term 
$2\beta_2\beta_3$ in the numerator. As discussed above Eq.~\eqref{r_0_sm}, 
this term is much smaller than $\beta_2$ in the limit of a large number of atoms 
and has therefore been neglected in Eq.~\eqref{r_0_sm}. Hence, 
Eq.~\eqref{r_0_sm} can serve as an approximate expression for both $r_{0+}$ 
and $r_{0-}$. We also note that it can be shown that the transmission 
coefficient for any $2\times 2$ transfer matrix is independent of whether the 
field is incident from one side or the other (given by Eq.~\eqref{t_T_n} for 
$t_0$). However, we still account for the possible difference in the 
scattering coefficients in the numerical calculations (both for $2\times 2$ 
and $4\times 4$ transfer matrices) by using the matrix element 
$M_{\text{Sagnac},22}$ from Eq.~\eqref{M_Sagnac} to define the scattering 
coefficients  $R_1$ and $R_0$ needed in
Eqs.~\eqref{phi_out_1_definition}~and~\eqref{CJ_fidelity_computational_basis_final},
i.e.
\begin{align}
\label{R_0_definition_two_sided}
&R_0=-\frac{1}{2}\del{r_{0+}e^{2ik_0 l_1}-(t_{0+}+t_{0-})e^{ik_0 (l_1+l_2)}+r_{0-}e^{2ik_0 l_2}},\\
\label{R_1_definition_two_sided}
&R_1=-\frac{1}{2}\del{r_{1+}e^{2ik_0 l_1}-(t_{1+}+t_{1-})e^{ik_0 (l_1+l_2)}+r_{1-}e^{2ik_0 l_2}},
\end{align}
where the scattering coefficients with ``$+$'' in the subscript assume incident 
photon from the left of the ensemble, and the scattering coefficients with ``$-$'' 
in the subscript assume incident photon from the right of the ensemble.
In most of the calculations, we assume that 
$e^{2ik_0 l_1}=e^{2ik_0 l_2}=e^{ik_0 (l_1+l_2)}=1$. This is satisfied if $l_1$ 
and $l_2$ are integer multiples of the wavelength $2\pi/k_0$. To find the 
error introduced by $l_1$ or $l_2$ deviating from these values, we both do a 
numerical and an analytical calculation in Sec.~\ref{sec_analytical_fidelity} below.

Since the distance between the atoms is fixed in our calculations, as we 
change the number of atoms, we also change the length of the ensemble and 
consequently also the total round trip length of the Sagnac interferometer. 
This introduces additional phase factors which need to be accounted for. There 
is an overall 
phase factor $(-1)^{n-1}$ in Eqs. \eqref{t_0_sm}, \eqref{t_1_sm_z}, and 
\eqref{t_1_sm_Omega} compared with Eqs. \eqref{t_ensemble_N} and 
\eqref{t_imp_ensemble_N} of the main article. This phase factor 
reflects the fact the ensemble length changes by half a wavelength (i.e. 
$2d=\pi/k_0$) whenever a unit cell is added to the ensemble. When 
$r_{0\pm}\approx 0$ and $|t_{0\pm}|\approx 1$, the factor $(-1)^{n-1}$ 
directly appears as the overall phase of $R_0$. Hence, without the adjustment, 
the ideal value $R_0=1$ can only be obtained for odd $n$. There are similar 
phase factor considerations for $R_1$ due to the phase factor of the 
reflection coefficient $r_{1-}$ ($r_{1+}$ is assumed to be at $z=0$ and hence 
does not acquire phase factors with changing ensemble length). To address this 
issue, we require that the total round trip of the interferometer to be 
equal to an odd number of half wavelengths and thus independent of the precise 
atom number. In the calculations, this is accomplished by adding suitable 
length of free propagation $d_\text{extra}$ such that 
$\exp(ik_0 (L+l_1+l_2+d_\text{extra}))=-1$. Experimentally, stabilizing the 
total length of the interferometer will suffice.

In general (e.g. for the dual-V scheme), we can choose any inter-atomic spacing $d$ that is 
not a multiple of $\pi/(2k_0)$ (see Fig.~\ref{fig_cphase_fidelity_kd} below). 
For this general case, we expect that instead of the overall phase factor 
$(-1)^{n-1}$, the transmission coefficients have the phase factor 
$\exp(ik_0 L+\pi)$. As noted above, we remove this phase factor in the 
calculations by adding a distance of free propagation $d_\text{extra}$ to the 
right of the ensemble chosen such that 
$\exp(ik_0 d_\text{extra})=\exp(-ik_0 L-\pi)$. Multiplying this extra matrix 
of free propagation modifies the scattering coefficients according to
\begin{gather}
\begin{aligned}
&r_{0+}\rightarrow r_{0+},\\
&r_{0-}\rightarrow r_{0-}\exp(2ik_0 d_\text{extra}),\\
&t_{0\pm}\rightarrow t_{0\pm}\exp(ik_0 d_\text{extra}).
\end{aligned}
\end{gather}
The physical interpretation of this mathematical result is that, since the 
free propagation was added on the right of the ensemble, then reflection for 
the field incident from the left ($r_{0+}$) is unaffected, while the 
reflection coefficient for fields incident from the right ($r_{0-}$) acquires 
twice the propagation phase. The transmission coefficients only acquire the 
propagation phase once.

\section{EIT storage and retrieval}
\label{eit_storage_retrieval}
\begin{figure}[bt]
\begin{center}
\includegraphics{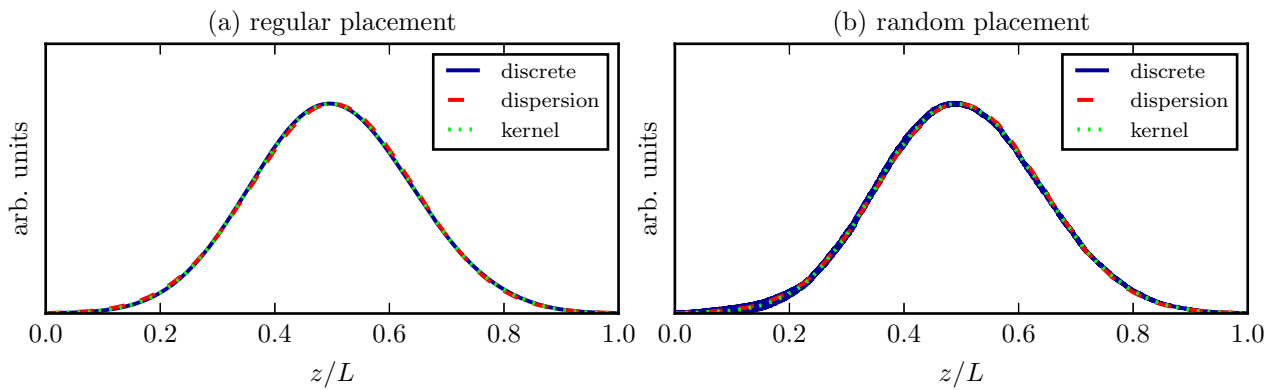}
\end{center}
\caption{The stored Gaussian spin wave computed using three different models for EIT storage: 
the dispersion relation of Sec.~\ref{dispersion_eit_storage_retrieval} 
(``dispersion''), the fully discrete theory of 
Sec.~\ref{discrete_eit_storage_retrieval} (``discrete''), and the storage 
kernel of Sec.~\ref{continuum_eit_storage_retrieval} (``kernel''). Contrary to the 
fidelity 
calculations, the field is incident from one side (left) only to show the influence of 
random placement of the atoms more clearly. The common parameters for the two subplots are 
$\Gamma_\text{1D}/\Gamma=0.05$, $N=10^4$, $\Omega_0/\Gamma=1$, $\sigma/L=0.1$ 
(width of the stored Gaussian spin wave). (a) Regularly placed atoms are assumed with 
inter-atomic distance $d=0.266\pi/k_0$. All three curves are nearly 
indistinguishable from each other. (b) Randomly placed atoms where the position 
of each atom is chosen from a uniform distribution over the whole ensemble. 
The average density is the same as in (a). The curve for the fully discrete 
storage differs slightly from the two others and exhibits rapid variation 
with position that appears as if the line itself is thicker. This rapid variation 
becomes more visible for higher values of $\Gamma_\text{1D}/\Gamma$.}
\label{fig_eit_storage_models_comparison}
\end{figure}

\subsection{Overview}
We model the EIT storage and retrieval process in three different ways:
\begin{enumerate}
\item Using the dispersion relation (see for instance 
Ref.~\cite{chang_njp11a_sm} and Sec.~\ref{dispersion_eit_storage_retrieval} below).
\item Using the fully discrete theory (see Ref.~\cite{caneva_discrete_model_sm} and 
Sec.~\ref{discrete_eit_storage_retrieval} below).
\item Using the storage and retrieval kernels (see Ref.~\cite{gorshkov_pra07_2_sm} and 
Sec.~\ref{continuum_eit_storage_retrieval} below).
\end{enumerate}

We consider the so-called adiabatic EIT storage \cite{gorshkov_pra07_2_sm}
where a single-photon wave packet is incident on the 
ensemble and is mapped onto a spin wave. Using the state labels in 
Fig.~\ref{fig_setup}(a) and Fig.~\ref{fig_setup}(b) of the 
main article, a 
spin wave is a superposition of states where a 
single atom is in state~$|c\rangle$ and the rest are in state~$|a\rangle$. We 
assume a constant Rabi frequency 
of the classical drive $\Omega(z)=\Omega_0$, but choosing a co-propagating 
classical drive with Rabi frequency $\Omega(z)=\Omega_0e^{\pm ik_0z}$ will only change the 
spatially-dependent phase factor of the stored spin wave.
In the limit of high storage efficiency, the temporal profile of the photon and the 
spatial profile of the stored spin wave will approximately have the same form. E.g. if a 
photon with Gaussian temporal wave packet is stored, the resulting spin wave 
will have a Gaussian spatial profile. This is a consequence of the 
time-independent Rabi frequency of the classical drive, and it allows us 
to use the EIT dispersion relation to describe the storage and retrieval.

The EIT dispersion relation is used to gain intuition about the storage 
and retrieval process and also for the analytical calculations. In the numerical 
calculations of the fidelities and success probability, the atoms are always 
modeled as being discrete. However, the fully discrete EIT storage and retrieval 
model~(Sec.~\ref{discrete_eit_storage_retrieval}) becomes very computationally 
demanding as the
number of atoms increases. Therefore, to be able to calculate fidelities with 
a large number of atoms, we instead use the less computationally demanding 
storage and retrieval kernels derived using the continuum model of the atomic ensemble and suitably 
discretized (Sec.~\ref{continuum_eit_storage_retrieval}). In 
Fig.~\ref{fig_eit_storage_models_comparison}(a), we show that the three models 
agree very well for the case of regularly placed atoms with large optical 
depth $d_\text{opt}=2N\Gamma_\text{1D}/\Gamma=1000$. For randomly placed 
atoms in Fig.~\ref{fig_eit_storage_models_comparison}(b), the fully discrete theory gives slightly different results compared to 
using the discretized continuum theories (dispersion relation or storage 
kernel). If $\Gamma_\text{1D}/\Gamma$ is 
increased, there are more significant differences that we believe to be caused by the continuum theory not accounting for the reflection of parts of the propagating excitation due to disorder. Since we consider a relatively small 
$\Gamma_\text{1D}/\Gamma=0.05$ in most of our numerical examples, we always 
use the 
discretized continuum kernels for storage and retrieval, even for 
randomly placed atoms (in Fig.~\ref{fig_cphase_fidelity_random_placement} below).

\subsection{Using the dispersion relation}
\label{dispersion_eit_storage_retrieval}
The adiabatic EIT storage and retrieval can be modeled in a particularly simple way if the 
influence of the interface between the atomic medium and vacuum is ignored. To 
use the EIT dispersion relation, we need to assume that the 
ensemble is of infinite extent. However, to compute the storage and 
retrieval efficiency, we need to assume propagation through a finite ensemble. 
In the calculations below, this is reflected in infinite bounds for 
the integration but a finite propagation length.
The only processes that happen in this model is that the stored photon 
wave packet broadens in space as it propagates, and its norm decays due to 
spontaneous emission. The EIT storage and retrieval efficiency will then be 
the norm of the wave packet that has propagated for the full length of the 
ensemble $L$ (with a stop at $L/2$ to allow for the second photon to be 
scattered off the ensemble).

The EIT dispersion relation is~\cite{chang_njp11a_sm}
\begin{gather}
\delta_k\approx v_\text{g} (k-k_0) + \frac{1}{2}\alpha (k-k_0)^2
\end{gather}
with
\begin{gather}\label{eit_group_velocity_and_attenuation_not_rescaled}
v_\text{g}
=\frac{2L|\Omega_0|^2}{N\Gamma_\text{1D}},
\quad
\alpha
=-i\frac{4L^2|\Omega_0|^2\Gamma'}{N^2\Gamma_\text{1D}^2}.
\end{gather}

In rescaled coordinates $\tilde{z}=z/L$ and wave vectors $\tilde{k}=kL$ the 
dispersion relation can be written
\begin{gather}
\delta_{\tilde{k}}\approx \tilde{v}_\text{g} (\tilde{k}-\tilde{k}_0) + \frac{1}{2}\tilde{\alpha} (\tilde{k}-\tilde{k}_0)^2
\end{gather}
with
\begin{gather}\label{eit_group_velocity_and_attenuation}
\tilde{v}_\text{g}
=\frac{v_\text{g}}{L}
=\frac{2|\Omega_0|^2}{N\Gamma_\text{1D}},
\quad
\tilde{\alpha}
=\frac{\alpha}{L^2}
=-i\frac{4|\Omega_0|^2\Gamma'}{N^2\Gamma_\text{1D}^2}.
\end{gather}

We only consider incident photons where the temporal profile is Gaussian. 
When such a photon is mapped onto a stored excitation, this results in 
an approximately Gaussian spatial profile of the form
\begin{gather}\label{gaussian_wavefunction}
S(\tilde{z})=\frac{1}{(2\pi\tilde{\sigma}^2)^{1/4}}
\exp\del{-\frac{(\tilde{z}-\tilde{\mu})^2}{4\tilde{\sigma}^2}}
e^{i\tilde{k}_0\tilde{z}},
\end{gather}
where the rescaled quantities are $\tilde{\sigma}=\sigma/L$, 
$\tilde{\mu}=\mu/L$, $\tilde{k}_0=\tilde{k}_0L$. The spatial profile at later 
times, can be found by Fourier transforming $S(z)$, multiplying the Fourier 
transform by $\exp(-i\delta_{\tilde{k}} t)$, and taking the inverse Fourier 
transform. We obtain
\begin{gather}\label{gaussian_wavefunction_arbitrary_time}
S(\tilde{z},t)=\frac{1}{(2\pi\tilde{\sigma}^2)^{1/4}}
\sqrt{\frac{1}{1+i\tilde{\alpha} t/(2\tilde{\sigma}^2)}}
\exp\del{-\frac{(\tilde{z}-\tilde{\mu}-\tilde{v}_\text{g} t)^2}
{4\tilde{\sigma}^2\del{1+i\tilde{\alpha} t/(2\tilde{\sigma}^2)}}}
e^{i\tilde{k}_0\tilde{z}}.
\end{gather}

The norm squared of the wave packet at a time $t\geq 0$ is given by
\begin{gather}\label{gaussian_wavefunction_norm_arbitrary_time}
\mathcal{N}_S^2(t)
=\int_{-\infty}^\infty |S(\tilde{z},t)|^2 \dif\tilde{z}
=\frac{1}{\sqrt{1+i\tilde{\alpha} t/(2\tilde{\sigma}^2)}}.
\end{gather}
The combined storage and retrieval efficiency $\eta_\text{EIT}$ is given by Eq. 
\eqref{gaussian_wavefunction_norm_arbitrary_time} with $t=1/\tilde{v}_\text{g}=L/v_\text{g}$, i.e. 
the time required to pass the whole ensemble. We thereby get
\begin{gather}\label{gaussian_wavefunction_norm_after_passing_ensemble}
\eta_\text{EIT}=\mathcal{N}_S^2(t=1/\tilde{v}_\text{g})
=\frac{1}{\sqrt{1+\frac{\Gamma'}{N\Gamma_\text{1D}\tilde{\sigma}^2}}}
\approx 1-\frac{1}{2}\frac{\Gamma'}{N\Gamma_\text{1D}\tilde{\sigma}^2}.
\end{gather}

\subsection{Using the fully discrete theory}
\label{discrete_eit_storage_retrieval}
Using the intuition about EIT from 
Sec.~\ref{dispersion_eit_storage_retrieval}, we can implement numerical 
simulations of EIT storage and retrieval 
accounting for the discrete nature of the atoms. This approach is very similar to 
the ``electric field elimination'' approach of 
Ref.~\cite{caneva_discrete_model_sm}. The main difference is that, since storage and 
retrieval of a single photon only requires calculating the dynamics in the 
atomic single-excitation manifold, we can eliminate the electric field 
directly in the Schr\"odinger picture instead of the Heisenberg picture in 
Ref.~\cite{caneva_discrete_model_sm}.

The electric field operator can be written 
$\hE(z)=\hE_+(z)e^{ik_0 z}+\hE_-(z)e^{-ik_0 z}$, 
where $\hE_+$ and $\hE_-$ are the parts of the field that 
propagate to the right (positive direction) and left (negative direction), respectively. The 
two parts are assumed to be completely separate fields, and their commutators 
are 
\begin{gather}
[\hE_\alpha(z),\hE^\dagger_\beta(z')]
=\delta_{\alpha\beta}\delta(z-z'),
\end{gather}
where $\alpha,\beta\in\cbr{+,-}$.

The Hamiltonian for the ensemble of $\Lambda$-type atoms coupled to the electric 
field is 
$\hat{H}=\hat{H}_\text{a}+\hat{H}_\text{i}+\hat{H}_\text{p}$, 
representing the atomic, interaction and photonic Hamiltonian, respectively. 
The three parts of the Hamiltonian are given by
\begin{align}\label{Lambda_Hamiltonian}
&\hat{H}_\text{a}=-\hbar\sum_{j}\sbr{\del{\Delta_0+i\frac{\Gamma'}{2}}\hat{\sigma}_{bb,j}
+\delta_0\hat{\sigma}_{cc,j}},\\
&\hat{H}_\text{i}
=-\hbar\sum_{j}\Bigg\{
\sbr{\hat{\sigma}_{bc,j}\Omega_0+\text{H.c.}}
+g\sqrt{2\pi}
\left[
\hat{\sigma}_{ba,j}\hE(z_j)+\text{H.c.}\right]\Bigg\},
\displaybreak[0]\\
&\hat{H}_\text{p}
=-i\hbar c\int \sbr{\hE_{+}^\dagger(z)\dpd{\hE_{+}(z)}{z}
-\hE_{-}^\dagger(z)\dpd{\hE_{-}(z)}{z}}\dif z,
\end{align}
where $c$ is the speed of light (group velocity in the waveguide). We assume 
that the detuning $\Delta_0$ is 
always set to zero during storage and retrieval (this was also assumed in 
Sec.~\ref{dispersion_eit_storage_retrieval} above). However, if desired, 
off-resonant ($\Delta_0\neq 0$) EIT storage and retrieval is also 
possible~\cite{gorshkov_pra07_2_sm}, and hence we keep the $\Delta_0$ 
term in the equations of motion below for completeness.

In the single-excitation manifold, the state 
can be written
\begin{gather}\label{Lambda_system_single_excitation_state}
\begin{aligned}
|\psi(t)\rangle
=\;&\sum_{j}\del{P_j(t)\hat{\sigma}_{ba,j}+S_j(t)\hat{\sigma}_{ca,j}}
|a\rangle^N|\text{vac}\rangle\\
&+
\del{\int \frac{\Phi_+(z,t)}{\sqrt{c}}\hE_+^\dagger(z)|\text{vac}\rangle\dif z
+\int \frac{\Phi_-(z,t)}{\sqrt{c}}\hE_-^\dagger(z)|\text{vac}\rangle\dif z}|a\rangle^N.
\end{aligned}
\end{gather}
From the Schr\"odinger equation, we get the equations of motion for the atomic 
coefficients
\begin{align}\label{Lambda_system_P_equation}
&\dpd{P_j}{t}
=\del{i\Delta_0-\frac{\Gamma'}{2}} P_j+i\Omega_0 S_j
+i\sqrt{\frac{\Gamma_\text{1D}}{2}}
\del{\Phi_+(z_j,t)e^{ik_0 z_j}
+\Phi_-(z_j,t)e^{-ik_0 z_j}},\\
\label{Lambda_system_S_equation}
&\dpd{S_j}{t}=i\delta_0 S_j+i\Omega_0^* P_j,
\end{align}
where $\Gamma_\text{1D}=4\pi g^2/c$. For the electric field coefficients $\Phi_\pm$, we 
have the equations
\begin{align}
&\del{\dpd{}{t}\pm c\dpd{}{z}}\Phi_{\pm}(z,t)
=ic\sqrt{\frac{\Gamma_\text{1D}}{2}}\sum_{j} \delta(z-z_j)P_je^{\mp ik_0 z_j}.
\end{align}
These equations can be formally integrated, resulting in
\begin{gather}\label{Lambda_system_sigma_E_field_eqs_sols}
\Phi_{\pm}(z,t)
=\Phi_{\pm,\text{in}}(z \mp ct)
+i\sqrt{\frac{\Gamma_\text{1D}}{2}}\sum_{j} 
\theta\delnospace{\pm(z-z_j)}P_j\delnospace{t\mp\frac{z-z_j}{c}}
e^{\mp ik_0 z_j},
\end{gather}
where $\Phi_{\pm,\text{in}}(z \pm ct)$ are the input fields, and $\theta$ is 
the Heaviside theta function. Inserting these solutions into 
Eq.~\eqref{Lambda_system_P_equation} and approximating 
$P_j\delnospace{t-|z-z_j|/c}\approx P_j(t)$~\cite{caneva_discrete_model_sm}, gives
\begin{gather}\label{Lambda_system_P_equation_field_eliminated}
\dpd{P_j}{t}
=\del{i\Delta_0-\frac{\Gamma'}{2}} P_j+i\Omega_0 S_j
-\frac{\Gamma_\text{1D}}{2}\sum_{j'} P_{j'} e^{ik_0 |z_j-z_{j'}|}
+i\sqrt{\frac{\Gamma_\text{1D}}{2}}
\del{\Phi_{+,\text{in}}(z_j-ct)e^{ik_0 z_j}
+\Phi_{-,\text{in}}(z_j+ct)e^{-ik_0 z_j}}.
\end{gather}

The fidelity calculations in Sec.~\ref{sec_fidelity_general} above are formulated in terms of 
$\phi_{A,\text{in}}(t)$ and $\phi_{A,\text{out}}(t)$ that are the 
input field to be stored and the retrieved output field, respectively. (There are two 
different output fields, $\phi_{A,\text{out},0}(t)$ and $\phi_{A,\text{out},1}(t)$, 
but for the discussion of storage and retrieval, the difference between them is not 
important.) We need to account for the beam splitter in the 
Sagnac interferometer. Hence, the relations between the fields in this section 
and Sec.~\ref{sec_fidelity_general} are
\begin{align}
&\Phi_{+,\text{in}}(z_j-ct)
=\frac{1}{\sqrt{2}}
\phi_{A,\text{in}}(t-z_j/c),\\
&\Phi_{-,\text{in}}(z_j+ct)
=\frac{1}{\sqrt{2}}
\phi_{A,\text{in}}(t-(L-z_j)/c),\\
\label{phi_out_in_terms_of_Phi_pm_sym}
&\phi_{A,\text{out}}(t)
=\frac{1}{\sqrt{2}}
\del{\Phi_+(L,t)+\Phi_-(0,t)}.
\end{align}
Note that the field is incident from two directions in order to ensure that no 
excitations are stored on atoms at the nodes of the standing wave of the 
classical drive applied during the scattering for the $\Lambda$-type 
scheme. The conditions for this may not necessarily be the same as the 
conditions derived for scattering from the Sagnac interferometer (see Sec.~\ref{Sec_sagnac_interferometer_and_phases}). If this is an issue, it can 
be compensated by adjusting the position of the atoms between storage and 
scattering, e.g., by adjusting the phases of the trapping lasers or of the 
classical drives. For the dual-V scheme, there is no phase requirement during 
storage and this is not a concern.

As the input wave function, we choose
\begin{gather}\label{gaussian_phi_A_in}
\phi_{A,\text{in}}(t)
=\frac{1}{(2\pi\sigma_\text{in}^2)^{1/4}}
\exp\del{-\frac{(t-\mu_\text{in})^2}{4\sigma_\text{in}^2}}
\end{gather}
where the width and central time,
\begin{align}
\label{gaussian_phi_A_in_sigma}
&\sigma_\text{in}=\sigma/\del{v_\text{g}\sqrt{1+i\alpha L/(4\sigma^2v_\text{g})}},\\
\label{gaussian_phi_A_in_mu}
&\mu_\text{in}=4\sigma_\text{in},
\end{align}
are defined in terms of the EIT group velocity 
\eqref{eit_group_velocity_and_attenuation_not_rescaled}.
The factor $\sqrt{1+i\alpha L/(4\sigma^2v_\text{g})}$ (a real number, since 
$\alpha$ is imaginary) in the definition of 
$\sigma_\text{in}$ is introduced to compensate for the 
spin wave 
becoming wider as it propagates inside the ensemble (see 
Eq.~\eqref{gaussian_wavefunction_arbitrary_time}). This particular factor is 
chosen such that the stored Gaussian spin wave (centered at the position $L/2$) has 
width $\sigma$. In the end, since we optimize over $\sigma$, this adjustment 
has no effect on the final values of the numerically calculated fidelities and 
success probability. However, it ensures that the optimal $\sigma$ in the 
numerical calculations is similar to the optimal value found by neglecting 
broadening of the spin wave under propagation.

In the fully discrete model, we do not explicitly calculate the storage and 
retrieval kernels that appear in Eqs. \eqref{phi_out_0_definition} and 
\eqref{phi_out_1_definition}. Instead, we calculate the action of these 
kernels on respectively a specific $\phi_{A,\text{in}}(t)$ or a spin wave given by 
the coefficients $S_j$. For storage, this amounts to numerically solving Eqs. 
\eqref{Lambda_system_S_equation} and 
\eqref{Lambda_system_P_equation_field_eliminated} for a given
$\phi_{A,\text{in}}(t)$ and the initial condition $P_j=S_j=0$ at $t=0$. We take 
the stored spin wave to be the coefficients $S_j$ at 
$t=\mu_\text{in}/c+L/(2v_\text{g})$. This final time is the sum of the time for 
propagation through vacuum and (half of) the EIT medium. For retrieval, Eqs. 
\eqref{Lambda_system_S_equation} and 
\eqref{Lambda_system_P_equation_field_eliminated} are solved with 
$\phi_{A,\text{in}}(t)=0$ under the initial conditions that at $t=0$ the 
coefficients $S_j(t=0)$ are set to the spin wave that is to be retrieved, and 
$P_j(t=0)=0$. At each time step, we calculate $\phi_{A,\text{out}}(t)$ using 
Eqs.~\eqref{Lambda_system_sigma_E_field_eqs_sols} with 
$P_j\delnospace{t\mp(z-z_j)/c}\approx P_j(t)$ along with Eq.~\eqref{phi_out_in_terms_of_Phi_pm_sym}. We assume that the retrieval 
happens until $t=L/v_\text{g}$, i.e. the time that it takes for the EIT 
polaritons to move through the whole ensemble.

\subsection{Using the storage and retrieval kernels}
\label{continuum_eit_storage_retrieval}
As an approximation to the fully discrete theory of 
Sec.~\ref{discrete_eit_storage_retrieval}, one can use the continuum theory of 
Ref.~\cite{gorshkov_pra07_2_sm}. The continuum approximation allows the 
derivation of 
explicit expressions for the linear maps (given in terms of integration with a particular kernel) describing storage and retrieval. 

Before doing the continuum approximation, we first rewrite the 
Hamiltonian~\eqref{Lambda_Hamiltonian} in terms of the collective 
operators
\begin{align}\label{collective_sigma_definition}
\hat{\sigma}_{\alpha\beta}(z)=\frac{1}{n_0}\sum_{j}\delta(z-z_j)\hat{\sigma}_{\alpha\beta,j},
\end{align}
where $\alpha,\beta\in\cbr{a,b,c}$, and $n_0=N/L$ is the average density. These collective operators have the 
equal time commutation relation
\begin{align}\label{collective_sigma_commutator}
[\hat{\sigma}_{\alpha\beta}(z),\hat{\sigma}_{\alpha'\beta'}(z')]
&=\frac{1}{n_0}\delta(z-z')
(\delta_{\beta,\alpha'}\hat{\sigma}_{\alpha\beta'}
-\delta_{\beta',\alpha}\hat{\sigma}_{\alpha'\beta}).
\end{align}
Using the collective atomic operators, the 
Hamiltonian~\eqref{Lambda_Hamiltonian} is
\begin{align}
&\hat{H}_\text{a}=-\hbar n_0\int \Big[\del{\Delta_0+i\frac{\Gamma'}{2}}\hat{\sigma}_{bb}(z)
+\delta_0\hat{\sigma}_{cc}(z)\Big]\dif z\\
&\hat{H}_\text{i}=-\hbar n_0\int \Bigg\{\left[\hat{\sigma}_{bc}(z)\Omega_0
+\text{H.c.}\right]
+g\sqrt{2\pi}\left[\hat{\sigma}_{ba}(z)\hE(z)
+\text{H.c.}\right]\Bigg\}\dif z\\
&\hat{H}_\text{p}=-i\hbar c\int\Bigg[\hE_{+}^{\dagger}(z)\dpd{\hE_{+}(z)}{z}
-\hE_{-}^{\dagger}(z)\dpd{\hE_{-}(z)}{z}\Bigg]\dif z.
\end{align}
Instead of the state \eqref{Lambda_system_single_excitation_state}, we use
\begin{gather}
\begin{aligned}
|\psi(t)\rangle
=\;&\int\del{\frac{\sqrt{N}}{L}P(z,t)\hat{\sigma}_{ba}(z)
+\frac{\sqrt{N}}{L}S(z,t)\hat{\sigma}_{ca}(z)}
|a\rangle^N|\text{vac}\rangle\dif z\\
&+
\del{\int \frac{\Phi_+(z,t)}{\sqrt{c}}\hE_+^\dagger(z)|\text{vac}\rangle\dif z
+\int \frac{\Phi_-(z,t)}{\sqrt{c}}\hE_-^\dagger(z)|\text{vac}\rangle\dif z}|a\rangle^N.
\end{aligned}
\end{gather}
Note that if, for example, the excitation is entirely in the 
metastable states at a time $t$, we have 
\begin{gather}\label{Lambda_system_single_excitation_state_continuum_S}
1
=\langle \psi(t)|\psi(t)\rangle
=\frac{N}{L^2}\int \int S^*(z,t)S(z',t)
[\hat{\sigma}_{ab}(z),\hat{\sigma}_{ba}(z')] \dif z
=\frac{1}{L}\int \int |S(z,t)|^2\dif z,
\end{gather}
where we have used the continuum approximation $\sum_{j}\delta(z-z_j)\approx n_0$~\cite{dispersion-relations_sm} together with the low excitation approximation 
$\sigma_{aa,j}\approx 1$ to get $\sigma_{aa}(z)\approx 1$ and 
$\sigma_{bb}\approx 0$. 
Equation~\eqref{Lambda_system_single_excitation_state_continuum_S} also gives 
the normalization condition for $S$ in the continuum model (and similarly for $P$). 

The equations of motion for the coefficients are
\begin{align}
\label{Lambda_system_Phi_equation_continuum}
&\del{\dpd{}{t}\pm c\dpd{}{z}}\Phi_{\pm}(z,t)
=ic\sqrt{\frac{\Gamma_\text{1D}N}{2}}\frac{1}{L}P(z,t),\\
\label{Lambda_system_P_equation_continuum}
&\dpd{}{t}P(z,t)
=\del{i\Delta_0-\frac{\Gamma'}{2}}P(z,t)+i\Omega_0 S(z,t)
+i\sqrt{\frac{\Gamma_\text{1D}N}{2}}
\del{\Phi_+(z,t)e^{ik_0 z}
+\Phi_-(z,t)e^{-ik_0 z}},\\
\label{Lambda_system_S_equation_continuum}
&\dpd{}{t}S(z,t)=i\delta_0 S(z,t)+i\Omega_0^* P(z,t).
\end{align}
As an extension to the theory of 
Ref.~\cite{gorshkov_pra07_2_sm}, we want to consider an input field that can be 
incident from both sides instead of only one. The approach that we 
use is to consider the parts of the single photon excitation incident from the opposite 
sides as being stored separately from each other. When doing this, we ignore 
the fact that the two parts have opposite spatial phases 
$e^{\pm ik_0 z}$ that interfere inside the ensemble to produce a spatially 
modulated spin wave with amplitude proportional to $\cos(k_0 z)$.

\begin{figure}[t]
\begin{center}
\includegraphics{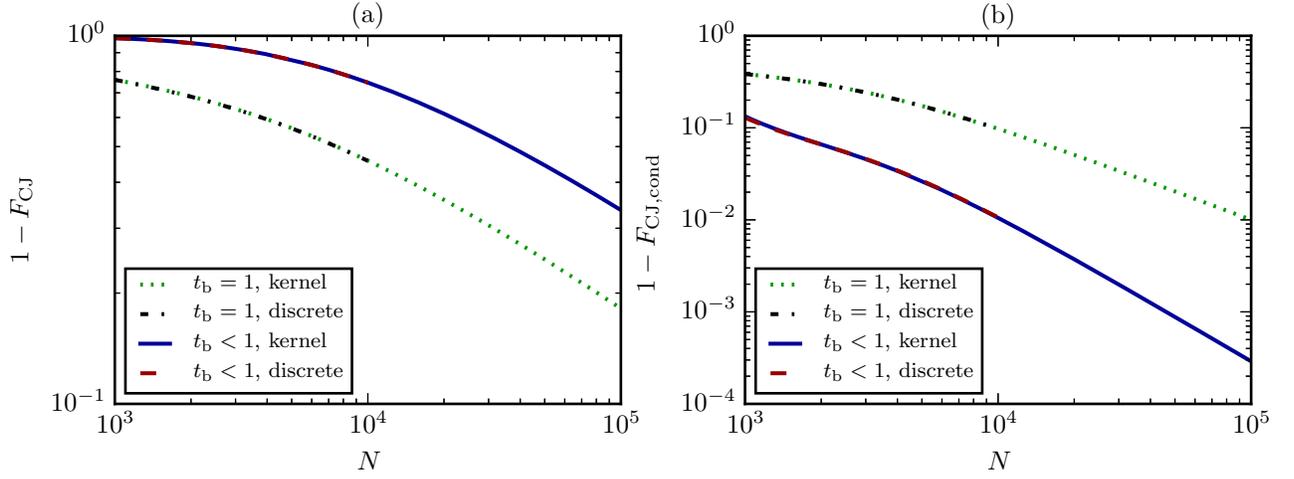}
\end{center}
\caption{Comparison of (a) unconditional and (b) conditional Choi-Jamiolkowski fidelities for 
the $\Lambda$\mbox{-}type scheme with different models used for EIT storage and 
retrieval plotted as functions of the number of atoms 
$N$ with fixed $\Gamma_\text{1D}/\Gamma=0.05$ and $\Omega_0/\Gamma=1$. Under EIT (storage and retrieval), $\Omega(z)=\Omega_0$. 
Under stationary light (scattering), $\Omega(z)=\Omega_0\cos(k_0 z)$. The ``discrete'' plots use 
the fully discrete theory of Sec.~\ref{discrete_eit_storage_retrieval}, and 
the ``kernel'' plots use the discretized continuum 
storage and retrieval kernels discussed in 
Sec.~\ref{continuum_eit_storage_retrieval}. In both cases, the optimal $\Delta_\text{c}$ 
and $\tilde{\sigma}$ (width of the stored Gaussian spin wave) are found by 
doing numerical optimization using the discretized continuum storage and retrieval kernels, 
since the fully discrete model is computationally much more demanding.
}
\label{fig_cphase_fidelity_lambda_check_discrete}
\end{figure}

Such spatial modulation of the stored spin wave is very important for the 
$\Lambda$\mbox{-}scheme, since the part of the excitation that is stored on 
the nodes of the standing wave of the classical drive does not significantly 
change the scattering properties of the ensemble (see 
Sec.~\ref{Lambda_r_t_with_stored_photon}). 
Expressed in terms of the notation introduced in the fidelity calculations 
(see Eq.~\eqref{phi_out_1_definition}), we 
have $R_{1,j}(\delta_B)\approx R_0(\delta_B)$ for odd $j$ with atoms placed at 
positions $z_j=j\pi/(2k_0)$ (where ${0\leq j\leq N-1}$), i.e. $k_0 z_j$ being an odd 
multiple of $\pi/2$. With the photon incident 
from both sides, there is no amplitude on these atoms, since $S(z_j)\propto \cos(k_0 z_j)=0$ for odd $j$. This is correctly reproduced 
by the fully discrete model of Sec.~\ref{discrete_eit_storage_retrieval}, 
since it always accounts for the phases of free propagation.
On the other hand, due to the removal of the rapidly varying spatial phases in the continuum theory, this factor $\cos(k_0 z_j)$ is not 
present in the two separate parts of the stored spin wave, incorrectly resulting in 
non-zero probability of storage into the atoms at positions $z_j=j\pi/(2k_0)$ 
with odd $j$. To prevent 
acquiring a reflection coefficient with a wrong phase ($R_0(\delta_B)$) at these 
atomic positions, we 
set $R_{1,j}(\delta_B)=R_{1,j-1}(\delta_B)$ for odd $j$, effectively redistributing
the stored photon such that it is only stored on atoms with even $j$. As we show in 
Fig.~\ref{fig_cphase_fidelity_lambda_check_discrete}, this phenomenological 
adjustment of the reflection coefficient in the continuum model 
gives results that are essentially indistinguishable from the results
produced by the fully 
discrete model of Sec.~\ref{discrete_eit_storage_retrieval}. For the dual\mbox{-}V scheme, no 
such adjustment of the reflection coefficients is needed neither in the 
continuum nor the discrete model, which also produce indistinguishable results.

The fields incident from the left and right couple to different 
components of the atomic coefficients that can be written
\begin{align}\label{P_from_P_pm}
&P(z,t)=P_+(z,t)e^{ik_0 z}+P_-(z,t)e^{-ik_0 z},\\
&S(z,t)=S_+(z,t)e^{ik_0 z}+S_-(z,t)e^{-ik_0 z}.
\end{align}
After inserting these 
definitions into Eqs. \eqref{Lambda_system_Phi_equation_continuum}, 
\eqref{Lambda_system_P_equation_continuum} and 
\eqref{Lambda_system_S_equation_continuum} and separating the components, we get
\begin{align}
&\del{\dpd{}{t}\pm c\dpd{}{z}}\Phi_\pm(z,t)
=ic\sqrt{\frac{\Gamma_\text{1D}N}{2}}\frac{1}{L}P_\pm(z,t),\\
\label{Lambda_system_P_equation_continuum_pm}
&\dpd{}{t}P_\pm(z,t)
=\del{i\Delta_0-\frac{\Gamma'}{2}}P_\pm(z,t)+i\Omega_0 S_\pm(z,t)
+i\sqrt{\frac{\Gamma_\text{1D}N}{2}}\Phi_\pm(z,t),\\
\label{Lambda_system_S_equation_continuum_pm}
&\dpd{}{t}S_\pm(z,t)=i\delta_0 S_\pm(z,t)+i\Omega_0^* P_\pm(z,t),
\end{align}

To solve for $\Phi_+$, $P_+$ 
and $S_+$, the approach in Ref.~\cite{gorshkov_pra07_2_sm} can be used directly. 
It consists of transforming into the coordinates $\tilde{z}=z/L$ and 
$\tilde{t}=t-z/c$, Laplace transforming in the spatial coordinate $\tilde{z}$, 
solving the
equations of the Laplace transforms (under the adiabatic approximation 
$\pd{}{t}P_+\approx 0$) and taking the inverse Laplace transform. As a 
minor modification, we also transform back from the co-propagating time 
coordinate $\tilde{t}=t-z/c$ to the original $t$ (by replacing all 
$\tilde{t}$ by $t$ in the final expressions). Solving for $\Phi_-$, $P_-$ 
and $S_-$ is simply a spatial reflection of the original problem 
around $\tilde{z}=1/2$. We find that the stored spin wave is
\begin{align}
\label{S_stored_with_kernel_plus}
&S_+(\tilde{z},t)
=\int_0^{t}K_\text{s}(\tilde{z},t-t')
\Phi_+(\tilde{z}=0,t')
\dif t',\\
\label{S_stored_with_kernel_minus}
&S_-(\tilde{z},t)
=\int_0^{t}K_\text{s}(1-\tilde{z},t-t')
\Phi_-(\tilde{z}=1,t')
\dif t',
\end{align}
where the storage kernel (in the adiabatic approximation) is
\begin{gather}\label{storage_kernel}
K_\text{s}(\tilde{z},t)
\approx -\frac{\sqrt{b}\Omega_0^* e^{i\delta_0 t}}{(\Gamma'/2)-i\Delta_0}
I_0\delnospace{2\frac{\sqrt{|\Omega_0|^2 t b\tilde{z}}}
{(\Gamma'/2)-i\Delta_0}}
\exp\delnospace{-\frac{|\Omega_0|^2 t+b\tilde{z}}
{(\Gamma'/2)-i\Delta_0}},
\end{gather}
written in terms of $b = (N\Gamma_\text{1D})/2$ (related to the resonant optical depth 
by $d_\text{opt}=4b/\Gamma$) and the modified Bessel 
function of the first kind $I_0$. The retrieved field is
\begin{align}
\label{E_retrieved_with_kernel_plus}
&\Phi_+(\tilde{z}=1,t)
=\int_0^1
K_\text{r}(\tilde{z},t)
S_+(\tilde{z},t=0)
\dif \tilde{z},\\
\label{E_retrieved_with_kernel_minus}
&\Phi_-(\tilde{z}=0,t)
=\int_0^1
K_\text{r}(1-\tilde{z},t)
S_-(\tilde{z},t=0)
\dif \tilde{z},
\end{align}
where the retrieval kernel is
\begin{gather}
\label{retrieval_kernel}
K_\text{r}(\tilde{z},t)
\approx -\frac{\sqrt{b}\Omega_0 e^{i\delta_0 t}}{(\Gamma'/2)-i\Delta_0}
I_0\delnospace{2\frac{\sqrt{|\Omega_0|^2 t b(1-\tilde{z})}}
{(\Gamma'/2)-i\Delta_0}}
\exp\delnospace{-\frac{|\Omega_0|^2 t+b(1-\tilde{z})}{(\Gamma'/2)-i\Delta_0}}.
\end{gather}

The stored spin waves are such that they are centered at $\tilde{z}=1/2$ 
and become narrower in space for increasing optical depth (see the fidelity 
derivations in Sec.~\ref{sec_analytical_fidelity} below). The input and output 
photon wave functions are centered around times $t\propto L/v_\text{g}=N\Gamma_\text{1D}/(2|\Omega_0|^2)=b/|\Omega_0|^2$. 
Inserting these mean values into 
Eqs.~\eqref{storage_kernel}~and~\eqref{retrieval_kernel}, we see that the 
argument of $I_0$ becomes very big, since 
$\sqrt{|\Omega_0|^2 t b\tilde{z}}
=\sqrt{|\Omega_0|^2 t b(1-\tilde{z})}\propto b$ and $\Delta_0=0$.
This allows us to use the asymptotic expansion
$I_0(x)\approx \exp(x)/\sqrt{2\pi x}$ that is 
valid for $|x|\gg 1$ and $\arg(x)<\pi/2$. In this limit, the kernels become
\begin{align}
\label{storage_kernel_approx}
&K_\text{s}(\tilde{z},t)
\approx -\frac{\sqrt{b}\Omega_0^*e^{i\delta t}}{2\sqrt{\pi}\sqrt{(\Gamma'/2)-i\Delta_0}}
\frac{1}
{\del{|\Omega_0|^2 t b\tilde{z}}^{1/4}}
\exp\delnospace{-\frac{\del{\sqrt{|\Omega_0|^2 t}-\sqrt{b\tilde{z}}}^2}
{(\Gamma'/2)-i\Delta_0}},\\
\label{retrieval_kernel_approx}
&K_\text{r}(\tilde{z},t)
\approx-\frac{\sqrt{b}\Omega_0 e^{i\delta t}}
{2\sqrt{\pi}\sqrt{(\Gamma'/2)-i\Delta_0}}
\frac{1}{\del{|\Omega_0|^2 t b(1-\tilde{z})}^{1/4}}
\exp\delnospace{-\frac{\del{\sqrt{|\Omega_0|^2 t}-\sqrt{b(1-\tilde{z})}}^2}
{(\Gamma'/2)-i\Delta_0}}.
\end{align}
These formulas have a better numerical behavior compared to 
Eqs.~\eqref{storage_kernel}~and~\eqref{retrieval_kernel}, since one does not 
need to multiply the value of the $I_0$ (exponentially large) 
with an exponentially small factor.
Hence, we always use 
Eqs.~\eqref{storage_kernel_approx}~and~\eqref{retrieval_kernel_approx} in the 
numerical calculations.

The relations between the fields in this section and 
Sec.~\ref{sec_fidelity_general} are
\begin{align}
&\Phi_+(\tilde{z}=0,t)
=\frac{1}{\sqrt{2}}
\phi_{A,\text{in}}(t),\\
&\Phi_-(\tilde{z}=1,t)
=\frac{1}{\sqrt{2}}
\phi_{A,\text{in}}(t),\\
\label{phi_out_in_terms_of_Phi_pm_sym_continuum}
&\phi_{A,\text{out}}(t)
=\frac{1}{\sqrt{2}}
\del{\Phi_+(\tilde{z}=1,t)
+\Phi_-(\tilde{z}=0,t)},
\end{align}
where $\phi_{A,\text{in}}(t)$ is given by Eq.~\eqref{gaussian_phi_A_in} like in 
the fully discrete model.

The results of this section assume that the atoms can be modeled as a continuum, but the 
scattering coefficients in Eq.~\eqref{phi_out_1_definition} are only given at 
the discrete atom positions. Hence, we need to sample the resulting continuum solutions at the discrete 
positions of the atoms. In the numerical calculations, the continuum solutions 
are always sampled as if the atoms were placed regularly independent of the 
actual placement.
To justify regular sampling, we note that instead of the rescaled position 
coordinate $\tilde{z}=z/L$, one could use 
$\tilde{z}=\int_0^z (n_0(z')/N) \dif z'$~\cite{gorshkov_pra07_2_sm}, where 
$n_0(z)$ is the local density of the atoms. For an average density $n_0=N/L$, 
this rescaled coordinate is equivalent to $\tilde{z}=z/L$. For the local 
density $n_0(z)=\sum_{j} \delta(z-z_j)$, the rescaled position becomes 
$\tilde{z}=\sum_{j} \theta(z-z_j)/N$, where $\theta$ is the Heaviside 
theta function. With the convention $\theta(0)=0$, each $z_j$ 
(where $0\leq j\leq N-1$) is transformed
into $\tilde{z}=j/N$ regardless of the actual value of $z_j$.

Having in mind both the separation of the spin waves into two independent 
parts and the sampling of the continuum solutions at regular intervals, we can 
define the storage and retrieval kernels that will be used in Eqs. 
\eqref{phi_out_0_definition} and \eqref{phi_out_1_definition}. We define 
the vector representing the spin wave to have $2N$ elements---for two 
separately stored spin waves that have fields incident either from the left or 
from the right as the input. Using the same storage time 
$\mu_\text{in}/c+L/(2v_\text{g})$ as for the discrete model in 
Sec.~\ref{discrete_eit_storage_retrieval}, the storage kernel is
\begin{align}
&K_{\text{s},j}(t_A)=K_\text{s}\delnospace{\tilde{z}=j/N,t=\mu_\text{in}/c+L/(2v_\text{g})-t_A}/\sqrt{N}
\text{ for }0\leq j\leq N-1,\\
&K_{\text{s},j}(t_A)=K_\text{s}\delnospace{\tilde{z}=1-(j-N)/N,t=\mu_\text{in}/c+L/(2v_\text{g})-t_A}/\sqrt{N}
\text{ for }N\leq j\leq 2N-1,
\end{align}
where we have assumed that the coefficients corresponding to $S_+$ are stored 
in the part of the vector with indices $0\leq j\leq N-1$, and the coefficients 
corresponding to $S_-$ are stored in the part of the vector with 
$N\leq j\leq 2N-1$. The retrieval kernel is
\begin{align}
&K_{\text{r},j}(t_A)=K_\text{r}(\tilde{z}=j/N,t=t_A)/\sqrt{N}
\text{ for }0\leq j\leq N-1,\\
&K_{\text{r},j}(t_A)=K_\text{r}(\tilde{z}=1-(j-N)/N,t=t_A)/\sqrt{N}
\text{ for }N\leq j\leq 2N-1.
\end{align}
As a consequence of having a spin wave vector with $2N$ elements, in 
Eq.~\eqref{phi_out_1_definition} we define 
$R_{1,j}(\delta_B)=R_{1,j-N}(\delta_B)$ for $N\leq j\leq 2N-1$.

\section{Dephasing and decay of the stored photon}
\label{sec_dephasing_decay_of_stored_photon}
Depending on 
the chosen implementation, the classical drive that creates the stationary 
light (involving states $|a\rangle$, $|b\rangle$ and $|c\rangle$) may also
couple state~$|d\rangle$ to some excited state. E.g. in the implementation shown in 
Fig.~\ref{fig_Rb_implementation},
the classical drive couples state $|d\rangle$ ($F=2$, $m_F=-1$) to state 
$|f\rangle$ ($F'=2$, $m_{F'}=0$) with a large detuning $\Delta_\text{hfs}$ 
approximately equal to the hyperfine splitting. This coupling induces both 
dephasing (when the excitation in state $|f\rangle$ incoherently 
decays back to state $|d\rangle$) and decay (when the 
excitation in state $|f\rangle$ incoherently decays to any other possible state).

All the states that $|f\rangle$ 
can decay into besides $|d\rangle$ can be treated as a single state $|g\rangle$ 
in a simplified model. In the implementation of 
Fig.~\ref{fig_Rb_implementation}, state $|f\rangle$ can also decay to state 
$|c\rangle$, but we ignore this in this simplified model for two reasons. 
First, if the $\pi$-pulse is perfect, the incoherent part in state $|c\rangle$ 
will get transferred to state $|d\rangle$ from which any retrieval is 
suppressed by the large detuning $\sim\Delta_\text{hfs}$ of the classical drive. 
Second, even if the $\pi$-pulse is imperfect, the remaining incoherent 
part in state $|c\rangle$ will have a very low retrieval efficiency due to 
mode mismatch (see the discussion below).

The Hamiltonian describing coupling of the classical drive to the 
$|d\rangle\leftrightarrow|f\rangle$ transition is
\begin{gather}
\hat{H}_\text{d}=-\hbar\sum_{j}\sbr{\Delta_\text{hfs}\hat{\sigma}_{ff,j}
+\hat{\sigma}_{fd,j}\Omega_0
+\hat{\sigma}_{df,j}\Omega_0^*}.
\end{gather}
The incoherent decay from state $|f\rangle$ is described by the operators
\begin{align}
&\hat{L}_{1,j}=\sqrt{\Gamma_1}\hat{\sigma}_{gf,j},\\
&\hat{L}_{2,j}=\sqrt{\Gamma_2}\hat{\sigma}_{df,j},
\end{align}
where we assume that $\Gamma'=\Gamma_1+\Gamma_2$. The evolution of the 
density matrix $\rho$ is given by the master equation
\begin{gather}\label{master_equation}
\dot{\rho}=-\frac{i}{\hbar}\sbr{H,\rho}
-\frac{1}{2}\sum_{j}\sum_{k=1}^{2}\del{
\hat{L}_{k,j}^\dagger\hat{L}_{k,j}\rho
+\rho\hat{L}_{k,j}^\dagger\hat{L}_{k,j}
-2\hat{L}_{k,j}\rho\hat{L}_{k,j}^\dagger}.
\end{gather}
If we adiabatically eliminate~\cite{reiter_pra12_sm} the excited state $|f\rangle$, 
the effective Hamiltonian and decay operators are instead
\begin{align}
\label{H_d_eff}
&\hat{H}_\text{d,eff}=\sum_j\frac{\Delta_\text{hfs}|\Omega_0|^2}{\Delta_\text{hfs}^2+\Gamma'^2/4}\hat{\sigma}_{dd,j},\\
\label{L_1_eff}
&\hat{L}_{1,j,\text{eff}}=\frac{\sqrt{\Gamma_2}\Omega_0}{\Delta_\text{hfs}+i\Gamma'/2}\hat{\sigma}_{gd,j},\\
\label{L_2_eff}
&\hat{L}_{2,j,\text{eff}}=\frac{\sqrt{\Gamma_1}\Omega_0}{\Delta_\text{hfs}+i\Gamma'/2}\hat{\sigma}_{dd,j}.
\end{align}
The effective dynamics is then given by
\begin{gather}\label{master_equation_eff}
\begin{aligned}
\dot{\rho}=-\frac{i}{\hbar}\sbr{H_\text{d,eff},\rho}
&-\frac{\Gamma_{1,\text{eff}}}{2}\sum_{j}\del{
\hat{\sigma}_{dg,j}\hat{\sigma}_{gd,j}\rho
+\rho\hat{\sigma}_{dg,j}\hat{\sigma}_{gd,j}
-2\hat{\sigma}_{gd,j}\rho\hat{\sigma}_{dg,j}}\\
&-\frac{\Gamma_{2,\text{eff}}}{2}\sum_{j}\del{
\hat{\sigma}_{dd,j}\hat{\sigma}_{dd,j}\rho
+\rho\hat{\sigma}_{dd,j}\hat{\sigma}_{dd,j}
-2\hat{\sigma}_{dd,j}\rho\hat{\sigma}_{dd,j}},
\end{aligned}
\end{gather}
where
\begin{align}
\label{Gamma_1_eff_expr}
&\Gamma_{1,\text{eff}}=\frac{\Gamma_1|\Omega_0|^2}{\Delta_\text{hfs}^2+\Gamma'^2/4},\\
\label{Gamma_2_eff_expr}
&\Gamma_{2,\text{eff}}=\frac{\Gamma_2|\Omega_0|^2}{\Delta_\text{hfs}^2+\Gamma'^2/4}.
\end{align}
In general, $\Gamma_{1,\text{eff}}$ and $\Gamma_{2,\text{eff}}$ are
respectively set by the $T_1$ and $T_2$ times of the system, and 
Eqs.~\eqref{Gamma_1_eff_expr}~and~\eqref{Gamma_2_eff_expr} describe the 
classical drive induced contributions to these decoherence processes.

We note that dephasing ($\Gamma_{2,\text{eff}}$) has the same effect 
as decay ($\Gamma_{1,\text{eff}}$) to a good approximation. This 
is a consequence of the fact that photon~$A$ has to be retrieved at the end of 
the gate protocol. Dephasing creates an incoherent part of state of the stored photon
(a classical mixture of excitations in different atoms) that 
cannot be retrieved in the same way as the 
coherent part (a superposition of excitations in different atoms) due to lack of mode matching. This can be
understood from the simple model of EIT storage described in 
Sec.~\ref{dispersion_eit_storage_retrieval} 
above. In particular, letting $\sigma\rightarrow 0$ in 
Eq.~\eqref{gaussian_wavefunction_norm_after_passing_ensemble} results in 
$\eta_\text{EIT}\rightarrow 0$.
Therefore, the 
retrieved state of an excitation that has experienced dephasing during storage 
in the ensemble is not different from the state that has experienced decay.

Thus, even if, in principle, dephasing is much worse than population 
decay for the stored photon (since it cannot be postselected by measuring 
photon number), the retrieval of the incoherent part of the state is strongly 
suppressed. Hence, both decay and dephasing can be taken to only affect the 
unconditional fidelity and success probability and not the conditional 
fidelity. A more precise analysis
of this error on the conditional fidelity may be needed for conditional 
fidelities that are very close to unity, but we will only limit ourselves to 
estimating the
effect on the unconditional fidelity and success probability. In 
Sec.~\ref{sec_analytical_fidelity} below, we 
show that scattering time $t_\text{s}$ (inverse of the bandwidth~$\sigma_B$) 
cannot be smaller than approximately 
$(\Gamma_\text{1D}^{3/2}N^{7/4})/(4\pi^{3/2}\sqrt{\Gamma'}|\Omega_0|^2)$ if 
high unconditional fidelity and success probability is to be 
maintained. Hence, the error for the unconditional fidelity and success 
probability is 
\begin{gather}
(\Gamma_{1,\text{eff}}+\Gamma_{2,\text{eff}})t_\text{s}\sim
\frac{\Gamma_\text{1D}^{3/2}\sqrt{\Gamma'}N^{7/4}}{4\pi^{3/2}\Delta_\text{hfs}^2}.
\end{gather}
For $\Delta_\text{hfs}/\Gamma\sim 10^3$, as in e.g. $^{87}$Rb~\cite{steck_rb87_sm}, 
and choosing $\Gamma_\text{1D}/\Gamma=0.05$, we get the error
$\Gamma_\text{eff}t_\text{s}\sim 5\cdot 10^{-10}\cdot N^{7/4}$ that is 
negligible for $N=10^4$ ($\Gamma_\text{eff}t_\text{s}\sim 0.005$), but becomes 
significant for $N=10^5$ ($\Gamma_\text{eff}t_\text{s}\sim 0.28$).

\section{$\pi$-pulses}\label{sec_pi_pulse}
Here, we discuss the $\pi$-pulses used to reversibly map states $|c\rangle$ and 
$|d\rangle$. In the implementation of Fig.~\ref{fig_Rb_implementation}, 
microwave pulses are assumed, but the mapping could, in principle, also be done with 
optical fields. The selection of particular $m_F$-levels is possible due to 
their splitting in energy by an applied magnetic field. Since the frequency 
splitting between the adjacent $m_F$-levels is $\sim\Delta_\text{c}$, 
this determines the time $t_\pi\gtrsim 1/\Delta_\text{c}$ needed for a $\pi$-pulse.

A possible imperfection in the controlled-phase gate can be caused by a wrong 
pulse area of the microwave pulse. To model this, we use the Hamiltonian
\begin{gather}
\hat{H}_\text{mw}
=-\hbar\sum_j\sbr{
\hat{\sigma}_{dc,j}|\Omega_\text{mw}|e^{i\phi_\text{mw}}
+\hat{\sigma}_{cd,j}|\Omega_\text{mw}|e^{-i\phi_\text{mw}}}.
\end{gather}
A perfect $\pi$-pulse happens when $|\Omega_\text{mw}|t=\pi/2$. We parametrize 
the deviation from a perfect $\pi$-pulse by an angle $\varphi$ and set 
$|\Omega_\text{mw}|t=\pi/2+\varphi$. To account for a possible deviation from 
a perfect $\pi$-pulse in the fidelity calculations, the analysis in 
Sec.~\ref{sec_fidelity_general} needs to be extended with assumptions about 
the evolution of states $|c_j a\rangle$ during scattering. For the 
computational basis states $|00\rangle$, $|01\rangle$, $|10\rangle$, there is 
no incident photon~$B$ on the atomic ensemble where a part of photon~$A$ is 
stored in the states $|c_j a\rangle$ due to a failed $\pi$-pulse. In this 
case, the evolution of photon~$A$ and photon~$B$ happens completely separately 
during scattering. The output wave packet for the part of photon~$A$ that was 
not transferred by both $\pi$-pulses can be written
\begin{gather}
\label{phi_out_a_definition}
\phi_{A,\text{out},a}(t_A)=\sum_{j,j'}\int K_{\text{r},j}(t_A)K_{\text{a},j,j'}K_{\text{s},j'}(t_A')
\phi_{A,\text{in}}(t_A')\dif t_A',
\end{gather}
where the kernel $K_{\text{a},j,j'}$ describes the evolution of the states 
$|c_j a\rangle$ during scattering of photon~$B$. For state $|11\rangle$, the evolution is, in principle, different, because both 
photons (stationary light polaritons) evolve in the two-excitation manifold of 
states (i.e. states like
$|c_j c_{j'} a\rangle=\hat{\sigma}_{ca,j}\hat{\sigma}_{ca,j'}|a\rangle^N$) 
simultaneously.
However, in the limit of few photons 
and many atoms, stationary light is a linear phenomenon, just like EIT. Thus, 
we can assume that the two photons evolve 
separately from each other even for state $|11\rangle$. We thereby arrive at 
the fidelity in the case of imperfect $\pi$-pulses
\begin{gather}\label{CJ_fidelity_computational_basis_final_imperfect_pi}
\begin{aligned}
F_\text{CJ}=
\frac{\eta_\text{EIT}}{16} \Bigg|
&\cos^2(\varphi)\del{
2t_\text{b}
+\int R_0(\delta_B)|\phi_B(\delta_B)|^2\dif \delta_B
-\frac{1}{\eta_\text{EIT}}
\iint \phi_{A,\text{out},0}^*(t_A)\phi_{A,\text{out},1}(t_A,\delta_B)
|\phi_B(\delta_B)|^2\dif t_A\dif \delta_B}\\
&+2\sin^2(\varphi)\int \phi_{A,\text{out},0}^*(t_A)\phi_{A,\text{out},a}(t_A)
\dif t_A
\Bigg|^2.
\end{aligned}
\end{gather}
Note that the factors $\cos(\varphi)$ and $\sin(\varphi)$ appear squared. 
This reflects the fact that the protocol involves two $\pi$-pulses. Hence, a 
photon is only retrieved if either both $\pi$-pulses succeed ($\cos^2(\varphi)$) 
or fail ($\sin^2(\varphi)$). 

The term proportional to $\sin^2(\varphi)$ in 
Eq.~\eqref{CJ_fidelity_computational_basis_final_imperfect_pi} corresponds to 
the unwanted process that could degrade the conditional 
fidelity. Since both $\pi$-pulses need to go wrong, it only enters to a higher 
order, however. The wave packet 
$\phi_{A,\text{out},a}(t_A)$ in this term could be calculated exactly with the 
fully discrete theory in
Sec.~\ref{discrete_eit_storage_retrieval} (and using $\Omega(z)=\Omega_0\cos(k_0 z)$ 
instead of $\Omega(z)=\Omega_0$). However, 
we will follow a different approach and assume that the excitation in 
states~$|c_j a\rangle$ is simply lost during scattering. This will, to a large 
extent, happen automatically due evolution of the excitation in 
states~$|c_j a\rangle$ under the conditions of stationary light that has a 
much higher dissipation rate compared to EIT. If this is not sufficient, an 
additional EIT retrieval sequence could be done before the second $\pi$-pulse to 
completely remove the part of the excitation still remaining in 
states~$|c_j a\rangle$. Mathematically, this results in 
$\phi_{A,\text{out},a}(t_A)=0$, so that the term proportional to 
$\sin^2(\varphi)$ in 
Eq.~\eqref{CJ_fidelity_computational_basis_final_imperfect_pi} vanishes.

Thus, the only difference between Eq.~\eqref{CJ_fidelity_computational_basis_final_imperfect_pi} and 
Eq.~\eqref{CJ_fidelity_computational_basis_final} is the overall factor 
$\cos^4(\varphi)$. This factor merely reflects that an excitation is lost if it was not successfully 
transferred by both $\pi$-pulses. Hence, the same overall factor appears in the 
success probability $P_\text{suc}$, and the conditional fidelity 
$F_\text{CJ,cond}=F_\text{CJ}/P_\text{suc}$ is unaffected by a wrong 
pulse area of the microwave pulse.

\section{Analytical fidelity of the $\Lambda$-type scheme}
\label{sec_analytical_fidelity}
Here, we derive
\begin{itemize} 
\item \cref{F_CJ_1,F_CJ_cond_1,F_CJ_2,F_CJ_cond_2} of the main 
article.
\item The error terms due to imperfectly balanced Sagnac interferrometers.
\item The bandwidth dependent correction to 
Eq.~\eqref{F_CJ_1} of the main article that is used in the discussion of 
the gate time.
\end{itemize}

From 
\cref{r_ensemble_N,t_ensemble_N,r_imp_ensemble_N,t_imp_ensemble_N} 
of the main article (including the $\Omega_0$ dependent terms, as shown in 
\cref{r_0_sm,t_0_sm,r_1_sm_Omega,t_1_sm_Omega}) and 
Eqs.~\eqref{R_0_definition_two_sided}~and~\eqref{R_0_definition_two_sided} 
above (assuming $r_{0+}=r_{0-}=r_0$, $t_{0+}=t_{0-}=t_0$, $r_{1+}=r_{1-}=r_1$, $t_{1+}=t_{1-}=t_1$),
we find the approximate reflection coefficients of the 
combined system of the atomic ensemble and the Sagnac interferometer. They are
\begin{align}
&\begin{aligned}
R_0
&=-\frac{1}{2}\del{r_0\del{e^{2ik_0 l_1}+e^{2ik_0 l_2}}-2t_0 e^{ik_0(l_1+l_2)}}\\
&\approx-\frac{1}{2}\del{\del{\frac{\Gamma_\text{1D}\Gamma'N}{16\Delta_\text{c}^2}
+\frac{\pi^2\Gamma'|\Omega_0|^2}{2\Delta_\text{c}^2\Gamma_\text{1D}N}}
\del{e^{2ik_0 l_1}+e^{2ik_0 l_2}}
-2\del{1-\frac{\Gamma_\text{1D}\Gamma'N}{16\Delta_\text{c}^2}
-\frac{\pi^2\Gamma'|\Omega_0|^2}{2\Delta_\text{c}^2\Gamma_\text{1D}N}}e^{ik_0(l_1+l_2)}},\\
\end{aligned}\\
&\begin{aligned}
R_{1,\text{s}}(\tilde{z})
&=\frac{1}{2}(R_1(\tilde{z})+R_1(1-\tilde{z}))\\
&=-\frac{1}{4}\del{\del{r_1(\tilde{z})+r_1(1-\tilde{z})}\del{e^{2ik_0 l_1}+e^{2ik_0 l_2}}
-2\del{t_1(\tilde{z})+t_1(1-\tilde{z})}e^{ik_0(l_1+l_2)}}\\
&\approx-\frac{1}{2}\left(\del{1-\frac{4\pi^2\Delta_\text{c}^2\Gamma'}
{\Gamma_\text{1D}^3N^2}
+\frac{32\pi^4 \Delta_\text{c}^2 \Gamma' |\Omega_0|^2}{\Gamma_\text{1D}^5 N^4}
-\frac{4\pi^4\Delta_\text{c}^2(2\Gamma_\text{1D}+\Gamma')}
{\Gamma_\text{1D}^3N^2}\left(\tilde{z}-\frac{1}{2}\right)^2}\del{e^{2ik_0 l_1}+e^{2ik_0 l_2}}\right.\\
&\left.-2\del{\frac{4\pi^2\Delta_\text{c}^2\Gamma'}
{\Gamma_\text{1D}^3N^2}
-\frac{32\pi^4 \Delta_\text{c}^2 \Gamma' |\Omega_0|^2}{\Gamma_\text{1D}^5 N^4}
+\frac{4\pi^4\Delta_\text{c}^2\Gamma'}
{\Gamma_\text{1D}^3N^2}\left(\tilde{z}-\frac{1}{2}\right)^2}e^{ik_0(l_1+l_2)}\right).
\end{aligned}
\end{align}
Note that in the symmetrized reflection coefficient $R_{1,\text{s}}(\tilde{z})$, 
the linear terms proportional to $\pm(\tilde{z}-1/2)$ (which are present in Eqs. 
\eqref{r_imp_ensemble_N} and \eqref{t_imp_ensemble_N} of the 
main article) cancel each other. Using the expression for the Gaussian spin wave given by 
Eq.~\eqref{gaussian_wavefunction} with $\tilde{\mu}=1/2$ we get
\begin{gather}\label{R_11_expression}
\begin{aligned}
R_{1,1}&\approx \int R_{1,\text{s}}(\tilde{z})|S(\tilde{z})|^2
\dif \tilde{z}\\
&\approx-\frac{1}{2}\left(\del{1-\frac{4\pi^2\Delta_\text{c}^2\Gamma'}
{\Gamma_\text{1D}^3N^2}
+\frac{32\pi^4 \Delta_\text{c}^2 \Gamma' |\Omega_0|^2}{\Gamma_\text{1D}^5 N^4}
-\frac{4\pi^4\Delta_\text{c}^2(2\Gamma_\text{1D}+\Gamma')}
{\Gamma_\text{1D}^3N^2}\tilde{\sigma}^2}\del{e^{2ik_0 l_1}+e^{2ik_0 l_2}}\right.\\
&\left.-2\del{\frac{4\pi^2\Delta_\text{c}^2\Gamma'}
{\Gamma_\text{1D}^3N^2}
-\frac{32\pi^4 \Delta_\text{c}^2 \Gamma' |\Omega_0|^2}{\Gamma_\text{1D}^5 N^4}
+\frac{4\pi^4\Delta_\text{c}^2\Gamma'}
{\Gamma_\text{1D}^3N^2}\tilde{\sigma}^2}e^{ik_0(l_1+l_2)}\right).
\end{aligned}
\end{gather}
Note that by using Eq.~\eqref{gaussian_wavefunction} instead of 
Eq.~\eqref{gaussian_wavefunction_arbitrary_time} with $t=L/(2v_\text{g})$, it 
may seem that we have neglected broadening of the spin wave during storage 
(propagation through the EIT medium for the length $L/2$). However, the 
definition of $\sigma$ is taken to be the width of the spin wave after 
storage, and we make the same choice for the numerical calculations (see the 
discussion below the 
Eqs.~\eqref{gaussian_phi_A_in_sigma}~and~\eqref{gaussian_phi_A_in_mu}). The 
dissipation (decreasing of the norm of 
Eq.~\eqref{gaussian_wavefunction_arbitrary_time} given by 
Eq.~\eqref{gaussian_wavefunction_norm_arbitrary_time}) is accounted for 
separately by the factors $\eta_\text{EIT}$ in 
Eq.~\eqref{F_CJ_general} of the main article (or 
Eq.~\eqref{CJ_fidelity_computational_basis_final}). Additionally, 
approximating $R_{1,1}=(1/\eta_\text{EIT})
\int \phi_{A,\text{out},0}^*(t)\phi_{A,\text{out},1}(t)\dif t$ by
$R_{1,1}\approx \int R_{1,\text{s}}(\tilde{z})|S(\tilde{z})|^2
\dif \tilde{z}$ neglects broadening and distortion of the 
pulse during retrieval. This approximation is valid, since the retrieved wave 
packet of photon~$A$ is expected to be changed in the same way by these 
effects, regardless of whether photon~$B$ was scattered off the ensemble 
between storage and retrieval or not. This assumption (and all the others 
required to derive the analytical expressions) is ultimately verified by 
Fig.~\ref{fig_cphase_fidelity_big_NAtoms} below.

\begin{figure}[hbt]
\begin{center}
\includegraphics{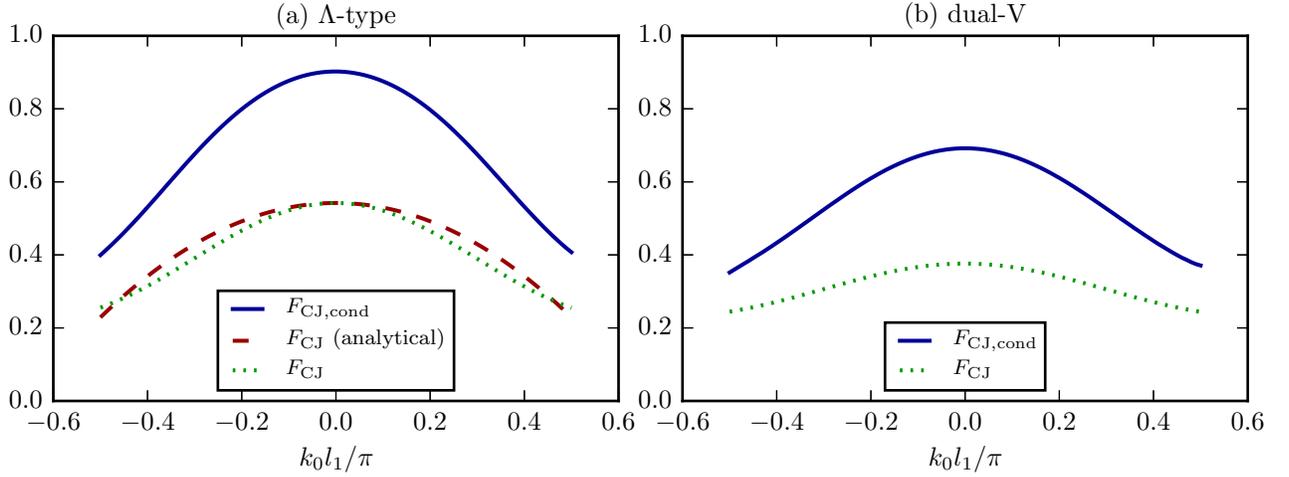}
\end{center}
\caption{(a) Unconditional and conditional Choi-Jamiolkowski 
fidelities for
the $\Lambda$\mbox{-}type scheme plotted as functions of the length of one of 
the arms of the Sagnac interferometer $l_1$ (modulo $2\pi/k_0$), while the 
other length is held constant $l_2=0$ (modulo $2\pi/k_0$) for a fixed number of atoms 
$N=10^4$, $\Gamma_\text{1D}/\Gamma=0.05$, $\Omega_0/\Gamma=1$, and interatomic 
distance $d=0.5\pi/k_0$. Under EIT (storage and retrieval), $\Omega(z)=\Omega_0$. 
Under stationary light (scattering), $\Omega(z)=\Omega_0\cos(k_0 z)$. The fidelities are plotted only 
for $t_\text{b}=1$. The analytical formula for $F_\text{CJ}$ is 
$F_\text{CJ}(k_0 l_1) = F_\text{CJ}(0)-(1/2-(5\pi\Gamma')/(8\Gamma_\text{1D}\sqrt{N}))(k_0 l_1)^2$.
(b) The same but for the dual-V scheme with $d=0.266\pi/k_0$.
Under stationary light (scattering), $\Omega_\pm(z)=\Omega_0e^{\pm ik_0 z}$.
For storage and retrieval both in (a) and (b), we use the discretized continuum storage and 
retrieval kernels discussed 
in Sec.~\ref{continuum_eit_storage_retrieval}.}
\label{fig_cphase_fidelity_kL1}
\end{figure}

To find the influence of an error in the alignment of the Sagnac 
interferometer, we set $k_0 l_2=0$ and expand around $k_0 l_1=0$. The 
opposite 
situation (set $k_0 l_1=0$ and expand around $k_0 l_2=0$) yields the same 
result due to symmetry of the expressions under the exchange of $l_1$ and $l_2$.
Using Eq.~\eqref{F_CJ_general} of the main article, an approximation 
for the unconditional CJ fidelity is
\begin{gather}\label{unconditional_CJ_fidelity}
\begin{aligned}
F_\text{CJ}
\approx\;&
1
-\epsilon_\text{b}
-\frac{\Gamma_\text{1D}\Gamma'N}{16\Delta_\text{c}^2}
-\frac{4\pi^2\Delta_\text{c}^2\Gamma'}{\Gamma_\text{1D}^3 N^2}
-\frac{\pi^2\Gamma'|\Omega_0|^2}{2\Delta_\text{c}^2\Gamma_\text{1D}N}
+\frac{32\pi^4 \Delta_\text{c}^2 \Gamma' |\Omega_0|^2}{\Gamma_\text{1D}^5 N^4}
-\frac{4\pi^4\Delta_\text{c}^2(\Gamma_\text{1D}+\Gamma')}
{\Gamma_\text{1D}^3 N^2}\tilde{\sigma}^2
-\frac{1}{2}\frac{\Gamma'}{N\Gamma_\text{1D}}\frac{1}{\tilde{\sigma}^2}\\
&+\del{-\frac{1}{2}+\frac{3}{8}\epsilon_\text{b}
+\frac{5\Gamma_\text{1D}\Gamma'N}{128\Delta_\text{c}^2}
+\frac{5\pi^2\Delta_\text{c}^2\Gamma'}{2\Gamma_\text{1D}^3 N^2}
+\frac{5\pi^2\Gamma'|\Omega_0|^2}{16\Delta_\text{c}^2 \Gamma_\text{1D} N}
-\frac{20\pi^4\Delta_\text{c}^2\Gamma'|\Omega_0|^2}{\Gamma_\text{1D}^5 N^4}
+\frac{7\pi^4\Delta_\text{c}^2\tilde{\sigma}^2}{2\Gamma_\text{1D}^2 N^2}
+\frac{5\pi^4\Delta_\text{c}^2\Gamma'\tilde{\sigma}^2}{2\Gamma_\text{1D}^3N^2}}
(k_0 l_1)^2
\end{aligned}
\end{gather}
where $\epsilon_\text{b}=1-t_\text{b}$, and all error terms (including 
$\epsilon_\text{b}$ and $k_0 l_1$) are assumed to be small. In 
Fig.~\ref{fig_cphase_fidelity_kL1}, we plot numerically and analytically 
calculated fidelities for the imperfectly balanced Sagnac interferometers.

\begin{figure}[hbt]
\begin{center}
\includegraphics{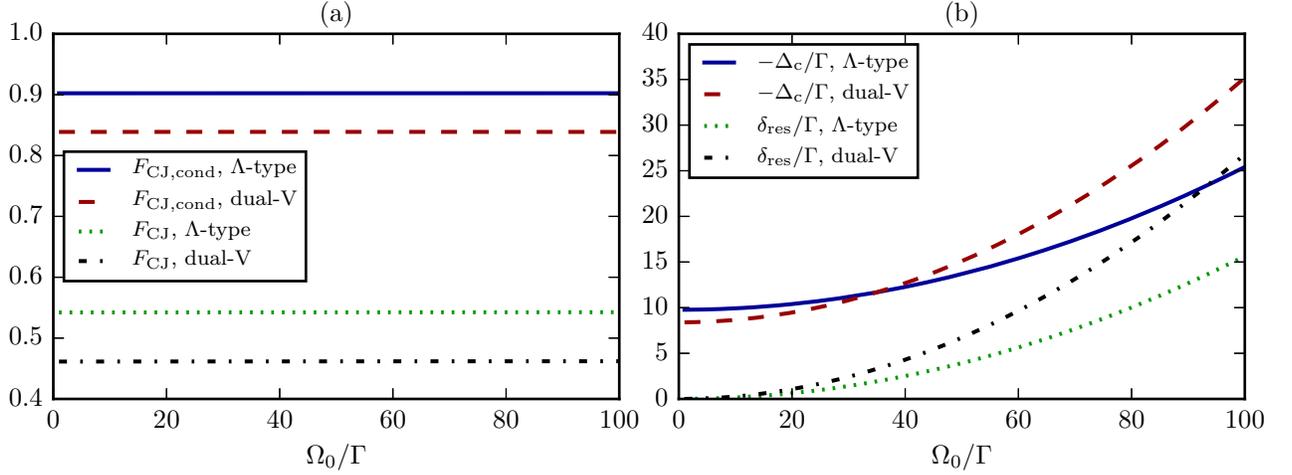}
\end{center}
\caption{(a) Unconditional and conditional Choi-Jamiolkowski 
fidelities for
the $\Lambda$\mbox{-}type and dual\mbox{-}V schemes plotted as functions of the Rabi frequency of 
the classical drive $\Omega_0$ for a fixed number of atoms 
$N=10^4$ and $\Gamma_\text{1D}/\Gamma=0.05$. Under EIT (storage and retrieval), $\Omega(z)=\Omega_0$. 
Under stationary light (scattering), $\Omega(z)=\Omega_0\cos(k_0 z)$ and 
$\Omega_\pm(z)=\Omega_0e^{\pm ik_0 z}$ for $\Lambda$\mbox{-}type and 
dual\mbox{-}V, respectively. The fidelities are plotted only 
for $t_\text{b}=1$. For $t_\text{b}<1$, they are likewise constant for 
the shown values of $\Omega_0/\Gamma$. The bandwidth of the scattered 
photon~$B$ is ignored in the calculations, and we take the Sagnac 
interferometer to be perfectly aligned ($k_0 l_1=k_0 l_2=0$).
For storage and retrieval, we use the discretized continuum storage and 
retrieval kernels discussed 
in Sec.~\ref{continuum_eit_storage_retrieval}. (b) Plot of the optimal
$\Delta_\text{c}$ and $\delta_\text{res}$ for the setups of (a).}
\label{fig_cphase_fidelity_OmegaScattering}
\end{figure}

In principle, if we want to optimize the above expression, we should optimize 
with respect to $\Delta_\text{c}$ and $\tilde{\sigma}$ simultaneously (we do this in the numerical calculations). Here, we use an 
approximate optimization procedure that ignores the fact that some 
error terms depend on the product of $\Delta_\text{c}$ and $\tilde{\sigma}$. As we 
will see, however, these error terms are smaller than the error terms that only 
depend on $\Delta_\text{c}$ for fixed $\Gamma_\text{1D}/\Gamma$ and large $N$. Therefore, we 
first optimize $F_\text{CJ}$ over $\Delta_\text{c}$ separately and then use 
the optimal value of $\Delta_\text{c}$ to optimize over $\tilde{\sigma}$. For 
small $|\Omega_0|$ and $k_0 l_1$, the optimal 
value of $\Delta_\text{c}$ is determined by the condition that the third and 
fourth error terms on the right hand side of Eq.~\eqref{unconditional_CJ_fidelity} are equal, i.e.
\begin{gather}
\frac{\Gamma_\text{1D}\Gamma' N}{16\Delta_\text{c}^2}
=\frac{4\pi^2\Delta_\text{c}^2\Gamma'}{\Gamma_\text{1D}^3 N^2}.
\end{gather}
This results in 
\begin{gather}\label{opt_Deltac}
\Delta_\text{c}^2=\frac{\Gamma_\text{1D}^2N^{3/2}}{8\pi}.
\end{gather}
Inserting this value of $\Delta_\text{c}$ into 
Eq.~\eqref{unconditional_CJ_fidelity} we obtain
\begin{align}\label{unconditional_CJ_fidelity_opt_Deltac}
F_\text{CJ}
\approx\;&
1
-\epsilon_\text{b}
-\frac{\pi\Gamma'}{\Gamma_\text{1D}\sqrt{N}}
-\frac{\pi^3(\Gamma_\text{1D}+\Gamma')}
{2\Gamma_\text{1D} \sqrt{N}}\tilde{\sigma}^2
-\frac{1}{2}\frac{\Gamma'}{N\Gamma_\text{1D}}\frac{1}{\tilde{\sigma}^2}
+\del{-\frac{1}{2}+\frac{3}{8}\epsilon_\text{b}
+\frac{5\Gamma'\pi}{8\Gamma_\text{1D}\sqrt{N}}
+\frac{\pi^3(7\Gamma_\text{1D}+5\Gamma')}{16\Gamma_\text{1D}\sqrt{N}}\tilde{\sigma}^2}
(k_0 l_1)^2.
\end{align}
Note that the $\Omega_0$ dependent terms cancel. In principle, we would need 
to look at higher order $\Omega_0$ dependent terms to estimate the error from 
having strong driving (big Rabi frequency $\Omega_0$). For simplicity, however, we restrict ourselves to a numerical investigation. As shown in Fig.~\ref{fig_cphase_fidelity_OmegaScattering}, even higher order 
$\Omega_0$ dependent terms can be canceled by adjusting the value of 
$\Delta_\text{c}$. The result is that both the unconditional and conditional fidelities (ignoring the bandwidth 
of the scattered photon~$B$) are independent of the 
value of $\Omega_0$ within a wide range. Additionally, 
Fig.~\ref{fig_cphase_fidelity_OmegaScattering} shows that the correction to 
the optimal $\Delta_\text{c}$ is negligible for $\Omega_0$ up to $30\Gamma$, so that we 
can safely use $\Omega_0=10\Gamma$ for the discussion of gate time 
in the main article 
for the parameters $\Gamma_\text{1D}/\Gamma=0.05$ and $N=10^4$, while keeping
$\Delta_\text{c}$ given by Eq.~\eqref{opt_Deltac} that is optimal for weak classical drives.
However, if stronger drive fields are available, it should be 
possible to perform even faster gate operation, provided that there are no other 
limitations to the strength of the classical drive (e.g. the requirements on frequency 
separation for the dual-color scheme discussed in 
Sec.~\ref{sec_rb_implementation} above).

We use Eq.~\eqref{unconditional_CJ_fidelity_opt_Deltac} to optimize over $\tilde{\sigma}$. The optimal $\tilde{\sigma}$ 
is obtained when 
\begin{gather}
\frac{\pi^3(\Gamma_\text{1D}+\Gamma')}
{2\Gamma_\text{1D} \sqrt{N}}\tilde{\sigma}^2
=\frac{1}{2}\frac{\Gamma'}{N\Gamma_\text{1D}}\frac{1}{\tilde{\sigma}^2}.
\end{gather}
From this condition we get
\begin{gather}\label{opt_sigma}
\tilde{\sigma}^2=\frac{1}{\pi^{3/2}N^{1/4}}\sqrt{\frac{\Gamma'}{\Gamma_\text{1D}+\Gamma'}}
\end{gather}
and with this value of $\tilde{\sigma}$, 
Eq.~\eqref{unconditional_CJ_fidelity_opt_Deltac} becomes
\begin{align}\label{unconditional_CJ_fidelity_opt_Deltac_sigma}
F_\text{CJ}
\approx\;&
1
-\epsilon_\text{b}
-\frac{\pi\Gamma'}{\Gamma_\text{1D}\sqrt{N}}
-\frac{\pi^{3/2}\sqrt{\Gamma_\text{1D}+\Gamma'}\sqrt{\Gamma'}}
{\Gamma_\text{1D} N^{3/4}}
+\del{-\frac{1}{2}+\frac{3}{8}\epsilon_\text{b}
+\frac{5\Gamma'\pi}{8\Gamma_\text{1D}\sqrt{N}}
+\frac{\pi^{3/2}(7\Gamma_\text{1D}+5\Gamma')}{16\Gamma_\text{1D}N^{3/4}}\sqrt{\frac{\Gamma'}{\Gamma_\text{1D}+\Gamma'}}}
(k_0 l_1)^2.
\end{align}
To obtain Eqs. \eqref{F_CJ_1} and \eqref{F_CJ_2} 
of the main article, we set $\epsilon_\text{b}=0$ 
and $\epsilon_\text{b}=1-R_0
=(\Gamma_\text{1D}\Gamma'N)/(8\Delta_\text{c}^2)
=(\pi\Gamma')/(\Gamma_\text{1D}\sqrt{N})$, respectively.
We see that for $N\rightarrow\infty$, the terms proportional to $N^{-3/4}$ 
approach zero faster than the terms proportional to $N^{-1/2}$. 
Hence, we have omitted the former terms in the main article. Furthermore, the $k_0 l_1$ 
dependent terms are also omitted, since we discuss them 
separately at the end of the main article. For parameters where the gate has a 
good performance, the parenthesis multiplying $(k_0 l_1)^2$ in Eq.~\eqref{unconditional_CJ_fidelity_opt_Deltac_sigma} is dominated by 
the constant term $-1/2$. In the main article, the total error
$Å\sim(k_0l)^2$ was quoted, where we have omitted the technical detail about 
two different fluctuating lengths ($l_1$ and $l_2$) and used $l$ to refer to 
both of them.

Next, we calculate the conditional fidelity $F_\text{CJ,cond}$. 
Here, we neglect the $\Omega_0$ dependent terms in 
\cref{r_0_sm,t_0_sm,r_1_sm_Omega,t_1_sm_Omega}, since we expect them to cancel 
in the same way as for the unconditional fidelity according to the 
numerical calculations in Fig.~\ref{fig_cphase_fidelity_OmegaScattering}.
The imperfection of the 
Sagnac interferometer alignment is also neglected, since the error terms are expected to be roughly the same 
as for the unconditional fidelity according to the numerical calculations in
Fig.~\ref{fig_cphase_fidelity_kL1}.
Since the conditional fidelity is 
given by $F_\text{CJ,cond}=F_\text{CJ}/P_\text{suc}$, the expansion of the ratio 
will contain higher order error terms than the expansion of the 
unconditional fidelity $F_\text{CJ}$. Hence, we need an expansion of 
$F_\text{CJ}$ with more terms than in Eq.~\eqref{unconditional_CJ_fidelity}. 
Including the second order terms and dividing out $\eta_\text{EIT}$ (since it 
gets canceled in $F_\text{CJ,cond}$), we get
\begin{gather}\label{unconditional_CJ_fidelity_second_order}
\begin{aligned}
\frac{F_\text{CJ}}{\eta_\text{EIT}}
\approx\;&
1
-\epsilon_\text{b}
-\frac{\Gamma_\text{1D}\Gamma'N}{16\Delta_\text{c}^2}
-\frac{4\pi^2\Delta_\text{c}^2\Gamma'}{\Gamma_\text{1D}^3 N^2}
-\frac{4\pi^4\Delta_\text{c}^2(\Gamma_\text{1D}+\Gamma')}
{\Gamma_\text{1D}^3 N^2}\tilde{\sigma}^2\\
&+\del{
\frac{1}{2}\epsilon_\text{b}
+\frac{\Gamma_\text{1D}\Gamma'N}{32\Delta_\text{c}^2}
+\frac{2\pi^2\Delta_\text{c}^2\Gamma'}{\Gamma_\text{1D}^3 N^2}
+\frac{2\pi^4\Delta_\text{c}^2(\Gamma_\text{1D}+\Gamma')}
{\Gamma_\text{1D}^3 N^2}\tilde{\sigma}^2
}^2
\end{aligned}
\end{gather}

Using Eq.~\eqref{P_suc_general}, we also get the success probability
\begin{align}
\frac{P_\text{suc}}{\eta_\text{EIT}}
=\frac{1}{4}\del{2(1-\epsilon_\text{b})^2
+\del{1-\frac{\Gamma_\text{1D}\Gamma'N}{8\Delta_\text{c}^2}}^2
+R_{1,2}},
\end{align}
where
\begin{gather}
\begin{aligned}
R_{1,2}\approx\;& \int |R_{1,\text{s}}(\tilde{z})|^2|S(\tilde{z})|^2
\dif \tilde{z}\\
=\;&1
-\frac{16\pi^2\Delta_\text{c}^2\Gamma'}
{\Gamma_\text{1D}^3N^2}
-\frac{16\pi^4\Delta_\text{c}^2(\Gamma_\text{1D}+\Gamma')}
{\Gamma_\text{1D}^3N^2}\tilde{\sigma}^2
+\del{\frac{8\pi^2\Delta_\text{c}^2\Gamma'}
{\Gamma_\text{1D}^3N^2}}^2
+\del{\frac{8\pi^4\Delta_\text{c}^2(\Gamma_\text{1D}+\Gamma')}
{\Gamma_\text{1D}^3N^2}}\del{3\tilde{\sigma}^4}\\
&+2\del{\frac{8\pi^2\Delta_\text{c}^2\Gamma'}
{\Gamma_\text{1D}^3N^2}}
\del{\frac{8\pi^4\Delta_\text{c}^2(\Gamma_\text{1D}+\Gamma')}
{\Gamma_\text{1D}^3N^2}}\tilde{\sigma}^2.
\end{aligned}
\end{gather}

Using the above expressions, the conditional fidelity can be written
\begin{gather}
F_\text{CJ,cond}
\approx 1-\epsilon_\text{cond,1}-\epsilon_\text{cond,2},
\end{gather}
where
\begin{align}
&\epsilon_\text{cond,1}
=\frac{1}{4}\del{
2\epsilon_\text{b}^2
+\del{\frac{\Gamma_\text{1D}\Gamma'N}{8\Delta_\text{c}^2}}^2
+\del{\frac{8\pi^2\Delta_\text{c}^2\Gamma'}
{\Gamma_\text{1D}^3N^2}}^2}
-\frac{1}{16}\del{
2\epsilon_\text{b}
+\frac{\Gamma_\text{1D}\Gamma'N}{8\Delta_\text{c}^2}
+\frac{8\pi^2\Delta_\text{c}^2\Gamma'}{\Gamma_\text{1D}^3 N^2}
}^2,\\
&\epsilon_\text{cond,2}=\frac{44\pi^8\Delta_\text{c}^4(\Gamma_\text{1D}+\Gamma')^2}{\Gamma_\text{1D}^6N^4}\tilde{\sigma}^4.
\end{align}
Using the $\Delta_\text{c}$ from Eq.~\eqref{opt_Deltac}, we get
\begin{align}
&\epsilon_\text{cond,1}
=\frac{1}{2}\del{
\epsilon_\text{b}^2
+\del{\frac{\pi\Gamma'}{\Gamma_\text{1D}\sqrt{N}}}^2
}
-\frac{1}{4}\del{
\epsilon_\text{b}
+\frac{\pi\Gamma'}{\Gamma_\text{1D}\sqrt{N}}
}^2,\\
&\epsilon_\text{cond,2}=\frac{11\pi^6(\Gamma_\text{1D}+\Gamma')^2}{\Gamma_\text{1D}^2N}\tilde{\sigma}^4.
\end{align}
If we choose $\epsilon_\text{b}=0$, $\epsilon_\text{cond,1}$ is the dominant 
error term with the value
\begin{gather}
\epsilon_\text{cond,1}
=\frac{1}{4}\del{\frac{\pi\Gamma'}{\Gamma_\text{1D}\sqrt{N}}}^2,
\end{gather}
and $\epsilon_\text{cond,2}$ can be neglected. If we choose 
$\epsilon_\text{b}=(\pi\Gamma')/(\Gamma_\text{1D}\sqrt{N})$, we get 
$\epsilon_\text{cond,1}=0$, and we need to keep $\epsilon_\text{cond,2}$. The 
value of $\epsilon_\text{cond,2}$ depends on the width $\tilde{\sigma}$ of the 
stored Gaussian spin wave. For simplicity, we use the value of $\tilde{\sigma}$ given 
by Eq.~\eqref{opt_sigma}, which makes the unconditional fidelity maximal. 
With this choice, the conditional fidelity is given by Eq.~\eqref{F_CJ_cond_2} 
of the main article.
A comparison of the analytical formulas and the numerical results for 
$F_\text{CJ}$ and $F_\text{CJ,cond}$ is shown in Fig.~\ref{fig_cphase_fidelity_big_NAtoms}.

\begin{figure}[hbt]
\begin{center}
\includegraphics{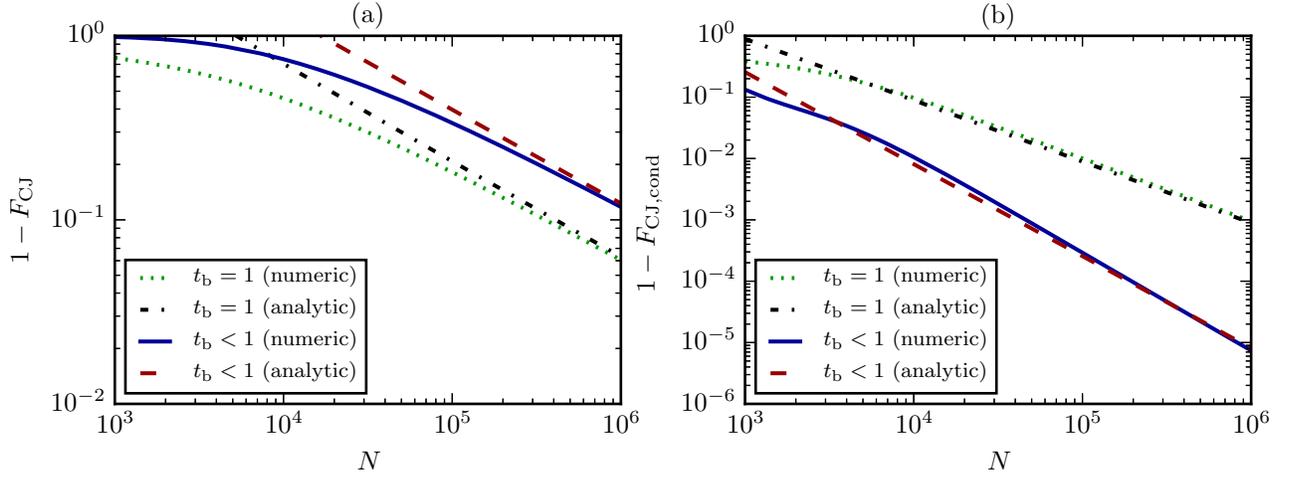}
\end{center}
\caption{Comparison of (a) unconditional and (b) conditional Choi-Jamiolkowski fidelities for 
the $\Lambda$\mbox{-}type scheme plotted as functions of the number of atoms 
$N$ with fixed $\Gamma_\text{1D}/\Gamma=0.05$ and $\Omega_0/\Gamma=1$. Under EIT (storage and retrieval), $\Omega(z)=\Omega_0$. 
Under stationary light (scattering), $\Omega(z)=\Omega_0\cos(k_0 z)$. Dotted green curves are the 
numerically calculated fidelities with $t_\text{b}=1$. Dash-dotted black 
curves are the approximate analytical results given by Eq.~\eqref{F_CJ_1} (unconditional) 
and Eq.~\eqref{F_CJ_cond_1} (conditional) of the main article. Solid blue curves are the numerically 
calculated fidelities with $t_\text{b}$ chosen such that the entanglement swap 
fidelity (which is approximately equal to the conditional Choi-Jamiolkowski fidelity as 
shown in Fig.~\ref{fig_cphase_fidelity_F_swap}) is maximal. The dashed red curves are the approximate analytical 
results given by Eq.~\eqref{F_CJ_2} (unconditional) and 
Eq.~\eqref{F_CJ_cond_2} (conditional) of the main article.
For storage and retrieval in the numerical calculations, we use the 
discretized continuum storage and retrieval kernels discussed in 
Sec.~\ref{continuum_eit_storage_retrieval}.}
\label{fig_cphase_fidelity_big_NAtoms}
\end{figure}

Next, we derive the correction to the unconditional fidelity due to non-zero 
bandwidth of the scattered photon~$B$. The general expression for the 
fidelity is given by Eq.~\eqref{CJ_fidelity_computational_basis_final}. As 
discussed in the main article, we ignore the non-zero bandwidth in the last 
term since the variation of $r_1$ and $t_1$ with frequency around 
the resonance detuning $\delta_\text{res}$ is smaller than variation of $r_0$ 
and $t_0$. Hence, we approximate
\begin{gather}
\frac{1}{\eta_\text{EIT}}
\iint \phi_{A,\text{out},0}^*(t_A)\phi_{A,\text{out},1}(t_A,\delta_B)
|\phi_B(\delta_B)|^2\dif t_A\dif \delta_B
\approx
(1/\eta_\text{EIT})
\int \phi_{A,\text{out},0}^*(t)\phi_{A,\text{out},1}(t)\dif t
=R_{1,1}
\end{gather}
such that the expression for the fidelity becomes
\begin{gather}\label{CJ_fidelity_computational_basis_final_approx}
F_\text{CJ}=
\frac{\eta_\text{EIT}}{16} \envert{
2t_\text{b}
+\int R_0(\delta_B)|\phi_B(\delta_B)|^2\dif \delta_B
-R_{1,1}
}^2.
\end{gather}
Using Eqs. \eqref{r_0_expansion_around_delta_res} and \eqref{delta_res_width} 
together with the expressions $t_0\approx 1-r_0$ and $R_0=-(r_0-t_0)$ and defining the 
spectral width of photon~$B$ by 
$\sigma_B^2=\int (\delta-\delta_\text{res})^2 |\phi_B(\delta_B)|^2\dif \delta_B$, we 
get (using $k_0 l_1 = k_0 l_2=0$ for brevity)
\begin{gather}
\int R_0(\delta)|\phi_B(\delta_B)|^2\dif \delta_B
\approx R_0(\delta_\text{res})-(4/w^2)\sigma_B^2
\approx 1-\frac{\Gamma_\text{1D}\Gamma'N}{8\Delta_\text{c}^2}
-\frac{\pi^2\Gamma'|\Omega_0|^2}{\Delta_\text{c}^2\Gamma_\text{1D}N}
-\frac{\Gamma_\text{1D}^6N^6}{512\Delta_\text{c}^4|\Omega_0|^4\pi^4}\sigma_B^2.
\end{gather}
Using the optimal $\Delta_\text{c}$ from Eq.~\eqref{opt_Deltac}, this becomes
\begin{gather}\label{R_0_expression_sigma_B_opt_Deltac}
\int R_0(\delta)|\phi_B(\delta_B)|^2\dif \delta_B
\approx 1-\frac{\pi\Gamma'}{\Gamma_\text{1D}\sqrt{N}}
-\frac{8\pi^3\Gamma'|\Omega_0|^2}{\Gamma_\text{1D}^3N^{5/2}}
-\frac{\Gamma_\text{1D}^2N^3}{8|\Omega_0|^4\pi^2}\sigma_B^2.
\end{gather}
Using the same optimal $\Delta_\text{c}$, Eq.~\eqref{R_11_expression} becomes
\begin{gather}\label{R_11_expression_opt_Deltac}
R_{1,1}\approx \int R_{1,\text{s}}(\tilde{z})|S(\tilde{z})|^2
\dif \tilde{z}
=-\del{1
-\frac{\pi\Gamma'}{\Gamma_\text{1D}\sqrt{N}}
+\frac{8\pi^3\Gamma'|\Omega_0|^2}{\Gamma_\text{1D}^3N^{5/2}}
-\frac{\pi^3(\Gamma_\text{1D}+\Gamma')}
{\Gamma_\text{1D} \sqrt{N}}\tilde{\sigma}^2}
\end{gather}
We see that the terms in 
Eqs.~\eqref{R_0_expression_sigma_B_opt_Deltac}~and~\eqref{R_11_expression_opt_Deltac}
that are proportional to $|\Omega_0|^2$ cancel upon insertion into 
Eq.~\eqref{CJ_fidelity_computational_basis_final_approx} (as already noted 
above for the case with $\sigma_B=0$). With the optimal value of $\tilde{\sigma}$~\eqref{opt_sigma}, 
we get
\begin{align}\label{unconditional_CJ_fidelity_opt_Deltac_sigma_bandwidth}
F_\text{CJ}
\approx\;&
1
-\epsilon_\text{b}
-\frac{\pi\Gamma'}{\Gamma_\text{1D}\sqrt{N}}
-\frac{\pi^{3/2}\sqrt{\Gamma_\text{1D}+\Gamma'}\sqrt{\Gamma'}}
{\Gamma_\text{1D} N^{3/4}}
-\frac{\Gamma_\text{1D}^2N^3}{16|\Omega_0|^4\pi^2}\sigma_B^2.
\end{align}
Compared to Eq.~\eqref{unconditional_CJ_fidelity_opt_Deltac_sigma}, there 
is an extra error term that depends on $\sigma_B$ (and the $k_0l_1$ dependent 
term is omitted).

To find the allowed value of $\sigma_B$, we require the error in 
Eq.~\eqref{unconditional_CJ_fidelity_opt_Deltac_sigma} due to non-zero 
$\sigma_B$ to be equal to the biggest $\sigma_B$ independent one, i.e.
\begin{gather}
\frac{\pi\Gamma'}{\Gamma_\text{1D}\sqrt{N}}
=\frac{\Gamma_\text{1D}^2N^3}{16|\Omega_0|^4\pi^2}\sigma_B^2.
\end{gather}
This gives
\begin{gather}\label{scattering_time}
\frac{1}{\sigma_B}=\frac{\Gamma_\text{1D}^{3/2}N^{7/4}}{4\pi^{3/2}\sqrt{\Gamma'}|\Omega_0|^2}.
\end{gather}
This is the expression used to determine the scattering time in the main article.

\section{Dependence on positions of the atoms}
\begin{figure}[hbt]
\begin{center}
\includegraphics{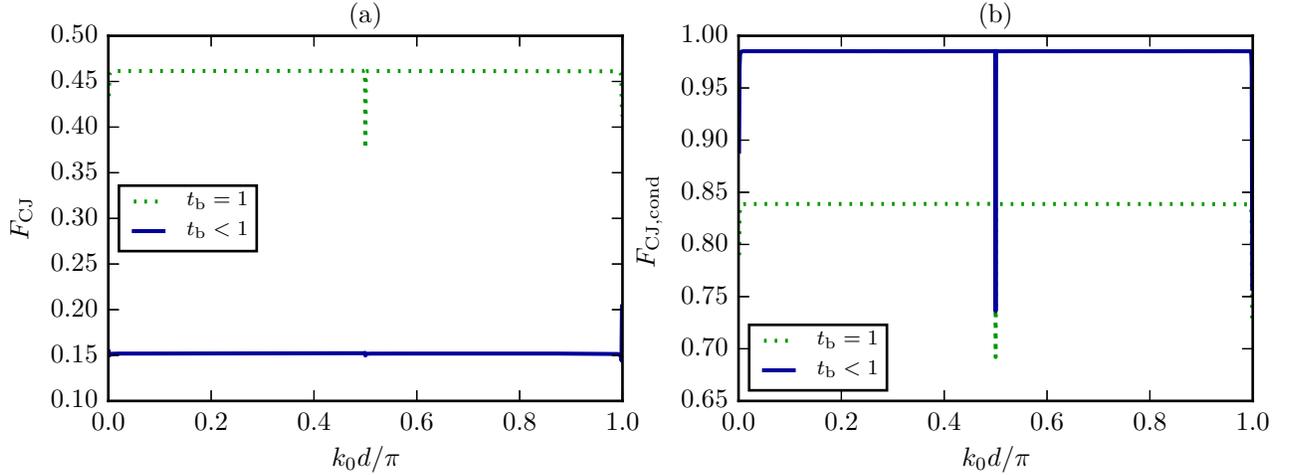}
\end{center}
\caption{(a) Unconditional and (b) conditional Choi-Jamiolkowski 
fidelities for
the dual\mbox{-}V scheme plotted as functions of the interatomic spacing $d$ 
times the photon wave vector $k_0$ for fixed number of atoms 
$N=10^4$, $\Gamma_\text{1D}/\Gamma=0.05$, and $\Omega_0/\Gamma=1$. Under EIT (storage and retrieval), $\Omega(z)=\Omega_0$. 
Under stationary light (scattering), $\Omega_\pm(z)=\Omega_0e^{\pm ik_0 z}$. 
For storage and retrieval, we use the discretized continuum storage and 
retrieval kernels discussed
in Sec.~\ref{continuum_eit_storage_retrieval}.}
\label{fig_cphase_fidelity_kd}
\end{figure}

While the $\Lambda$-type scheme is highly sensitive to the exact placement of 
the atoms, the dual-V scheme is much less sensitive. To verify this, we 
numerically evaluate the gate performance for various placements of the atoms. 
First, we investigate the dependence for regularly placed dual-V atoms with 
different interatomic spacings $d$. In Fig.~\ref{fig_cphase_fidelity_kd}, we 
see that for $k_0 d$ different from multiples of $\pi/2$, both the conditional 
and unconditional fidelities are approximately constant.

\begin{figure}[t]
\begin{center}
\includegraphics{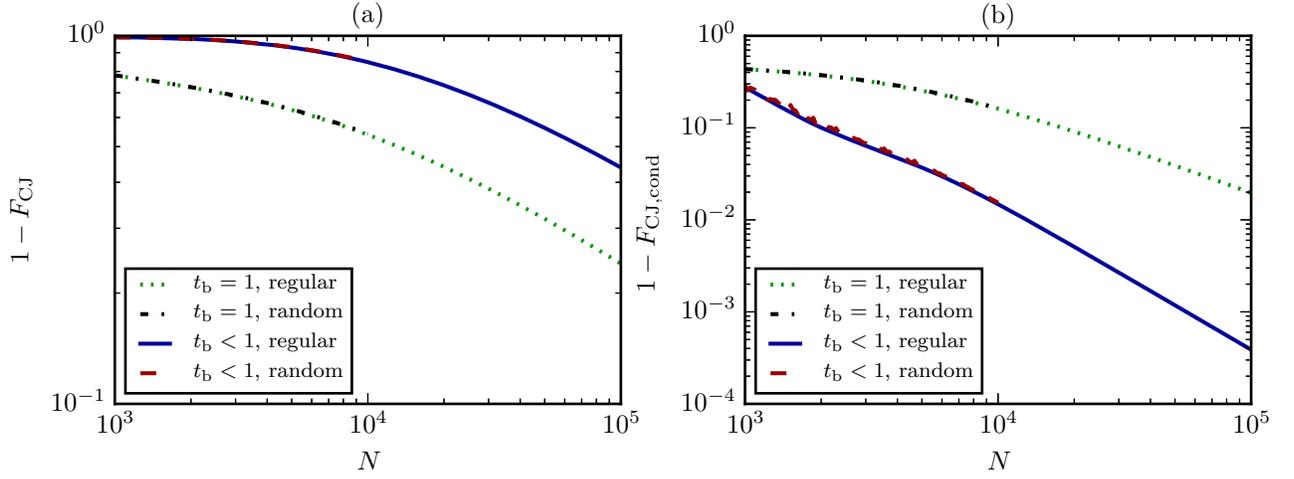}
\end{center}
\caption{Comparison of (a) unconditional and (b) conditional Choi-Jamiolkowski 
fidelities for
the dual\mbox{-}V scheme with different placement of the atoms (regular or random) 
plotted as functions of the number of atoms~$N$ with fixed 
$\Gamma_\text{1D}/\Gamma=0.05$ and $\Omega_0/\Gamma=1$. Under EIT (storage and retrieval), $\Omega(z)=\Omega_0$. 
Under stationary light (scattering), $\Omega_\pm(z)=\Omega_0e^{\pm ik_0 z}$. The regularly placed atoms have 
have positions $z_j=jd$ for $d=0.266\pi/k_0$ and $0\leq j \leq N-1$. The 
positions of the randomly
placed atoms are chosen from the uniform distribution over the whole ensemble 
and then sorted in increasing order. For random placement, we average over 
100 ensemble realizations.
For storage and retrieval, we use the discretized continuum storage and 
retrieval kernels discussed
in Sec.~\ref{continuum_eit_storage_retrieval}. 
}
\label{fig_cphase_fidelity_random_placement}
\end{figure}

Second, we consider randomly placed dual-V atoms. In 
Fig.~\ref{fig_cphase_fidelity_random_placement}, the fidelities with regular 
and random placement are seen to have almost the same behavior. For larger 
values of $\Gamma_\text{1D}/\Gamma$, the agreement between the two different 
placements becomes less good. They still have qualitatively the same behavior, 
but the random placement has slightly worse performance.

\end{document}